\documentclass[12pt]{article}
\usepackage{amsmath}
\usepackage{amsfonts}
\usepackage{amssymb}
\usepackage{mathrsfs}
\usepackage{eucal}
\usepackage{graphicx}

\newif{\ifcomentarios}
\comentariosfalse

\newtheorem{theorem}{Theorem}

\newtheorem{claim}[theorem]{Claim}

\newtheorem{corollary}[theorem]{Corollary}

\newtheorem{definition}[theorem]{Definition}

\newtheorem{lemma}[theorem]{Lemma}

\newtheorem{remark}[theorem]{Remark}

\newtheorem{proposition}[theorem]{Proposition}

\renewcommand{\mathcal}{\mathscr}
\renewcommand{\mathbf}{\boldsymbol}

\begin{document}

\author{David Brydges\thanks{%
Email: \texttt{db5d@math.ubc.ca}} \\
%EndAName
Departament of Mathematics\\
University of British Columbia \\
Vancouver, B.C., Canada V6T 1Y4 \and ~Domingos H. U. Marchetti\thanks{%
Email: \texttt{marchett@if.usp.br}} \\
%EndAName
Instituto de F\'{\i}sica\\
Universidade de S\~{a}o Paulo\\
Caixa Postal 66318\\
05314-970 S\~{a}o Paulo, SP, Brasil}
\title{On the Virial Series for a Gas of Particles with Uniformly Repulsive
Pairwise Interaction}
\date{}
\maketitle

\begin{abstract}
The pressure of a gas of particles with a uniformly repulsive pair
interaction in a finite container is shown to satisfy (exactly as a formal
object) a \textquotedblleft viscous\textquotedblright\ Hamilton-Jacobi (H-J)
equation whose solution in power series is recursively given by the
variation of constants formula. We investigate the solution of the H-J and
of its Legendre transform equation by the Cauchy-Majorant method and provide
a lower bound to the radius of convergence on the virial series of the gas
which goes beyond the threshold established by Lagrange's inversion formula.
A comparison between the behavior of the Mayer and virial coefficients for
different regimes of repulsion intensity is also provided.
\end{abstract}

\section{Introduction: Motivations and Results}

\setcounter{equation}{0} \setcounter{theorem}{0}

\paragraph{Context}

We shall answer and bring to attention some questions regarding the
Kamerlingh Onnes virial series of a system of particles interacting with
two--body pairwise repulsive potentials. The model of hard spheres, despite
its apparent simplicity and usefulness in modeling fluids, has not been
solved except in the limit of infinitely many dimensions ($d=\infty $). In
this limit, the equation of state 
\begin{equation}
\beta P=\rho +\frac{1}{2}\rho ^{2}  \label{eq-state}
\end{equation}%
truncates at the second virial coefficient and the fact there is no
premonitory signs of a phase transition in the fluid phase is attributed to
a diminished importance of fluctuations which, according to the theory of
critical phenomena, suggests the presence of phase transition (not
manifested by its virial series (\ref{eq-state}) or by its Mayer series,
which has no singularity at physical values of activities).

The classical system of hard spheres -- as well as the Ford model\cite%
{Uhlenbeck-Ford}, that realizes the upper bound for the radius of
convergence of its Mayer series -- shows, by Pad\'{e} approximation analysis
of its virial series, no pressure maximum\cite{Aguillera-Navarro-et-al},
indicating the failure of the equation in the fluid phase to supply any
information about the freezing phase transition. For a recent discussion on
this and other related issues on phase transition of hard-spheres see
Clisby-McCoy \cite{Clisby-McCoy} and references therein.

As early as the study by Riddell and Uhlenbeck\cite{Riddell-Uhlenbeck}, the
problem of finding asymptotic behavior of the Mayer and virial coefficients $%
b_{n}$ and $B_{n}$ for high $n$ intended to address the (same) question: 
\textit{is there (for the gas of hard--spheres) a transition point }$\rho
^{\ast }$\textit{\ -- predicted to be smaller by a factor }$0.806\,$\textit{%
\ than the close--packing density }$\rho _{\mathrm{cp}}$\ (see e.g. \textit{%
\cite{Rogers})\ by Kirkwood-Monroe\cite{Kirkwood-Monroe}?} (we refer to Sec.
4 of \cite{Clisby-McCoy} for a review on numerical studies, approximate
equations and sceneries for the position of the leading singularities). This
question remains unanswered as far as the present knowledge on the radius of
convergence of the Mayer and virial series is concerned (see e.g. \cite%
{Ruelle}). Further asymptotic investigation of the $b_{n}$ and $B_{n}$,
carried by Uhlenbeck and Ford \cite{Uhlenbeck-Ford} for the so--called
Gaussian model, had not reached any definite answer although the
mathematical problems that came out from their study inspired the present
investigation.

In the present paper we study a gas of point particles interacting via a
uniformly repulsive pairwise potential, in a finite volume $\Lambda $ of
size $\left\vert \Lambda \right\vert =1/(2\varepsilon )$. We prove,
employing a system of equations satisfied by the Ursell functions\cite%
{Brydges-Kennedy}, that the pressure $p=p(t,\mu )$, as a function of\ $t$, a
parameter (\textquotedblleft inverse temperature\textquotedblright ) that
interpolates the ideal to the real gas, and the chemical potential $\mu $,
satisfies (exactly, as a formal object) a \textquotedblleft
viscous\textquotedblright\ Hamilton--Jacobi equation%
\begin{equation}
p_{t}+\varepsilon \left( p_{\mu \mu }-p_{\mu }\right) +\frac{1}{2}\left(
p_{\mu }\right) ^{2}=0  \label{pde}
\end{equation}%
with $p(0,\mu )=e^{\mu }$. The repulsive interaction, expressed by the
\textquotedblleft wrong\textquotedblright\ sign in front of the Laplacian,
avoids the collapse of particles into a single point (equilibrium stability).

We are looking for solutions of (\ref{pde}) in the form of a power series in
the activity $z=\allowbreak e^{\mu }$:%
\begin{equation}
p(t,\mu )=\sum_{n=1}^{\infty }b_{n}(t)z^{n}~:=\wp (t,z)  \label{ptmu}
\end{equation}%
which are regular at $\varepsilon =0$ (infinite volume limit). The
so--called Mayer solution exists globally, i.e., for all $t\geq 0$ as a
holomorphic function in a neighborhood of $z=0$, uniformly in $\varepsilon $
(see Theorem \ref{convergence} for precise statement, which includes (\ref%
{pde}) among other equations and it is not intended to give optimal
estimation on the convergence radius). A refined version, Theorem \ref%
{bestmate}, optimizes the estimate on the radius of convergence up to one
for which (\ref{ptmu}), being the generating function for labelled
enumeration of simply connected Mayer graphs, is majorized by the
corresponding sum over labelled trees.

\paragraph{Motivations}

The convergence of Mayer series for the class of regular and stable
potentials (including nonnegative ones)\cite[Chapt. 4]{Ruelle}, is affected
by the presence, in its majorant, of a (movable and nonphysical) singularity
in the negative real line of the complex $z$--plane (see e.g. \cite%
{Guidi-Marchetti} and Appendix \ref{CVSO}) which would be inherited by the
virial series 
\begin{equation}
P(t,\rho )=\sum_{n=1}^{\infty }B_{n}\rho ^{n}~  \label{Pp}
\end{equation}%
whether Lagrange's inversion formula is applied to estimate the pressure: $%
\wp (t,Z(t,\rho ))$, where $Z=Z(t,\rho )$ solves $z\wp _{z}(t,z)=\rho $ for $%
z$ (see Appendix \ref{CVSO}), as in the classic paper by Lebowitz--Penrose 
\cite{Lebowitz-Penrose}.

Alternatively, the virial coefficients $\left( B_{n}\right) _{n\geq 1}$ (or
the \textbf{irreducible cluster integrals} $\left( \beta _{n}\right) _{n\geq
1}$, which is related to by 
\begin{equation}
B_{n}=-\frac{n-1}{n}\beta _{n-1}  \label{B}
\end{equation}%
for $n\geq 2$, with $B_{1}=1$) may be obtained from the Helmholtz free
energy density (see Proposition \ref{legendre}): 
\begin{eqnarray*}
F(t,\rho ) &=&\sup_{\mu }\left( \mu \rho -p(t,\mu )\right) \\
&=&\rho \log \rho -\rho -\mathfrak{\beta }(t,\rho )\ ,
\end{eqnarray*}%
whose derivative $\mathfrak{\beta }_{\rho }(t,\rho )$, excluding the free
energy of ideal gas $F(0,\rho )=\rho \log \rho -\rho \equiv F^{\mathrm{ideal}%
}$, generates the labelled \textquotedblleft enumeration\textquotedblright\
(total weight of species) of irreducible (2--connected) Mayer graphs: 
\begin{equation}
\varphi (t,\rho )=\sum_{n=1}^{\infty }\beta _{n}(t)\rho ^{n}~.  \label{sol1}
\end{equation}%
Consequently, with (\ref{B}), the radii of convergence $\mathcal{R}_{%
\mathfrak{\beta }}(t)$, $\mathcal{R}_{P}(t)$ and $\mathcal{R}_{\varphi }(t)$
of the power series $\mathfrak{\beta }(t,\rho )$, $P(t,\rho )$ and $\varphi
(t,\rho )$ about $\rho =0$ are all the same. The Helmholtz free energy
(extracted the ideal contribution) $F-F^{\mathrm{ideal}}=-\mathfrak{\beta }$
has been recently addressed by cluster expansion \cite%
{Pulvirenti-Tsagkarogiannis} and Morais--Procacci\cite{Morais-Procacci} have
shown, by means of an expression already known by Mayer (see Appendix \ref%
{ICI} for a derivation using Lagrange's inversion formula), that $\mathcal{R}%
_{\mathfrak{\beta }}$ satisfies Lebowitz--Penrose's lower bound on the
radius of convergence of the virial series. It thus seems quite opportune to
inquire whether the referred singularity on the Mayer series could somehow
be prevented.

The present work has been motivated by the following open problem:

\noindent \textit{Is there a system of interacting particles for which }$%
\mathfrak{\beta }$\textit{, }$P$\textit{\ and/or }$\varphi $\textit{\ can be
directly assessed, Lagrange's inversion formula be avoided and
Lebowitz-Penrose's lower bound on }$\mathcal{R}_{P}(t)$\textit{\ be
improved? }

\paragraph{Results}

We prove (Proposition \ref{PDEs}.c) that if $p(t,\mu )$ satisfies the
initial value problem (\ref{pde}) then $\varphi (t,\rho )$ satisfies 
\begin{equation}
\varphi _{t}+\rho +\varepsilon \left( \frac{1+\rho ^{2}\varphi _{\rho \rho }%
}{(1-\rho \varphi _{\rho })^{2}}-1\right) =0  \label{gfeq}
\end{equation}%
with $\varphi (0,\rho )=0$ and an affirmative answer to the
\textquotedblleft model\textquotedblright\ described by equation (\ref{gfeq}%
) has been provided. We establish (Theorem \ref{unique}), via the
Cauchy--Majorant method, a lower bound $\mathcal{R}_{\Phi }(t)$ for $%
\mathcal{R}_{\varphi }(t)$, where $\Phi (t,\rho )=\displaystyle\sum_{n\geq
1}\Phi _{n}(t)\rho ^{n}$ majorizes $\varphi (t,\rho )$ in the sense that $%
\left\vert \beta _{n}(t)\right\vert \leq \Phi _{n}(t)$ holds for all $n\in 
\mathbb{N}$ and $t\in \mathbb{R}_{+}$, uniformly in $\varepsilon $:

\begin{equation}
\lambda \mathcal{R}_{\varphi }(t)\geq \lambda \mathcal{R}_{\Phi
}(t)=\varkappa  \label{RR}
\end{equation}%
where $\varkappa =$ $\varkappa (\eta )$ with $\eta =e^{-2\varepsilon t}$ is
given by (\ref{varkappa}) and $\lambda =\left( 1-\eta \right) /(2\varepsilon
)$ is the $L_{1}$--norm of the Mayer $f$--function $f=e^{-\phi }-1$ with $%
\phi $ the uniformly repulsive potential (i.e., $\lambda =2B_{2}$). The
curve defined by the r.h.s. of (\ref{RR}) stays above the Lebowitz-Penrose's
lower bound $\mathcal{R}_{P}(t)>0.144767/\lambda $ (eq. (3.11) of \cite%
{Lebowitz-Penrose} with $u=1$ and $B=\lambda $) for all $t$ while, for $%
\varepsilon t\leq 0.00538$, the pre--factor $\varkappa $ goes beyond the
threshold $0.278465$, established for nonnegative potentials\ $\mathcal{R}%
_{P}(t)>0.278465/\lambda $ (see \cite{Ruelle}, Theorem 4.3.2 \textit{et seq.}%
). $\varkappa $ as a function of $\eta $ is plotted in Fig. \ref{fig3}. By (%
\ref{RR}), $\mathcal{R}_{\varphi }(t)>0$ for all $(t,\varepsilon )\in 
\mathbb{R}_{+}\times \mathbb{R}_{+}$ and the $t$ and $\varepsilon
=1/(2\left\vert \Lambda \right\vert )$ for which $\varphi $ converges in the
domain below the hyperbole $\varepsilon t=0.00538$ represent the regime of
\textquotedblleft high temperatures\textquotedblright\ and/or large volumes.
The proof of Theorem \ref{unique} introduces some novelties with respect to
the traditional majorant method and involves technical difficulties (see
Proposition \ref{Integral}).

Stronger results can actually be proven for equation (\ref{pde}). It follows
from Lieb's inequalities (see e.g. \cite{Ruelle}, Sect. 4.5, and references
therein) that alternating sign property (a.s.p.): $(-1)^{n-1}b_{n}>0$ and
upper and lower bounds%
\begin{equation}
e^{-1}\leq \lambda \mathcal{R}_{\wp }\leq 1~  \label{lambdaR}
\end{equation}%
on the radius of convergence $\mathcal{R}_{\wp }$ of the Mayer series for
the pressure (or density), hold for any nonnegative potential. In Theorem %
\ref{bestmate} we prove, in addition, that \textbf{(i)} $\lambda \mathcal{R}%
_{\wp }$ is monotone increasing in $t$, \textbf{(ii)} the inequalities (\ref%
{lambdaR}) saturate in both sides at $t=0$ and $\infty $ and \textbf{(iii)} $%
\lim_{\varepsilon \rightarrow 0}\lambda \mathcal{R}_{\wp }=e^{-1}$ for any $%
0<t<\infty $.

Regarding equation (\ref{gfeq}), we rely on few additional results and
explicit computation using Mathematica. We prove (Theorem \ref{asympt}) that 
\begin{equation}
\frac{\beta _{n}}{\lambda ^{n}}=(-1)^{n+1}\varepsilon t(1+O(\varepsilon t))\
,\qquad \text{as\ }\varepsilon t\rightarrow 0  \label{betan}
\end{equation}%
for $n\geq 2$ ($\beta _{1}/\lambda \equiv -1$) and, by continuity, the $%
\beta _{n}$'s satisfy both a.s.p. and $\lambda \mathcal{R}_{\varphi }=1$ if $%
t\varepsilon $ is sufficiently small. For $t\varepsilon \ll 1$, $\lambda 
\mathcal{R}_{P}=1$ holds even though, as explained in Appendix \ref{CVS}, we
have $\lambda \mathcal{R}_{P}\geq W(e^{-1})=0.278465...$, by Lagrange's
inversion formula. On the other hand, in the limit of $t$ large, $%
\lim_{t\rightarrow \infty }n\beta _{n}/\lambda ^{n}=-1$ and, as the computer
calculations indicate, $\lambda \mathcal{R}_{\varphi }=\lambda \mathcal{R}%
_{P}=1$ is expected to hold for all $(t,\varepsilon )$.

\paragraph{Related issues}

There are two other issues regarding the virial coefficients of systems with
repulsive potentials that can be addressed by the present (mean field)
model. The ninth and tenth order virial coefficients for hard spheres have
been calculated (numerically) in dimensions $2\leq d\leq 8$ by Clisby--McCoy 
\cite{Clisby-McCoy}. Collecting and reviewing a great deal of information,
the authors found that even coefficients start to be negative when $d\geq 5$
and provided strong evidence that the leading singularity for the virial
series lies away from the positive real line. We compute the $\beta _{n}$ ($%
=-(n+1)B_{n+1}/n$) from an exact recursion relation satisfied for the
\textquotedblleft model\textquotedblright\ (\ref{gfeq}) and show that, as
function of $n$, they oscillate about the axis (Fig. \ref{fig7} depicts $%
n\beta _{n}/\lambda $, $n=1$, ..., $17$ for different values of $t$). The
calculation indicate that the period increases from $1$ (the alternating
sign behavior (\ref{betan})) to infinity as $t$ varies from $0$ to $\infty $
. Each $n\beta _{n}/\lambda $, as a function of $t$, oscillates wildly,
apparently uncorrelated from the others, until it reaches a maximum at $%
t_{n}^{\ast }$ and from there on it converges rapidly to $-1$ (see Fig. \ref%
{fig6}), indicating that the leading singularity of the virial series
remains away from the positive real line, at least for $t<t_{\infty }^{\ast
}=\lim_{n\rightarrow \infty }t_{n}^{\ast }$.\footnote{%
Other models whose leading singularities are out of the real line include
the Gaussian model already mentioned (see \cite{Clisby-McCoy} and reference
therein) and hard--core lattice gases in two--dimensions, particularly the
hard--hexagon model whose radius of convergence of the virial series has
been determined exactly by Joyce \cite{Joyce} (see eq. (12.30) therein) and
is less than the critical density $\rho _{c}$ of this model.}

The results of our investigation are, to say the least, intriguing and
extension of the present analysis will be presented separately. The present
paper covers, in addition, some preliminary materials to make it
self--contained. We observe that, despite the volume $\Lambda $ is kept
finite, the \textquotedblleft macroscopic limit\textquotedblright\ is
already realized for a gas of point-like particles. From the point of view
of Lee--Yang theory, the partition function $\Xi _{\Lambda }=\Xi _{\Lambda
}(z)$ of a gas of hard-spheres is a polynomial in the activity $z$ whose
zeros (singularities of the pressure) may accumulate (eventually) at the
infinite volume limit. The singularities of the \textquotedblleft
macroscopic functions\textquotedblright\ for point--like gas of particles,
even at finite $\left\vert \Lambda \right\vert $, may be seen from the
Taylor series of the logarithm of partition function $\wp =\log \Xi
_{\Lambda }/\left\vert \Lambda \right\vert $ about $z=0$. According to
Jentzsch theorem (see e.g. \cite{Titchmarsh}), each point on the circle of
convergence of $\wp $ is a limit point of zeros of the $n$--th Taylor
polynomial $\wp _{n}$, $n=1,2,\ldots $.

The second issue, concerning with the Mayer and virial series at low
temperature, has been addressed recently by Jansen\cite{Jansen} for a class
of potentials satisfying a $n$--particle ground state geometry condition and
some of her results extend to nonnegative potentials as well. We should
mention that her results on the radii of convergence of the virial (\ref{Pp}%
) and the series of $\wp \circ Z(\rho )$ in power of $\rho $ (see e.g.
Theorem 3.8 of \cite{Jansen}) go, however, in the opposite direction of ours
for the uniformly repulsive potential. We should warn that, since we are
fixing $\beta =1$ (activity and density are given by $z=e^{\mu }$ and $\rho
=p_{\mu }$, without $\beta $) the limit $t\rightarrow \infty $ does not
really mean low temperature limit, although we sometimes abuse of language.

Equation (\ref{gfeq}) has a stationary solution: $\varphi _{0}(\rho )=\log
\left( 1-\rho /(2\varepsilon )\right) $ (solves (\ref{gfeq}) with $\varphi
_{t}=0$), from which one obtains the pressure (see eqs. (\ref{FmuP})-(\ref%
{Pan})) 
\begin{equation*}
P_{0}(\rho )=\int_{0}^{\rho }\left( 1-\zeta \varphi _{0}^{\prime }(\zeta
)\right) d\zeta =-2\varepsilon \log \left( 1-\rho /(2\varepsilon )\right)
\end{equation*}%
for the hard core lattice gas or Ford model (\ref{Ford}) in the low density
regime. The lower bound (\ref{RR}) on the radius of convergence $\mathcal{R}%
_{\varphi }$ implies that $\varphi _{0}(\rho )$ is attained, as $t$ goes to $%
\infty $, at least inside the domain $\left\vert \rho /\lambda \right\vert
<\varkappa _{\infty }$ where $\varkappa _{\infty }=3-2\sqrt{2}=0.171573$
(numerically improved to $0.275451$, see Remark \ref{phi0}). Since the power
series $-\displaystyle\sum_{n\geq 1}\left( \rho /2\varepsilon \right) ^{n}/n$
of $\varphi _{0}(\rho )$ converges in a domain $\left\vert \rho /\lambda
\right\vert <1$ larger than the former, we investigate, in our second paper,
the power series solution of (\ref{gfeq}) in exponential time variable $\eta
=e^{-2\varepsilon t}$ (transeries, see e.g. \cite{Costin}): 
\begin{equation}
\varphi (t,\rho )=\varphi _{0}(\rho )+\sum_{n=1}^{\infty }\varphi _{n}(\rho
)\eta ^{n}  \label{fifi}
\end{equation}%
asymptotic to $\varphi _{0}$ as $t\rightarrow \infty $. The results we
obtained using a more involved version of the Cauchy--majorant method, which
has its own interests, may be summarized as follows:

\textit{The }$\varphi _{n}$\textit{'s in (\ref{fifi}) can be written as }%
\begin{equation}
\varphi _{n}(\rho )=\left( k(\rho /2\varepsilon )\right) ^{n}Q_{n}(\rho
/2\varepsilon )\mathit{\ }  \label{kQ}
\end{equation}%
\textit{where }$k(\rho )$\textit{\ is the Kobe function }$\rho /(1-\rho
)^{2} $\textit{\ and }$Q_{n}(\rho )$\textit{\ is a polynomial of order }$n$%
\textit{\ in }$\rho $\textit{. The series (\ref{fifi}), written as }$%
\displaystyle\sum_{n\geq 1}Q_{n}(k\eta )^{n}$\textit{\ by (\ref{kQ}) and
with }$2\varepsilon =1$\textit{\ (w.l.g.), converges uniformly in the domain 
}$\Omega \subset \lbrack 0,1)\times \mathbb{D}_{1}$\textit{\ such that }$%
\left\vert \rho \right\vert \leq \rho _{0}<1$\textit{\ and }$\left\vert \eta
\right\vert <\eta _{0}$\textit{\ where }$\eta _{0}=\eta _{0}(\rho _{0})$%
\textit{\ tends to }$0$\textit{\ as }$\rho _{0}$\textit{\ goes to }$1$%
\textit{. }

These results, together with (\ref{RR}) and the global--existence/uniqueness
of the initial value problem (\ref{gfeq}), imply that, for large $t$, two
solutions of (\ref{gfeq}) coexist in the some domain $\left\vert \rho
\right\vert <\rho _{0}$, $\rho _{0}=\rho _{0}(t)>0$ (although (\ref{fifi})%
\textit{\ does not satisfy the (ideal gas) initial value }$\varphi (0,\rho
)=0$ and it might not even be seem from the partition function); as long as $%
\varepsilon >0$, they belong to different branches, (\ref{fifi})
corresponding to the \textquotedblleft low temperature\textquotedblright\
solution.

\paragraph{Outline}

The layout of the present paper is as follows. In Section \ref{URP} we
introduce our potential model and establish a relationship between
macroscopic functions and their corresponding PDE's. Sections \ref{CMM} and %
\ref{CPS} contain our main contributions: Theorems \ref{unique} and \ref%
{asympt} on the Cauchy--majorant and asymptotic methods applied to (\ref%
{gfeq}); Theorem \ref{bestmate} on the radius of convergence of Mayer
solution (\ref{ptmu}) of (\ref{pde}). Our conclusions and open problems are
summarized in Section \ref{CPS}. Four appendices are included to insert our
results into the context. In Appendix \ref{ICI} we review the theory of
imperfect gas due to Mayer. Appendix \ref{CVSO} provides an overview on the
convergence of virial series. Appendix \ref{HJE} proves Proposition \ref%
{PDEs}.a, the derivation of (\ref{pde}) for the pressure of a gas of
particles interacting via a uniformly repulsive potential, introduced in
Section \ref{URP}. Appendix \ref{CVS} establishes global existence and
uniqueness of a class of PDEs satisfying initial condition of Mayer type,
which includes (\ref{pde}), (\ref{gfeq}) and another equation related to
irreducible--edges Mayer graphs. Finally, Appendix \ref{PPI} contains our
technical results: Propositions \ref{Integral} and \ref{contra}, used in the
proof of Theorems \ref{unique} and \ref{asympt}, respectively.

\section{Thermodynamic Functions and their co\-rres\-pon\-ding
  PDE's\label{URP}} 

\setcounter{equation}{0} \setcounter{theorem}{0}

\paragraph{Two parameters potential model}

We consider an equilibrium system of \textbf{point-particles} in a finite
container $\Lambda \subset \mathbb{R}^{d}$ interacting via a uniformly
repulsive two--body potential%
\begin{equation}
\phi _{ij}\equiv \phi _{\Lambda }(t;x_{i}-x_{j})=\frac{t}{\left\vert \Lambda
\right\vert }  \label{ip}
\end{equation}%
for every $x_{i}$, $x_{j}\in \Lambda $.

Our interacting potential $\phi _{\Lambda }=\phi _{\Lambda
}(t;x)=t/\left\vert \Lambda \right\vert $ depends on the size $\left\vert
\Lambda \right\vert $ of the container and on a parameter $t$ which plays a
double role of \textquotedblleft time\textquotedblright\ (evolution)
variable and \textquotedblleft inverse temperature\textquotedblright\ or
repulsive intensity (so, we set $\beta =1$ in (\ref{Bfactor}) and change
variable $\beta $ by $t$ for all functions defined in Appendix \ref{ICI}).
Among all features of a reasonable physical potential, $\phi _{\Lambda }$
retains only the property of being repulsive (nonnegative). As the
so--called Gaussian model, introduced by Montroll et all \cite%
{Montroll-Berlin-Hart} (see \cite{Uhlenbeck-Ford}), $\phi _{\Lambda }$ is
able to isolate the combinatorial problem, arising when thermodynamic
functions are represented in power series (see (\ref{bb}) and (\ref{betan-1}%
)), from the integral on the configuration space, which can be easily
performed. Because the point-like particles are moving in a continuum space
(container $\Lambda $), the system is already in the \textquotedblleft
macroscopic limit\textquotedblright , despite its volume $\left\vert \Lambda
\right\vert $ is kept finite.

Note that $\phi _{\Lambda }$ is positive ($\phi _{\Lambda }(t;x)>0$) and of
positive-definite type: 
\begin{equation*}
\sum_{1\leq i,j\leq N}z_{i}\phi _{\Lambda }(t;x_{i}-x_{j})\bar{z}_{j}=\frac{t%
}{\left\vert \Lambda \right\vert }\left\vert \sum_{i=1}^{N}z_{i}\right\vert
^{2}\geq 0~,
\end{equation*}%
for any collection of points $\left( x_{i}\right) _{i=1}^{N}$ in $\Lambda $
and complex numbers $\left( z_{i}\right) _{i=1}^{N}$, satisfying therefore 
\textbf{stability}:%
\begin{equation*}
\sum_{1\leq i<j\leq N}\phi _{\Lambda }(t;x_{i}-x_{j})\geq \frac{-t}{%
2\left\vert \Lambda \right\vert }N
\end{equation*}%
and \textquotedblleft \textbf{integrability}\textquotedblright\ (in the $%
L_{1}(\Lambda ;d^{d}x)$--sense):%
\begin{equation*}
\left\Vert \phi _{\Lambda }(t;\cdot )\right\Vert _{1}=\int_{\Lambda }\phi
_{\Lambda }(t;y)d^{d}y=t
\end{equation*}%
and%
\begin{equation}
\left\Vert e^{-\phi _{\Lambda }(t;\cdot )}-1\right\Vert _{1}=\int_{\Lambda
}\left\vert e^{-\phi _{\Lambda }(t;y)}-1\right\vert d^{d}y=\left\vert
\Lambda \right\vert \left( 1-e^{-t/\left\vert \Lambda \right\vert }\right)
\equiv \lambda (t,\left\vert \Lambda \right\vert )~,  \label{ft}
\end{equation}%
where $\lambda (t,\left\vert \Lambda \right\vert )$ is monotone increasing
function of $t$ and $\left\vert \Lambda \right\vert $ such that $\lambda (%
\mathbb{R}_{+},\left\vert \Lambda \right\vert )=[0,\left\vert \Lambda
\right\vert ]$ and $\lambda (t,\mathbb{R}_{+})=[0,t]$ remain bounded for $%
\left\vert \Lambda \right\vert ,t>0$.

We refer to Appendix \ref{ICI} for basic properties and formal expressions
of the thermodynamic functions pertaining to this section.

\paragraph{Weight of a connected graph $G$}

The problem of the evaluation of $B_{n}$, the $n$--th coefficient (\ref%
{Bbeta}) of Kamerlingh Onnes virial series (\ref{vseries}), may be divided
into two distinct ones. The combinatorial problem, whose an introductory
presentation is given in Appendix \ref{ICI}, is independent of the law of
interacting forces between any two particles. The second problem for the
evaluation of $\beta _{n}=-(n+1)B_{n+1}/n$ is the integration over the
configuration space $\Lambda ^{n}$. For interacting potential (\ref{ip}),
the weight $w_{\Lambda }(G)$ of a irreducible (2--connected) graph $G$ with $%
n$ vertices and $l=\left\vert E(G)\right\vert $ edges involved in the sum (%
\ref{betan-1}), satisfies exactly 
\begin{eqnarray}
w_{\Lambda }(G) &=&\frac{1}{\left\vert \Lambda \right\vert }\int_{\Lambda
^{n}}\prod_{(ij)\in E(G)}(e^{\phi _{\Lambda
}(x_{i}-x_{j})}-1)d^{d}x_{1}\cdots d^{d}x_{n}  \notag \\
&=&\left( -1\right) ^{l}\left\vert \Lambda \right\vert ^{n-l-1}~\lambda
^{l}(t,\left\vert \Lambda \right\vert )~  \label{I}
\end{eqnarray}%
by (\ref{ft}), reducing the two problems to a purely combinatorial one --
this should be contrasted with the \textquotedblleft soft
repulsion\textquotedblright\ Gaussian model whose explicit evaluation, $%
w(G)=(-1)^{l}(\pi /\alpha )^{3(n-1)/2}\gamma ^{-3/2}$ depends on the graph
complexity $\gamma =\gamma (G)$ of $G$ in addition to the two other
dependences, on the number of vertices and edges, already in (\ref{I}) (see 
\cite{Uhlenbeck-Ford,Leroux}, for definition, evaluation and motivations).

The number of edges $l=\left\vert E(G)\right\vert $ of a connected graph $G$
with $n$ vertices satisfies 
\begin{equation*}
l\geq n-1~,
\end{equation*}%
with equality holding only for tree graphs $T$. Since the only $2$%
--connected tree is, by definition, the graph with two vertices connected by
a single edge $T_{2}$, and since for any connected graph $G$, we have, as $%
\left\vert \Lambda \right\vert $ goes to infinity, 
\begin{equation*}
w^{\infty }(G)=\lim_{\left\vert \Lambda \right\vert \rightarrow \infty
}w_{\Lambda }(G)=\left\{ 
\begin{array}{lll}
\left( -t\right) ^{l} &  & \text{if\ }G=T \\ 
0 &  & \text{otherwise}%
\end{array}%
\right. ~,
\end{equation*}%
by $\lambda (t,\infty )=t$, the limit weight $w^{\infty }(G)$ vanishes for
all irreducible ($2$--connected) graphs $G$ different from $T_{2}$. In this
limit, the virial series (\ref{vseries}) reads%
\begin{equation*}
P^{\infty }(t,\rho )=\rho +\frac{t}{2}\rho ^{2}~
\end{equation*}%
by dissymmetry theorem (see e.g. \cite{Bergeron-Labelle-Leroux}), which
agrees with the pressure of a system of hard spheres in $d=\infty $.

Another limit is attained when $t$ tends to $\infty $ (the low temperature
limit) with $\left\vert \Lambda \right\vert $ finite. For this limit, the
weight of a 2--connected graph $G$ reads 
\begin{equation*}
\lim_{t\rightarrow \infty }w_{\Lambda }(G)=(-1)^{l}\left\vert \Lambda
\right\vert ^{n-1}
\end{equation*}%
and, by a subtle cancellation (exactly as in the hard-core lattice gas, for
which $f_{ij}=e^{-\phi _{ij}}-1=-1$ if $x_{i}=x_{j}$ and $f_{ij}=0$ if $%
x_{i}\neq x_{j}$, except that the sum of a $n$--particle configuration gives 
$\left\vert \Lambda \right\vert $ instead $\left\vert \Lambda \right\vert
^{n}$),%
\begin{equation*}
P(\infty ,\rho )=\frac{-1}{\left\vert \Lambda \right\vert }\log \left(
1-\left\vert \Lambda \right\vert \rho \right) \ ,\qquad \rho <1/\left\vert
\Lambda \right\vert
\end{equation*}%
agreeing, this time, with the pressure (\ref{Ford}) of Ford's model in the
low density regime. Both limit functions will be shown to be attained by our
investigation of macroscopic functions (pressure and the Helmholtz free
energy) as power series solution of related partial differential equations.

\paragraph{Partial differential equations}

As a consequence of (\ref{I}), we have

\begin{proposition}
\item \label{PDEs}

\begin{enumerate}
\item[(a)] The pressure $p=p_{\Lambda }(t,\mu )$ of a uniformly repulsive
interacting system, as a function of $t$ and the chemical potential $\mu
=\log z$, for any fixed $\Lambda $, satisfies a partial differential
equation (PDE)%
\begin{equation}
p_{t}+\varepsilon (p_{\mu \mu }-p_{\mu })+\frac{1}{2}\left( p_{\mu }\right)
^{2}=0~,  \label{pde1}
\end{equation}%
$\varepsilon =1/\left( 2\left\vert \Lambda \right\vert \right) $, with
initial condition $p_{\Lambda }(0,\mu )=e^{\mu }$.

\item[(b)] The Helmholtz free energy density $F=F_{\Lambda }(t,\rho )$
defined by the (formal) Legendre transform of $p_{\Lambda }(t,\mu )$ w.r.t. $%
\mu $ (see Proposition \ref{legendre}) satisfies 
\begin{equation}
F_{t}-\varepsilon (\frac{1}{F_{\rho \rho }}-\rho )-\frac{1}{2}\rho ^{2}=0
\label{Feq}
\end{equation}%
with $F_{\Lambda }(0,\rho )=\rho \log \rho -\rho $.

\item[(c)] The function $\varphi =\varphi (t,\rho )$, defined by 
\begin{equation}
F_{\rho }=\log \rho -\varphi ~,  \label{Frho}
\end{equation}%
satisfies 
\begin{equation}
\varphi _{t}+\rho +\varepsilon \left( \frac{1+\rho ^{2}\varphi _{\rho \rho }%
}{\left( 1-\rho \varphi _{\rho }\right) ^{2}}-1\right) =0  \label{eq}
\end{equation}%
with $\varphi \left( 0,\rho \right) =0$ and generates, by Proposition \ref%
{legendre}, the irreducible cluster integrals $\beta _{n}$, $n=1,2,\ldots $,
defined by (\ref{beta}), i.e., the $\beta _{n}$'s are the coefficients of a
formal power series in $\rho $ (\ref{sol1}) of $\varphi $.
\end{enumerate}
\end{proposition}

We defer the proof of item \textbf{(a)} of Proposition \ref{PDEs} to
Appendix \ref{HJE} and anticipate that (\ref{pde1}) is deduced from the
Brydges--Kennedy equations\cite{Brydges-Kennedy} for the Ursell functions,
which are briefly introduced there.

\textbf{\ \medskip }

\noindent \textit{Proof of Proposition \ref{PDEs}, parts \textbf{(b)} and 
\textbf{(c)}.} We follow here \cite{Courant-Hilbert}. The Legendre transform
of a formal power series $p(t,\mu )$ in $e^{\mu }$, with respect to $\mu $,
is 
\begin{equation}
F(t,\rho )=\rho \mu ^{\ast }-p(t,\mu ^{\ast })  \label{FP}
\end{equation}%
where $\mu ^{\ast }=\mu ^{\ast }(t,\rho )$ solves (formally)%
\begin{equation}
\rho =p_{\mu }(t,\mu )  \label{rhopmu}
\end{equation}%
for $\mu $; the Legendre transform of $F(t,\rho )$ with respect to $\rho $
is 
\begin{equation}
p(t,\mu )=\rho ^{\ast }\mu -F(t,\rho ^{\ast })  \label{pLF}
\end{equation}%
where $\rho ^{\ast }=\rho ^{\ast }(t,\mu )$ solves (formally)%
\begin{equation}
\mu =F_{\rho }(t,\rho )  \label{muFrho}
\end{equation}%
for $\rho $. Differentiating (\ref{pLF}) w.r.t. $\mu $, yields%
\begin{equation}
p_{\mu }=\rho ^{\ast }+\left( \mu -F_{\rho }\right) \rho _{\mu }^{\ast
}=\rho ^{\ast }~,  \label{pmu}
\end{equation}%
by (\ref{muFrho}). Differentiating (\ref{pmu}) w.r.t. $\mu $, yields%
\begin{eqnarray}
p_{\mu \mu } &=&2\rho _{\mu }^{\ast }+\left( \mu -F_{\rho }\right) \rho
_{\mu \mu }^{\ast }-F_{\rho \rho }\left( \rho _{\mu }^{\ast }\right) ^{2} 
\notag \\
&=&2\rho _{\mu }^{\ast }-F_{\rho \rho }\left( \rho _{\mu }^{\ast }\right)
^{2}=\rho _{\mu }^{\ast }  \label{pmumu}
\end{eqnarray}%
by (\ref{muFrho}) again. The last equality reads: $\rho _{\mu }^{\ast
}-F_{\rho \rho }\left( \rho _{\mu }^{\ast }\right) ^{2}=0$ whose nontrivial
solution $\rho _{\mu }^{\ast }\neq 0$ satisfies%
\begin{equation}
\rho _{\mu }^{\ast }=\frac{1}{F_{\rho \rho }}=p_{\mu \mu }  \label{rhomu}
\end{equation}%
by (\ref{pmumu}). We deduce from (\ref{pLF}), together with (\ref{muFrho}),
that%
\begin{equation}
p_{t}=-F_{t}~.  \label{pF}
\end{equation}%
(\ref{pF}), (\ref{rhopmu}) and (\ref{rhomu}) substituted into (\ref{pde1}),
yields (\ref{Feq}) and concludes the proof of part \textbf{(b)}.

Differentiating (\ref{Feq}) w.r.t. $\rho $ together with (\ref{Frho}) yields
(\ref{eq}). We observe that the operations involved in the Legendre
transform, together with (\ref{Frho}), (sum, multiplication, derivatives,
inverse, composition, reciprocal, ...) apply over the ring $\mathcal{C}%
^{1}[[z]]$ of formal power series in $z=e^{\mu }$ with $\mathcal{C}^{1}$
coefficients $b_{n}(t)$, $n\geq 1$.

For the statement of item \textbf{(c)} after (\ref{eq}), we follow
Proposition \ref{legendre}. By differentiating (\ref{F}) w.r.t. $\rho $, we
conclude that $\varphi $, defined by (\ref{Frho}) and in Theorem \ref{T2} by
(\ref{phi}), have the same power series, whose coefficients are the
irreducible integrals $\beta _{n}$. The proof of Proposition \ref{PDEs} is
now complete.

\hfill $\Box $

An alternate proof of the statement \textbf{(c)} starts from (\ref{FP}):%
\begin{equation}
F(t,\rho )=\rho \mu ^{\ast }(t,\rho )-P(t,\rho )~  \label{FmuP}
\end{equation}%
where $p\left( t,\mu ^{\ast }(t,\rho )\right) =P(t,\rho )$. Differentiating (%
\ref{FmuP}) w.r.t. $\rho $, together with $\mu ^{\ast }(t,\rho )=F_{\rho
}(t,\rho )=\log \rho -\varphi (t,\rho )$, yields%
\begin{equation*}
P_{\rho }(t,\rho )=1-\rho \varphi _{\rho }(t,\rho )~.
\end{equation*}%
Assuming that $\varphi $ has a power series (\ref{sol1}), this equation can
be formally integrated, with $P(t,0)=0$: 
\begin{eqnarray}
P(t,\rho ) &=&\int_{0}^{\rho }\left( 1-\eta \varphi _{\rho }(t,\eta )\right)
d\eta  \notag \\
&=&\rho -\sum_{n=1}^{\infty }\frac{n}{n+1}\beta _{n}(t)\rho ^{n+1}
\label{Pan}
\end{eqnarray}%
from which we conclude that the $\beta _{n}$ are the irreducible cluster
integrals, by (\ref{P}). The Kamerlingh Onnes virial series (\ref{Pp}) is
thus given by (\ref{Pan}) so, the power series (\ref{sol1}) of $\varphi
(t,\rho )$ converges if, and only if, the virial series converges. We
address in the following the former convergence to conclude about the latter.

\section{Convergence of Virial Series: Cauchy--Ma\-jo\-rant Method \label%
{CMM}}

\setcounter{equation}{0} \setcounter{theorem}{0}

We shall prove, among other results, the following

\begin{theorem}
\label{unique}The initial value problem (\ref{eq}) admits a unique solution (%
\ref{sol1}) in power series of $\rho $ which converges uniformly in $%
\varepsilon $, at least, inside the domain: $\left( t,\rho \right) \in 
\mathbb{R}_{+}\times \mathbb{C}$ such that 
\begin{equation}
\left\vert \rho \right\vert <~2\varepsilon \frac{2+\sqrt{1-e^{-2\varepsilon
t}}-2\sqrt{1-e^{-2\varepsilon t}/2+\sqrt{1-e^{-2\varepsilon t}}}}{%
1-e^{-4\varepsilon t}}~.~  \label{rhoepsilon}
\end{equation}
\end{theorem}

\begin{corollary}
\label{vs}The radius of convergence $\mathcal{R}_{P}(t)$ of the virial
series (\ref{Pan}) for a system of point particles interacting via the
(uniformly repulsive) two--body potential $\phi _{\Lambda }$, given by (\ref%
{ip}), satisfies%
\begin{equation}
\mathcal{R}_{P}(t)\geq \varkappa \frac{1}{\lambda }  \label{Rt}
\end{equation}%
where 
\begin{equation}
\varkappa =\frac{2+\sqrt{1-\eta }-2\sqrt{1-\eta /2+\sqrt{1-\eta }}}{1+\eta }
\label{varkappa}
\end{equation}%
$\eta =\eta (t,\left\vert \Lambda \right\vert )=e^{-t/\left\vert \Lambda
\right\vert }$ and $\lambda =$ $\lambda (t,\left\vert \Lambda \right\vert
)=\left\vert \Lambda \right\vert \left( 1-\eta \right) $ is the function
defined in (\ref{ft}).
\end{corollary}

\begin{figure}[tbp]
\centering\includegraphics[scale=0.55]{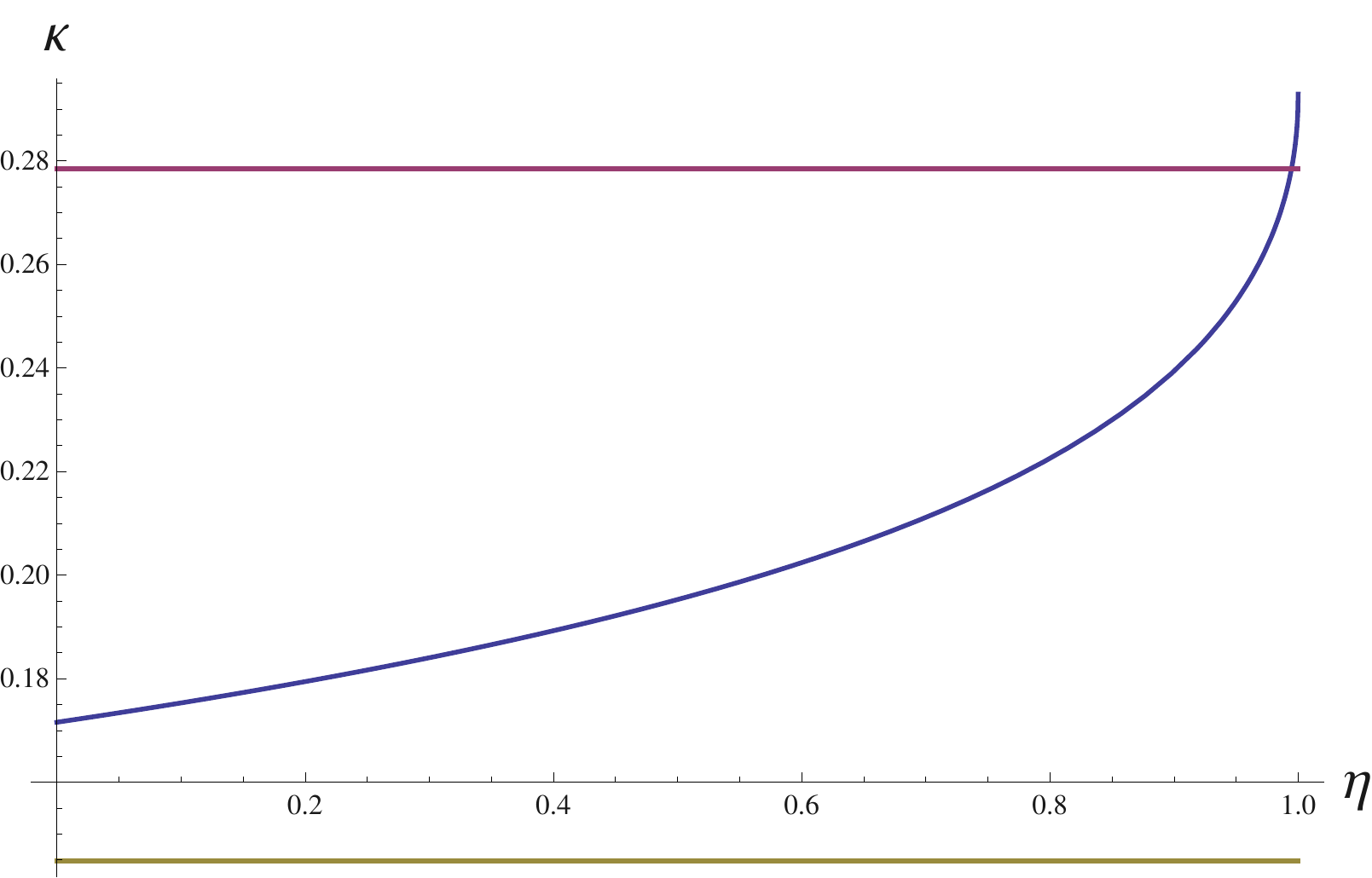}
\caption{$\varkappa $ as a function of $\protect\eta $}
\label{kappa}
\end{figure}

For fixed $\Lambda $ the prefactor $\varkappa $ varies from $1-1/\sqrt{2}%
=0.292893$ to $3-2\sqrt{2}=\allowbreak 0.171573$ as $t$ varies from $0$ to $%
\infty $. Figure \ref{kappa} plots $\varkappa $ as a function of $\eta $
where the two constants are Lebowitz-Penrose's lower bound ($\varkappa
=0.144767$) and the threshold ($\varkappa =0.278465$) established for
nonnegative potentials by Lagrange inversion formula (see Sect. \ref{CVS}).

The proof of Corollary \ref{vs} is an immediate consequence of Theorem \ref%
{unique} together with the observation after equation (\ref{Pan}).

\medskip

\noindent \textit{Proof of Theorem \ref{unique}.} The proof will be divided
into two parts. Firstly, the Cauchy-majorant method is applied to (\ref{eq})
without preoccupation with the radius of convergence. We then push the
method further -- not as far it can go, but enough to go beyond the
threshold.

\paragraph{Solution of (\protect\ref{eq}) in power series}

We have introduced in Appendix \ref{CVS} a basic majorant method capable of
dealing with equation of the form $u_{t}+\mathcal{A}\left(
t,u_{x},u_{xx}\right) =0$ (see (\ref{PDE})-(\ref{eta})). We observe that for
a solution (\ref{sol}) of this kind of equation, in power series of $z=e^{x}$%
, $\partial /\partial x$ acts over a function of $z$ as $z\partial /\partial
z$. In equation (\ref{eq}), $\rho $ plays the role of the $z$ variable and
we write $\mathcal{A}=\mathcal{A}(\rho ,a,b)$ with $a$ and $b$ in the place
of $\rho \varphi _{\rho }$ and $\rho (\rho \varphi _{\rho })_{\rho }$,
respectively. Note that (\ref{eq}) differs slightly from (\ref{PDE}) (it has
trivial initial condition, $A_{0,0}=\mathcal{A}\left( \rho ,0,0\right) $
depends explicitly on $\rho $ and $\mathcal{A}$ doesn't depend on $t$) and
some additional features allow it to be analyzed more carefully.

Since $\rho ^{2}\varphi _{\rho \rho }=\rho (\rho \varphi _{\rho })_{\rho
}-\rho \varphi _{\rho }=b-a$, equation (\ref{eq}) can be written in the form
of a conservation law: 
\begin{equation}
\varphi _{t}+\left( \mathcal{J}(\rho ,\rho \varphi _{\rho })\right) _{\rho
}=0  \label{form}
\end{equation}%
with%
\begin{eqnarray}
\mathcal{J}(\rho ,a) &=&\frac{\rho ^{2}}{2}+\varepsilon \rho \left( \frac{1}{%
1-a}-1\right)  \notag \\
&=&\frac{\rho ^{2}}{2}+\varepsilon \rho \sum_{n=1}^{\infty }a^{n}~,
\label{J}
\end{eqnarray}%
(i.e., $\mathcal{A}=\mathcal{J}_{\rho }+\mathcal{J}_{a}\cdot (\rho \varphi
_{\rho })_{\rho }=\rho +\varepsilon \left( (1+b-a)/(1-a)^{2}-1\right) $) and
initial condition $\varphi (0,\rho )\equiv 0$. The proof that there exist a
unique formal power series solution for equations of the form (\ref{form})
is given in Theorem \ref{convergence} or may be concluded from the following
calculations. We refer to Appendix \ref{CVS} for details.

\paragraph{Integral equation for the $\protect\beta _{k}$'s}

For convenience, we introduce a sequence $\mathbf{\gamma }=\left( \gamma
_{k}\right) _{k\geq 1}$, with $\gamma _{k}=k\beta _{k}$, and the convolution
product $\mathbf{\gamma }\ast \mathbf{\delta }$ of two sequences $\mathbf{%
\gamma }$ and $\mathbf{\delta }$ is defined by (\ref{prod}): $\left( \mathbf{%
\gamma }\ast \mathbf{\delta }\right) _{k}=\displaystyle\sum_{j=1}^{k-1}%
\gamma _{k-j}\delta _{j}$. The power series (\ref{sol1}) of $\varphi $
substituted into the integral (w.r.t. $\rho $, from $0$ to $\rho $) of (\ref%
{form}), together with 
\begin{equation*}
\frac{1}{\rho }\int_{0}^{\rho }\varphi _{t}(t,\bar{\rho})d\bar{\rho}%
=\sum_{k=1}^{\infty }\dfrac{1}{k+1}\dot{\beta}_{k}(t)\rho ^{k}
\end{equation*}%
and the fundamental theorem of calculus, yield a system of first order
differential equations 
\begin{equation}
\frac{1}{k(k+1)}\dot{\gamma}_{k}+\varepsilon \gamma _{k}=-h_{k}(\gamma
_{1},\ldots ,\gamma _{k-1};t)\ ,\qquad k\geq 1  \label{ak}
\end{equation}%
with $\gamma _{k}(0)=0$, where the $h_{k}$'s are given by $h_{1}=1/2$ and,
for $k\geq 2$,%
\begin{equation}
h_{k}(\gamma _{1},\ldots ,\gamma _{k-1};t)=\varepsilon \sum_{n=2}^{k}\left( 
\underset{n}{\underbrace{\mathbf{\gamma }\ast \cdots \ast \mathbf{\gamma }}}%
\right) _{k}~.  \label{fk}
\end{equation}

The solution of (\ref{ak}) for $k=1$: 
\begin{equation*}
\frac{1}{2}\dot{\gamma}_{1}+\varepsilon \gamma _{1}=-\frac{1}{2}
\end{equation*}%
with $\gamma _{1}(0)=0$ is%
\begin{eqnarray}
\gamma _{1}(t) &=&e^{-2\varepsilon t}\int_{0}^{t}e^{2\varepsilon s}\left(
-1\right) ds  \notag \\
&=&\frac{-1}{2\varepsilon }(1-e^{-2\varepsilon t})=-\lambda  \label{a1}
\end{eqnarray}%
with $\lambda =\lambda (t,\left\vert \Lambda \right\vert )$ the function
defined by (\ref{ft}) (recall $\varepsilon =1/(2\left\vert \Lambda
\right\vert )$).

By the variation of constants formula, (\ref{ak}) is equivalent to a system
of integral equations 
\begin{equation}
\gamma _{k}(t)=-k(k+1)\int_{0}^{t}e^{-\varepsilon k(k+1)(t-s)}h_{k}(\gamma
_{1},\ldots ,\gamma _{k-1};s)ds~,\qquad k\geq 2  \label{akt}
\end{equation}%
which can be evaluate recursively starting from $\gamma _{1}(t)=-\lambda $.
For the first few $k$'s, e.g., for $k=2$, $h_{2}=\varepsilon \gamma
_{1}^{2}(t)$ and 
\begin{equation}
\gamma _{2}(t)=-6\varepsilon e^{-6\varepsilon t}\int_{0}^{t}e^{6\varepsilon
s}\gamma _{1}^{2}(s)ds=-2\varepsilon \lambda ^{3},  \label{a2}
\end{equation}%
and for certain class of terms the integral (\ref{akt}) can be performed by
\textquotedblleft hand\textquotedblright\ (see Proposition \ref{Integral}
and its proof in Appendix \ref{PPI}).

\paragraph{An exact equation satisfied by the majorant}

To approach equation (\ref{eq}) by the Cauchy--Majorant method, we need the
following (stronger than usual)

\begin{definition}
\label{maj}A function%
\begin{equation}
\Phi (t,\rho )=\sum_{n=1}^{\infty }\Phi _{n}(t)\rho ^{n}  \label{majorant}
\end{equation}%
is a majorant of $\varphi (t,\rho )$, whose power series in $\rho $ is given
by (\ref{sol1}), if

\begin{enumerate}
\item each $\Phi _{n}(t)$ is positive, continuous and monotone increasing
function of $t$;

\item there exist a family of domains $\Omega _{t}=[0,t)\times \mathbb{D}%
_{r(t)}$, in which the series (\ref{majorant}) converges absolutely in $%
\mathbb{D}_{r(t)}$ and%
\begin{equation}
\left\vert \beta _{n}(s)\right\vert \leq \Phi _{n}(s)  \label{ans}
\end{equation}%
is satisfied for $0\leq s<t$, uniformly in $\varepsilon $.
\end{enumerate}

We write $\varphi \ll \Phi $ for the majorant relation.
\end{definition}

We observe that the majorant relation is preserved by derivative w.r.t. $%
\rho $, integration in both $t$ and $\rho $, convex combination,
multiplication and composition (see e.g. \cite{vanderHoeven}).

We start with 
\begin{equation}
\left\vert \gamma _{1}(t)\right\vert =\lambda \left( t,(2\varepsilon
)^{-1}\right) =\Phi _{1}(t)  \label{a1lambda}
\end{equation}%
and generate a recursive equation for $\left( \Phi _{k}\right) _{k\geq 1}$
through (\ref{akt}). For convenience, we introduce a majorant 
\begin{equation}
\Psi =\rho \Phi _{\rho }=\sum_{k=1}^{\infty }\Psi _{k}\rho ^{k},\qquad \Psi
_{k}=k\Phi _{k},  \label{majorant1}
\end{equation}%
of $\rho \varphi _{\rho }$ and observe that $\varphi \ll \Phi
\Longleftrightarrow \rho \varphi _{\rho }\ll \Psi $. In our first attempt of
constructing $\Psi $, we shall not push the method to its limit.

Assuming (\ref{ans}) holds for $1\leq n\leq k-1$, (\ref{akt}) can be bounded
as%
\begin{eqnarray}
\left\vert \gamma _{k}(t)\right\vert &\leq
&k(k+1)\int_{0}^{t}e^{-\varepsilon k(k+1)(t-s)}h_{k}(\left\vert \gamma
_{1}\right\vert ,\ldots ,\left\vert \gamma _{k-1}\right\vert ;s)ds  \notag \\
&\leq &h_{k}(\Psi _{1},\ldots ,\Psi
_{k-1};t)k(k+1)\int_{0}^{t}e^{-\varepsilon k(k+1)(t-s)}ds  \notag \\
&=&\frac{1}{\varepsilon }h_{k}(\Psi _{1},\ldots ,\Psi _{k-1};t):=\Psi
_{k}(t)~.  \label{Phik}
\end{eqnarray}%
For $k=2$, by (\ref{a2}), we can do better:%
\begin{equation}
\left\vert \gamma _{2}(t)\right\vert =2\varepsilon \lambda ^{3}(t)=\frac{1}{%
\varepsilon }h_{2}\left( \Psi _{1};t\right) -\eta (t)\lambda ^{2}:=\Psi
_{2}(t)~,  \label{a2b}
\end{equation}%
where we have used $1-2\varepsilon \lambda =e^{-2\varepsilon t}\equiv \eta
(t)$.

Summing the above recursive relation for the $\Psi _{k}$, multiplied by $%
\rho ^{k}$: 
\begin{equation*}
\sum_{k\geq 2}\Psi _{k}\rho ^{k}=\frac{1}{\varepsilon }\sum_{k\geq
2}h_{k}\left( \Phi _{1},\ldots ,\Phi _{k-1};t\right) \rho ^{k}-\eta \lambda
^{2}\rho ^{2}~,
\end{equation*}%
for $t$ fixed, results 
\begin{equation}
\Psi -\Psi _{1}\rho =\frac{1}{1-\Psi }-1-\Psi -\eta \lambda ^{2}\rho ^{2}
\label{AB}
\end{equation}%
by (\ref{majorant1}), (\ref{fk}), (\ref{J}) and Proposition \ref{fg}.

We observe that the r.h.s. of (\ref{AB}) is the function in (\ref{J})
between parenthesis: $1/(1-\Psi )-1$, subtracted by $\Psi +\eta \lambda
^{2}\rho ^{2}$, and (\ref{AB}) is equivalent to a quadratic polynomial
equation: $2\Psi ^{2}-(1+r-\eta r^{2})\Psi +r-\eta r^{2}=0$, with $r=\lambda
\rho $, whose solution yields 
\begin{equation}
\Psi (t,\rho )=H(\lambda \rho )~  \label{Phih}
\end{equation}%
with

\begin{equation}
H(r)=\frac{1}{4}\left( 1+r-\eta r^{2}-\sqrt{\left( 1+r-\eta r^{2}\right)
^{2}-8(r-\eta r^{2})}\right) ~.  \label{h}
\end{equation}%
We have chosen the branch of square root in (\ref{h}), for which every
coefficient $c_{k}$ of the power series $\displaystyle\sum_{k\geq
1}c_{k}r^{k}$ of $H$ in $r$ remains positive for $0\leq \eta <1$.\footnote{%
The sign can be checked using (\ref{Faa}) or Scott's formula\cite%
{Floater-Lyche} for higher order chain rule to the composite function $-%
\sqrt{p(r)}$, where $p(r)$ is the alternating sign fourth order
(discriminant) polynomial in $r$, inside the square root in (\ref{h}).}
Explicitly, 
\begin{eqnarray}
H(r) &=&r+(1-\eta )r^{2}+(3-2\eta )r^{3}+(11-9\eta +\eta
^{2})r^{4}+(45-44\eta +9\eta ^{2})r^{5}+  \notag \\
&&\qquad (197-225\eta +66\eta ^{2}-3\eta ^{3})r^{6}+\cdots  \notag
\end{eqnarray}%
and we observe that the $c_{k}$'s are monotone decreasing in $\eta $
(monotone increasing in $t$). The fact that (\ref{Phih}) is a function of $%
r=\lambda \rho $ is due to the homogeneity of the $h_{k}$'s:%
\begin{equation}
h_{k}(\lambda \delta _{1},\ldots ,\lambda ^{k-1}\delta _{k-1};t)=\lambda
^{k}h_{k}(\delta _{1},\ldots ,\delta _{k-1};t)  \label{hh}
\end{equation}%
for all $k\geq 2$, $\lambda $ and $\mathbf{\delta }=\left( \delta
_{k}\right) _{k\geq 1}$. Since $\Psi _{k}(t)=c_{k}\lambda ^{k}$ and $\lambda
(t)$ is (strictly) positive and monotone increasing in $t$, the same may be
concluded of the $\Psi _{k}$'s.

The discriminant polynomial $p(r)=\eta ^{2}r^{4}-2\eta r^{3}+(1+6\eta
)r^{2}-6r+1$ of the quadratic equation has four roots: 
\begin{equation*}
r_{\pm ,\pm }=\frac{1\pm \sqrt{1-(12\pm 8\sqrt{2})\eta }}{2\eta }
\end{equation*}%
the nearest to the origin $r_{-,-}=r_{-,-}(t,\varepsilon )$, determines the
radius of convergence of $H$. So, the power series of $H$ in $r$ converges
provided 
\begin{equation}
\left\vert r\right\vert <r_{-,-}(t,\varepsilon )~.  \label{rr}
\end{equation}

It follows from (\ref{Phih}) that, the radius of convergence of $\Psi $
(consequently, of $\Phi $ too) is $r_{-,-}/\lambda $. It is clear from
definition (\ref{Phik}), (\ref{a1lambda}), (\ref{a2b}) and property (\ref{hh}%
), that $\Psi _{k}(t)$ is monotone increasing function of $t$, which,
together with the inequality (\ref{Phik}), proves that $\Psi $ is a majorant
of $\rho \varphi _{\rho }$ in the sense of Definition \ref{maj} with $%
r(t)=r_{-,-}(t,\varepsilon )/\lambda (t,1/2\varepsilon )$. All these
properties hold, in addition, uniformly in $\varepsilon \in \mathbb{R}_{+}$.

\begin{figure}[tbp]
\centering
\includegraphics[scale=0.55]{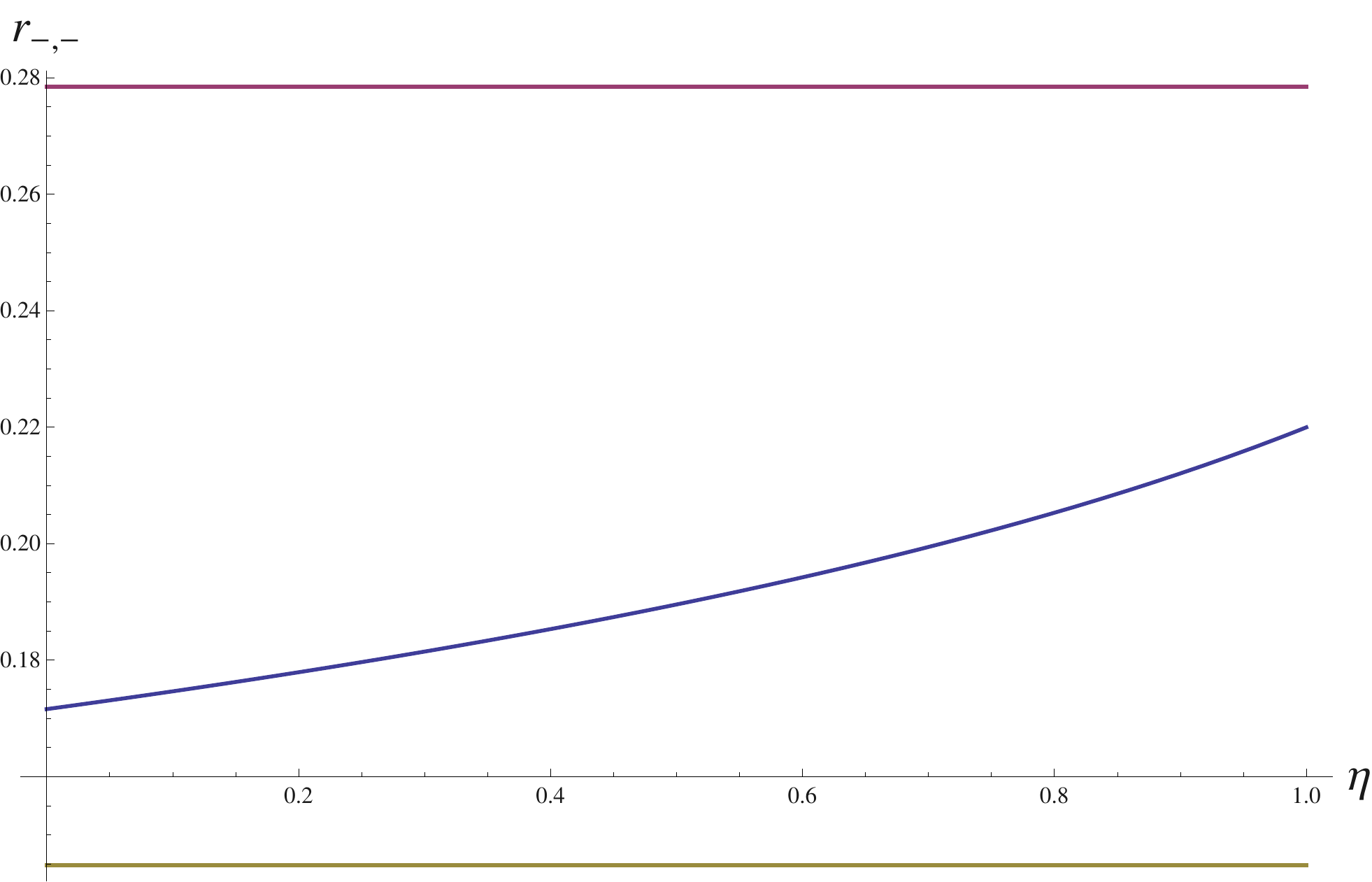}
\caption{Plot of $r_{-,-}$ as a function of $\protect\eta $.}
\label{fig3}
\end{figure}

As one can see from Fig. \ref{fig3}, the improved definition (\ref{a2b}) of $%
\Psi _{2}$ yields a radius of convergence lying above Lebowitz--Penrose's
lower--bound (eq. (\ref{R0}) with $\kappa =1$: $\mathcal{R}\geq
0.144767/\lambda $). If we have used (\ref{Phik}) also for $k=2$ (i.e., set $%
\eta =0$ from (\ref{a2b}) to (\ref{h})), then the root of the second order
polynomial $r^{2}-6r+1$, closest to the origin, $r_{-,-}(\infty ,\varepsilon
)=3-2\sqrt{2}=\allowbreak 0.171\,57$. Since $r_{-,-}(t,\varepsilon )$ hasn't
reached the threshold $0.278465$ for any $t$, $\varepsilon \in \mathbb{R}%
_{+} $, the improvement, however, isn't enough.

\paragraph{Improved majorant equation}

We come to the second part of the proof. The $k$--th and $(k-1)$--th terms
of $h_{k}(\gamma _{1},\ldots ,\gamma _{k-1};t)$,%
\begin{eqnarray}
l_{k}(\gamma _{1};t) &=&\varepsilon \left( \underset{k}{\underbrace{\mathbf{%
\gamma }\ast \cdots \ast \mathbf{\gamma }}}\right) _{k}=\varepsilon \gamma
_{1}^{k}  \notag \\
m_{k}(\gamma _{1},\gamma _{2};t) &=&\varepsilon \left( \underset{k-1}{%
\underbrace{\mathbf{\gamma }\ast \cdots \ast \mathbf{\gamma }}}\right)
_{k}=\varepsilon \left( k-1\right) \gamma _{1}^{k-2}\gamma _{2}  \label{lm}
\end{eqnarray}%
depend only on $\gamma _{1}=\beta _{1}=-\lambda $ and $\gamma _{2}=2\beta
_{2}=-2\varepsilon \lambda ^{3}$, by (\ref{a1}), (\ref{a2}) and Definition %
\ref{Conv}. They are separated as their integral 
\begin{eqnarray}
k(k+1)\frac{1}{\lambda ^{k}(t)}\int_{0}^{t}e^{-\varepsilon
k(k+1)(t-s)}(l_{k}(\left\vert \gamma _{1}\right\vert ;s)+m_{k}(\left\vert
\gamma _{1}\right\vert ,\left\vert \gamma _{2}\right\vert ;s))ds &=&  \notag
\\
2\varepsilon \lambda (t)\frac{k(k+1)}{2}\frac{1}{\lambda ^{k+1}(t)}%
\int_{0}^{t}e^{-\varepsilon k(k+1)(t-s)}\lambda ^{k}(s)(1+2\varepsilon
\lambda (s)(k-1))ds &\equiv &~\mathcal{S}_{k}(t)  \label{S}
\end{eqnarray}%
will be estimated more accurately than the respective integral for the
remaining terms%
\begin{equation}
\tilde{h}_{k}(\gamma _{1},\ldots ,\gamma _{k-1};t)=h_{k}(\gamma _{1},\ldots
,\gamma _{k-1};t)-l_{k}(\gamma _{1};t)-m_{k}(\gamma _{1},\gamma _{2};t)~.
\label{hktilde}
\end{equation}

We refer to (\ref{Phik}) and its refined procedure (\ref{a2b}) for $k=2$. To
improve (\ref{Phik}) for $k\geq 3$, let $\mathbf{\delta }=\left( \delta
_{k}\right) _{k\geq 1}$ be given by%
\begin{equation}
\gamma _{k}(t)=\lambda ^{k}(t)~\delta _{k}(t)\ ,\qquad \lambda (t)=\frac{%
1-\eta (t)}{2\varepsilon }\ \text{and\ }\eta (t)=e^{-2\varepsilon t}~,
\label{lambdaeta}
\end{equation}%
and note that $\displaystyle\sum_{k\geq 1}^{\infty }\delta _{k}r^{k}=\rho
\varphi _{\rho }(t,\rho )$, $r=\lambda \rho $. By (\ref{hh}), (\ref{akt})
can be written as $\delta _{1}=-1$, $\delta _{2}=-(1-\eta )$,%
\begin{equation}
\delta _{k}(t)=-\left( 1-\eta (t)\right) \frac{k(k+1)}{2}\frac{1}{\lambda
^{k+1}(t)}\int_{0}^{t}e^{-\varepsilon k(k+1)(t-s)}\lambda ^{k}(s)\frac{1}{%
\varepsilon }h_{k}(\delta _{1},\ldots ,\delta _{k-1};s)ds~,  \label{deltakt}
\end{equation}%
for $k\geq 3$, and a majorant $\Psi (t,\rho )=\displaystyle\sum_{k\geq
1}^{\infty }C_{k}r^{k}$ of $\rho \varphi _{\rho }(t,\rho )$ will be
constructed in the sense of Definition \ref{maj} for the variable $r$.
Supposing that $\left\vert \delta _{n}(s)\right\vert \leq C_{n}(s)$ holds
for $1\leq n\leq k-1$, equation (\ref{deltakt}) can be majorized by%
\begin{eqnarray}
\left\vert \delta _{k}(t)\right\vert &\leq &k(k+1)\frac{1}{\lambda ^{k}(t)}%
\int_{0}^{t}e^{-\varepsilon k(k+1)(t-s)}\lambda ^{k}(s)\tilde{h}%
_{k}(\left\vert \delta _{1}\right\vert ,\ldots ,\left\vert \delta
_{k-1}\right\vert ;s)ds+\mathcal{S}_{k}(t)  \notag \\
&\leq &\frac{1}{\varepsilon }\tilde{h}_{k}(C_{1},\ldots ,C_{k-1};t)+\mathcal{%
S}_{k}(t)  \notag \\
&\leq &\frac{1}{\varepsilon }h_{k}(C_{1},\ldots ,C_{k-1};t)-\mathcal{T}%
_{k}(t)~,  \label{bk}
\end{eqnarray}%
where, by (\ref{S}), (\ref{hktilde}) and (\ref{lambdaeta}),%
\begin{equation}
\mathcal{T}_{k}(t)=\mathcal{Q}_{k}(t)+(k-1)(1-\eta (t))\mathcal{Q}_{k+1}(t)
\label{Tk}
\end{equation}%
with 
\begin{equation}
\mathcal{Q}_{k}(t)=1-(1-\eta (t))\frac{k(k+1)}{2}\frac{1}{\lambda ^{k+1}(t)}%
\int_{0}^{t}e^{-\varepsilon k(k+1)(t-s)}\lambda ^{k}(s)ds~  \label{Qk}
\end{equation}%
is such that

\begin{lemma}
\label{Teta} For every $k\geq 3$, the function defined by (\ref{Tk}) and (%
\ref{Qk}), as a function of $\eta $: 
\begin{equation}
\mathcal{T}_{k}(t)=T_{k}\circ \eta (t)  \label{TTeta}
\end{equation}
\end{lemma}

is a positive (concave) polynomial of degree $k(k+1)/2+1$ which satisfies $%
T_{k}(0)=0$, $T_{k}(1)=1$ and%
\begin{equation}
T_{k}(\eta )>\eta \ ,\qquad 0<\eta <1~.  \label{T-eta}
\end{equation}

\noindent \textit{Proof.} Lemma \ref{Teta} follows immediately from
Proposition \ref{Integral}.\textbf{i}--\textbf{iv}, whose statements and
proof are deferred to Appendix \ref{PPI}. A positive polynomial is a
polynomial with positive coefficients. The degree and positivity of $T_{k}$
follows from items \textbf{i.} and \textbf{ii.} and inequality (\ref{T-eta})
is stated in item \textbf{iv.}

\hfill $\Box $

Proposition \ref{Integral}, the most technical result of the present work,
will be useful also in the proof of Theorem \ref{bestmate}.

\begin{remark}
\label{tk}$\mathcal{T}_{k}(t)$ for $k=2$ differs from the other values of $k$%
: $\mathcal{T}_{2}(t)=\mathcal{Q}_{2}(t)=\eta (t)$. Note that, by (\ref{fk}%
), $h_{2}$ has exactly one term and the second term in the r.h.s. of (\ref%
{Tk}) does not appear for $k=2$. As (\ref{Deltaeta}) deviates from the
linear function $\eta $ for $k>2$ (see Fig. \ref{fig5}), the majorant
sequence (\ref{bPhik}) is susceptible to be improved (see Remark \ref{phi0}
below).
\end{remark}

\begin{figure}[tbp]
\centering
\includegraphics[scale=0.55]{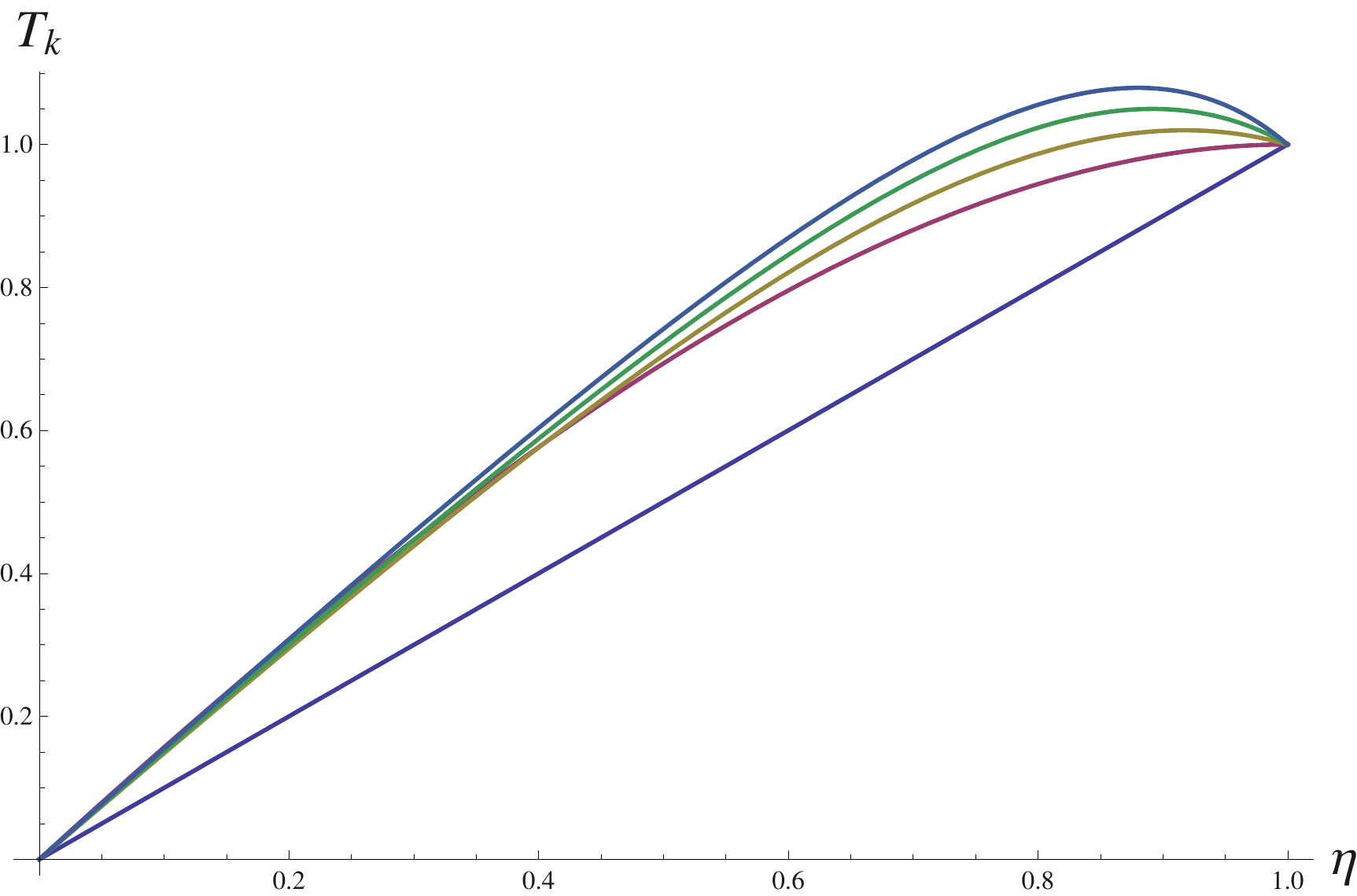}
\caption{The function $T_{k}(\protect\eta )$ for $k=2,\ldots ,6$.}
\label{fig5}
\end{figure}

Applying Lemma \ref{Teta} to (\ref{bk}), the majorant sequence $\left( \Psi
_{k}\right) _{k\geq 1}$, given by (\ref{Phik}), is redefined for $k\geq 2$
as: %\begin{subequations}
\begin{equation}
\left\vert \gamma _{k}(t)\right\vert \leq \frac{1}{\varepsilon }h_{k}\left(
\Psi _{1},\ldots ,\Psi _{k-1};t\right) -\eta \lambda ^{k}:=\Psi _{k}(t)~;
\label{bPhik}
\end{equation}%
summing (\ref{bPhik}) multiplied by $\rho ^{k}$, yields an algebraic
equation: %\end{subequations}
\begin{equation*}
\frac{1}{1-\Psi }-\left( 1+2\Psi \right) +\lambda \rho -\eta \frac{\lambda
^{2}\rho ^{2}}{1-\lambda \rho }=0
\end{equation*}%
which is equivalent to a slightly modified quadratic polynomial equation for 
$\Psi $ (see (\ref{AB}) et seq.), whose solution is given by (\ref{Phih})
with $H(r)$ replaced by%
\begin{equation*}
H_{1}(r)=\frac{1}{4}\left( 1+r-\eta \frac{r^{2}}{1-r}-\sqrt{\left( 1+r-\eta 
\frac{r^{2}}{1-r}\right) ^{2}-8\left( r-\eta \frac{r^{2}}{1-r}\right) }%
\right) ~.
\end{equation*}%
As long as $\left\vert r\right\vert <1$, the discriminant polynomial $%
p_{1}(r)=$ $\left( 1-r^{2}-\eta r^{2}\right) ^{2}-8(r(1-r)^{2}-\eta
r^{2}(1-r))$ has four real roots:%
\begin{equation}
R_{\sigma ,\sigma ^{\prime }}(t,\varepsilon )=\frac{2-\sigma ^{\prime }\sqrt{%
1-\eta }+2\sigma \sqrt{1-\eta /2-\sigma ^{\prime }\sqrt{1-\eta }}}{1+\eta }\
,\qquad \sigma ,\sigma ^{\prime }\in \left\{ -1,1\right\} ~,  \label{Rss}
\end{equation}%
the smallest one, $R_{-,-}$, together with the threshold: $r=0.28952$, has
been depicted in Fig. \ref{kappa} as a function of $\eta =e^{-2\varepsilon
t} $.

It follows from (\ref{bPhik}) that $\Psi _{k}(t)$ is positive and monotone
increasing function of $t$, which proves that $\Psi $ is a majorant of $\rho
\varphi _{\rho }$ in the sense of Definition \ref{maj}. Moreover, as $\left.
H_{1}\right\vert _{\eta =1}(r)=r$ for $0\leq r\leq 1-1/\sqrt{2}$, all
coefficients of the series for $\rho \Phi _{\rho }$ em power of $\lambda
\rho $, except the first one, vanish in this limit as they are proportional
to $(1-\eta )$.

This concludes the proof of Theorem \ref{unique}.

\hfill $\Box $

\section{Convergence of Power Series: Miscellaneous Methods\label{CPS}}

\setcounter{equation}{0} \setcounter{theorem}{0}

We summarize the result obtained so far and present some extensions,
conclusions and comparisons.

\begin{enumerate}
\item The radius of convergence $\mathcal{R}_{\Phi }(t)$ for the majorant $%
\Phi $ of $\varphi $ (r.h.s. of (\ref{rhoepsilon})) varies monotonously from 
$\infty $ ($=\left( 1-1/\sqrt{2}\right) /\lim_{t\rightarrow 0}\lambda (t)$)
to $2\varepsilon \cdot 0.171573$ ($=\left( 3/2-\sqrt{2}\right)
/\lim_{t\rightarrow \infty }\lambda (t)$) as $t$ varies from $0$ to $\infty $%
. We recall that $\lambda =2B_{2}$, where $B_{2}$ is the second virial
coefficient.

\item The radius of convergence for the pressure $P(t,\rho )$, the Helmholtz
free energy $F(t,\rho )-F^{\mathrm{ideal}}(\rho )=-\mathfrak{\beta }(t,\rho
) $ and the generating function $\varphi (t,\rho )$ satisfy%
\begin{equation*}
\mathcal{R}_{P}(t)=\mathcal{R}_{\mathfrak{\beta }}(t)=\mathcal{R}_{\varphi
}(t)\geq \mathcal{R}_{\Phi }(t)
\end{equation*}%
by (\ref{Pan}), Proposition \ref{PDEs} and (\ref{ans}). Inequality (\ref{Rt}%
) in Corollary \ref{vs} improves Lebowitz--Penrose lower bound (\ref{R0})
for nonnegative potentials (put $\kappa =1$ and $B=\lambda $ there) and
surpass the threshold $0.278465/\lambda $ (see (\ref{RW})), attainable via
Mayer series (Theorem 4.3.2 of \cite{Ruelle} et. seq.) for $\eta \geq
0.99463 $ (or $\varepsilon t\leq 0.00538$).
\end{enumerate}

\paragraph{Asymptotic solution as $\protect\varepsilon t$ tends to $0$}

As observed in Remark \ref{tk}, the majorant sequence (\ref{bPhik}) may be
modified to improve (\ref{Rt}) further but we shall not follows this route.
Instead, we re--address equations (\ref{lambdaeta}) et seq., satisfied by
the coefficients $\delta _{k}=k\beta _{n}/\lambda ^{k}$ of the power series
of $\rho \varphi _{\rho }$ in the variable $r$, in order to obtain their
asymptotic limit as $\varepsilon t$ tends to $0$.

Our third conclusion, due to Theorem \ref{asympt} below, together with (\ref%
{Z})-(\ref{RW}), is

\begin{enumerate}
\item[3.] For $t\varepsilon \ll 1$, $\lambda \mathcal{R}_{P}=1$ holds even
though $\lambda \mathcal{R}_{P}\geq W(e^{-1})=0.278465...$, by Lagrange's
inversion formula.
\end{enumerate}

\begin{theorem}
\label{asympt}Let (\ref{sol1}) be the solution to the initial value problem (%
\ref{eq}) in power series of $\rho $ and write $\delta _{k}=k\beta
_{k}/\lambda ^{k}$, where $\lambda $ is given by (\ref{ft}). Then, $\delta
_{1}(t)=-1$ and, for $k\geq 2$, $\delta _{k}(t)$ is $\mathcal{C}^{1}$ with $%
\lim_{t\rightarrow 0}\delta _{k}(t)=0$ and
\end{theorem}

\begin{equation}
\lim_{t\rightarrow 0}\frac{\delta _{k}(t)}{t}=\dot{\delta}%
_{k}(0)=(-1)^{k+1}k\varepsilon \ .\qquad  \label{derivative}
\end{equation}%
The irreducible cluster integrals is thus given by $\beta _{1}=-\lambda $
and 
\begin{equation}
\beta _{k}(t)=(-1)^{k+1}\lambda ^{k}(t)\varepsilon t(1+O(\varepsilon t))\
,\qquad k\geq 2~,\ \varepsilon t\rightarrow 0  \label{betak}
\end{equation}%
where the $O(\varepsilon t)$ term has negative sign and%
\begin{equation}
\varphi (t,\rho )\sim -t\rho -\frac{t^{2}\rho ^{2}}{1+t\rho }\varepsilon t
\label{pole}
\end{equation}%
is analytic in the domain $t\left\vert \rho \right\vert <1$, as $\varepsilon
t$ tends to $0$.

\noindent \textit{Proof.} Let $\mathbf{\delta }=\left( \delta _{k}\right)
_{k\geq 1}$ be the sequence of coefficients defined by (\ref{lambdaeta})-(%
\ref{deltakt}). The derivative of (\ref{deltakt}), reads%
\begin{equation}
\dot{\delta}_{k}=-k(k+1)h_{k}-\varepsilon k(k+1)\delta _{k}+k\frac{\dot{%
\lambda}}{\lambda }\delta _{k}\ .~  \label{a}
\end{equation}%
We observe that $\lambda =(1-\eta )/2\varepsilon \sim t$, $\dot{\lambda}%
=\eta \sim 1$ and, with $\delta _{1}=-1$ and $\lim_{t\rightarrow 0}\delta
_{k}=0$, for $k\geq 2$, (which will be proven by induction), 
\begin{equation}
\frac{h_{k}}{\varepsilon }\sim \left( \underset{k}{\underbrace{\mathbf{%
\delta }\ast \cdots \ast \mathbf{\delta }}}\right) _{k}=(-1)^{k}~.
\label{aa}
\end{equation}%
These, together with (\ref{deltakt}), imply 
\begin{equation}
\lim_{t\rightarrow 0}\frac{1}{1-\eta }\delta _{k}=-\frac{k(k+1)}{2}\frac{1}{%
t^{k+1}}\int_{0}^{t}\left( -1\right) ^{k}s^{k}ds=\left( -1\right) ^{k}\frac{k%
}{2}~.  \label{aaa}
\end{equation}%
Substituting (\ref{aa}) and (\ref{aaa}) into (\ref{a}) gives (\ref%
{derivative}).

Now, we prove that (\ref{betak}), for $k\geq 2$, hold by induction. For $k=2$%
, $\delta _{2}=-(1-\eta )=-2\varepsilon t+O\left( (\varepsilon t)^{2}\right) 
$. Let us assume that $\delta _{j}(t)$ is a $\mathcal{C}^{1}$ function
satisfying%
\begin{equation}
\delta _{j}(t)=(-1)^{j+1}j\varepsilon t\left( 1+O\left( \varepsilon t\right)
\right) \ ,\qquad j=2,\ldots ,k-1~.  \label{deltaj}
\end{equation}%
Since%
\begin{equation}
\left( \underset{j}{\underbrace{\mathbf{\delta }\ast \cdots \ast \mathbf{%
\delta }}}\right) _{k}=(-1)^{k+1}j(k-j+1)\varepsilon t+O\left( (\varepsilon
t)^{2}\right) \ ,\qquad j=2,\ldots ,k-1  \label{deltadeltaj}
\end{equation}%
we have, by (\ref{fk}), together with (\ref{aa}),%
\begin{equation*}
h_{k}(\delta _{1},\ldots ,\delta _{k-1};s)=\left( -1\right) ^{k}\varepsilon
(1+O(\varepsilon s))
\end{equation*}%
is $\mathcal{C}^{1}$ and, by (\ref{deltakt}),%
\begin{eqnarray*}
\delta _{k}(t) &=&-2\varepsilon tk(k+1)\frac{1}{t^{k+1}}\int_{0}^{t}\left(
-1\right) ^{k}s^{k}\left( 1+O(\varepsilon s)\right) ds \\
&=&(-1)^{k+1}k\varepsilon t\left( 1+O\left( \varepsilon t\right) \right)
\end{eqnarray*}%
proving (\ref{deltaj}) for $j=k$. Observe that the error term depends
algebraically on $k$, has negative sign, and $\lim_{t\rightarrow 0}\delta
_{k}(t)=0$ holds for any $k$ fixed. Consequently, with $r=\lambda \rho <1$,%
\begin{eqnarray*}
\sum_{k=1}^{\infty }\beta _{k}(t)r^{k} &=&-r-\sum_{k=2}^{\infty
}(-1)^{k}r^{k}\varepsilon t\left( 1+O\left( \varepsilon t\right) \right) \\
&=&-r-\frac{r^{2}}{1-r}\varepsilon t\left( 1+O\left( \varepsilon t\right)
\right)
\end{eqnarray*}%
proving (\ref{pole}).

\hfill $\Box $

\paragraph{The stationary solution $\protect\varphi _{0}$ of (\protect\ref%
{eq}): upper bound for $\mathcal{R}_{\protect\varphi} (\infty )$}

We compare $\mathcal{R}_{\Phi }(\infty )=2\varepsilon \times 0.171573$ with
the radius of convergence $\mathcal{R}_{\psi _{0}}$ of the power series $%
\psi _{0}(\rho )=\sum_{k\geq 1}\tilde{\gamma}_{n}\rho ^{k}$ whose
coefficients $\tilde{\gamma}_{k}$'s solve the system of integral equations (%
\ref{akt}) in the limit as $t$ goes to $\infty $. Although the (limit)
system of integral equations can be solved explicitly, the limit in $\rho
\varphi _{\rho }(\infty ,\rho )=\lim_{t\rightarrow \infty }\sum_{k\geq
1}\gamma _{n}\rho ^{k}$ can be passed inside the sum in the domain $%
\left\vert \rho \right\vert <\mathcal{R}_{\Phi }(\infty )$, for which (\ref%
{sol1}) is known to be uniformly convergent, by Definition \ref{maj}.
Consequently, 
\begin{equation*}
\mathcal{R}_{\Phi }(\infty )\leq \mathcal{R}_{\varphi }(\infty )\leq 
\mathcal{R}_{\psi _{0}}
\end{equation*}%
with equality $\mathcal{R}_{\varphi }(\infty )=\mathcal{R}_{\psi _{0}}$
being satisfied if, and only if, the limit and sum in $\rho \varphi _{\rho
}(\infty ,\rho )$ can be interchanged.

The $\tilde{\gamma}_{k}$'s are shown in Proposition \ref{equiv} below to be
the coefficients of $\rho \varphi _{0}^{\prime }(\rho )$ in power series of $%
\rho $, with $\varphi _{0}$ the stationary solution (\ref{varphi0}) of (\ref%
{eq}). We shall first calculate $\psi _{0}=\rho \varphi _{0}^{\prime }$.
Setting $\varphi _{t}=0$ in (\ref{form}): $\mathcal{J}(\rho ,\rho \varphi
_{0}^{\prime })=0$ where $\mathcal{J}$ is given by (\ref{J}), yields%
\begin{equation}
\frac{\rho }{2}+\varepsilon \frac{\psi _{0}}{1-\psi _{0}}=0
\label{stationary}
\end{equation}%
whose solution%
\begin{equation}
\psi _{0}(\rho )=\frac{-\rho }{2\varepsilon }\frac{1}{1-\rho /(2\varepsilon )%
}~,  \label{psi0}
\end{equation}%
implies $\mathcal{R}_{\psi _{0}}=2\varepsilon =1/\lambda (\infty )$. In view
of (\ref{Pinfty}), we have

\begin{enumerate}
\item[4.] 
\begin{equation}
0.171573\leq \lim_{t\rightarrow \infty }\lambda (t)\mathcal{R}_{P}(t)\leq 1~
\label{kapainfty}
\end{equation}%
holds In the \textquotedblleft low--temperature\textquotedblright\ limit ($%
t\rightarrow \infty $).
\end{enumerate}

\begin{remark}
\label{phi0} Substituting in (\ref{bk}) the inequality $T_{k}(\eta )\geq 
\frac{67}{45}\eta $, valid for $k\geq 3$ as $\eta \rightarrow 0$ (see Fig. %
\ref{fig5}), the lower bound $0.171573$ in (\ref{kapainfty}) can be replaced
by $0.275451$, still far from $1$ (the upper bound). Since $\varphi
_{0}(\rho )$ converges in a domain $\left\vert \rho /(2\varepsilon
)\right\vert <1$ so large as the domain $\left\vert \rho \lambda
(t)\right\vert <1$ of convergence of $\varphi $ for $t\varepsilon $ small
enough, in our second paper we investigate the power series solution \textit{%
(\ref{fifi})} of (\ref{eq}) in exponential time variable $\eta
=e^{-2\varepsilon t}$ (transeries, see e.g. \cite{Costin}), asymptotic to $%
\varphi _{0}$ as $\varepsilon t\rightarrow \infty $, but we have found that
solution belongs to a branch different from the one obtained by solving (\ref%
{akt}) for $t<\infty $.
\end{remark}

\begin{proposition}
\label{equiv}Let the $\tilde{\gamma}_{k}$'s be recursively defined by 
\begin{equation}
\tilde{\gamma}_{k}=-k(k+1)\lim_{t\rightarrow \infty }e^{-\varepsilon
k(k+1)t}\int_{0}^{t}e^{\varepsilon k(k+1)s}h_{k}(\tilde{\gamma}_{1},\ldots ,%
\tilde{\gamma}_{k-1};s)ds~,\qquad k\geq 2  \label{integraleq}
\end{equation}%
with $\tilde{\gamma}_{1}=-1/(2\varepsilon )$. Then, the system of integral
equations is equivalent to%
\begin{equation}
\tilde{\gamma}_{k}=\frac{-1}{\varepsilon }h_{k}(\tilde{\gamma}_{1},\ldots ,%
\tilde{\gamma}_{k-1}),\qquad k\geq 2~,  \label{algebraic}
\end{equation}%
with $\tilde{\gamma}_{1}=-1/(2\varepsilon )$, whose solution%
\begin{equation*}
\tilde{\gamma}_{k}=-\left( \frac{1}{2\varepsilon }\right) ^{k},\qquad k\geq
1~,
\end{equation*}%
are the coefficients of (\ref{psi0}) in power series of $\rho $.
\end{proposition}

\noindent \textit{Proof.} The equivalence between (\ref{integraleq}) and (%
\ref{algebraic}) is shown by induction. Since $\tilde{\gamma}_{1}$ is
constant, $h_{2}(\tilde{\gamma}_{1};s)=$ $\tilde{\gamma}_{1}^{2}$ does not
depend on $s$ and (\ref{integraleq}) reads%
\begin{equation*}
\tilde{\gamma}_{2}=-h_{2}(\tilde{\gamma}_{1})\lim_{t\rightarrow \infty
}6e^{-6\varepsilon t}\int_{0}^{t}e^{6\varepsilon s}ds=\frac{-1}{\varepsilon }%
h_{2}(\tilde{\gamma}_{1})\lim_{t\rightarrow \infty }\left(
1-e^{-6\varepsilon t}\right) =\frac{-1}{\varepsilon }h_{2}(\tilde{\gamma}%
_{1})
\end{equation*}%
proving (\ref{algebraic}) for $k=2$. Assuming that $\tilde{\gamma}%
_{1},\ldots ,\tilde{\gamma}_{n-1}$ has been obtained by equation (\ref%
{algebraic}), then $h_{n}(\tilde{\gamma}_{1},\ldots ,\tilde{\gamma}_{n-1};s)$
does not depend on $s$, and 
\begin{eqnarray*}
\tilde{\gamma}_{n} &=&-n(n+1)\lim_{t\rightarrow \infty }e^{-\varepsilon
n(n+1)t}\int_{0}^{t}e^{\varepsilon n(n+1)s}h_{n}(\tilde{\gamma}_{1},\ldots ,%
\tilde{\gamma}_{n-1};s)ds \\
&=&\frac{-1}{\varepsilon }h_{n}(\tilde{\gamma}_{1}\ldots ,\tilde{\gamma}%
_{n-1})\lim_{t\rightarrow \infty }\left( 1-e^{-n(n+1)\varepsilon t}\right)
\end{eqnarray*}%
which proves (\ref{algebraic}) for $k=n$ and concludes the induction. For
the second part, multiplying by $\rho ^{k}$ and summing over $k\geq 2$ both
sides of (\ref{algebraic}), yields%
\begin{equation*}
\psi _{0}+\frac{\rho }{2\varepsilon }=\frac{-\psi _{0}^{2}}{1-\psi _{0}}~,
\end{equation*}%
by $\tilde{\gamma}_{1}=-1/(2\varepsilon )$, which is equivalent to (\ref%
{stationary}), concluding the proof.

\hfill $\Box $

Since $\varphi _{0}^{\prime }(\rho )=\psi _{0}(\rho )/\rho =\left( \log
(1-\rho /2\varepsilon )\right) ^{\prime }$ and $\varphi _{0}(0)=0$, we have%
\begin{equation}
\varphi _{0}(\rho )=\log \left( 1-\frac{\rho }{2\varepsilon }\right) ~.
\label{varphi0}
\end{equation}%
Replacing $\varphi _{0}^{\prime }(\rho )$ into the first line of (\ref{Pan}%
), taking into account that%
\begin{equation*}
\psi _{0}(\rho )=1-\frac{1}{1-\rho /(2\varepsilon )}=\left( \rho
+2\varepsilon \log \left( 1-\rho /(2\varepsilon )\right) \right) ^{\prime }
\end{equation*}%
yields, for fixed $\varepsilon >0$, the pressure at the stationary (ground)
state%
\begin{equation}
P_{0}(\rho )=-2\varepsilon \log \left( 1-\frac{\rho }{2\varepsilon }\right)
~.  \label{Pinfty}
\end{equation}

Note that the singularity at $\rho =2\varepsilon $ of $P_{0}(\rho )$ is on
the (positive) real line and its is of the same type as the one in Ford's
model (\ref{Ford}). Summarizing the conclusions of the last two paragraphs,
we distinguish three different limits for the pressure of our simple
particle system:%
\begin{eqnarray*}
\lim_{t\rightarrow 0}P(t,\rho ) &=&\rho ~ \\
\lim_{\varepsilon \rightarrow 0}P(t,\rho ) &=&\rho +\frac{t}{2}\rho ^{2} \\
\lim_{t\rightarrow \infty }P(t,\rho ) &=&-2\varepsilon \log \left( 1-\frac{%
\rho }{2\varepsilon }\right) ~
\end{eqnarray*}%
attained for $\rho $ on the domain of convergence (see items $3.$ and $4.$
above).

\begin{remark}

\begin{enumerate}
\item Mayer coefficients (\ref{bb}) satisfy $(-1)^{n-1}b_{n}\geq 0$ (see
e.g. Theorem 4.5.3 of \cite{Ruelle}) for any nonnegative potential (the
uniformly repulsive in consideration, in particular) which means that the
first (leading) singularity of $\wp (z)$ is at the negative real axis for
all $t>0$. Theorem \ref{asympt} shows that $(-1)^{n-1}\beta _{n}\geq 0$
holds in the limit as $\varepsilon t$ goes to $0$ whereas (\ref{Pinfty}),
the pressure as $t$ goes to $\infty $, is singular at $\rho =2\varepsilon >0$%
. The explicit computation of few coefficients $\delta _{n}=\gamma
_{n}/\lambda ^{n}=n\beta _{n}/\lambda ^{n}$, $n=1,\ldots ,29$, in Fig. \ref%
{fig6}, of the power series of $\rho \varphi _{\rho }$ in $r$, revels the
interpolation of these two distinct behaviors: the $\delta _{n}$'s, as a
function of $t$, oscillates very much for $t$ small but becomes more like
each other as $t$ increases. They will eventually converges to $-1$ (the
coefficients of $\psi _{0}=\rho \varphi _{0}^{\prime }$ -- see (\ref{psi0}%
)). The time $t_{n}^{\ast }$ for which $\delta _{n}$ starts to converge from
its maximum to $-1$ depends on $n$ and the convergence is very fast.
Oscillations and change of signs of the $\delta _{n}$'s, as function of $n$,
are shown in Fig. \ref{fig7} for different fixed values of $t$. Sinusoidal
oscillations occurs for most $t$ (excluding special values), the period of
which increasing with $t$.

\item \label{R2}(\ref{Pinfty}) can be rescaled to become independent of $%
\varepsilon $. As a matter of fact,%
\begin{equation*}
P_{\varepsilon }(t,\rho )=\varepsilon P_{1}\left( \varepsilon t,\frac{\rho }{%
\varepsilon }\right)
\end{equation*}%
holds for any $\left( t,\rho \right) \in \mathbb{R}_{+}\times \mathbb{C}$ by
(\ref{eq}) and (\ref{Pan}). We shall call (\ref{Pinfty}) equation of state
of Ford's type as it is similar to equation (\ref{Ford}), for $\left\vert
\rho \right\vert <1/2$, and to the equation of state of a hard--core lattice
gas.
\end{enumerate}
\end{remark}

\begin{figure}[tbp]
\centering \includegraphics[scale=0.57]{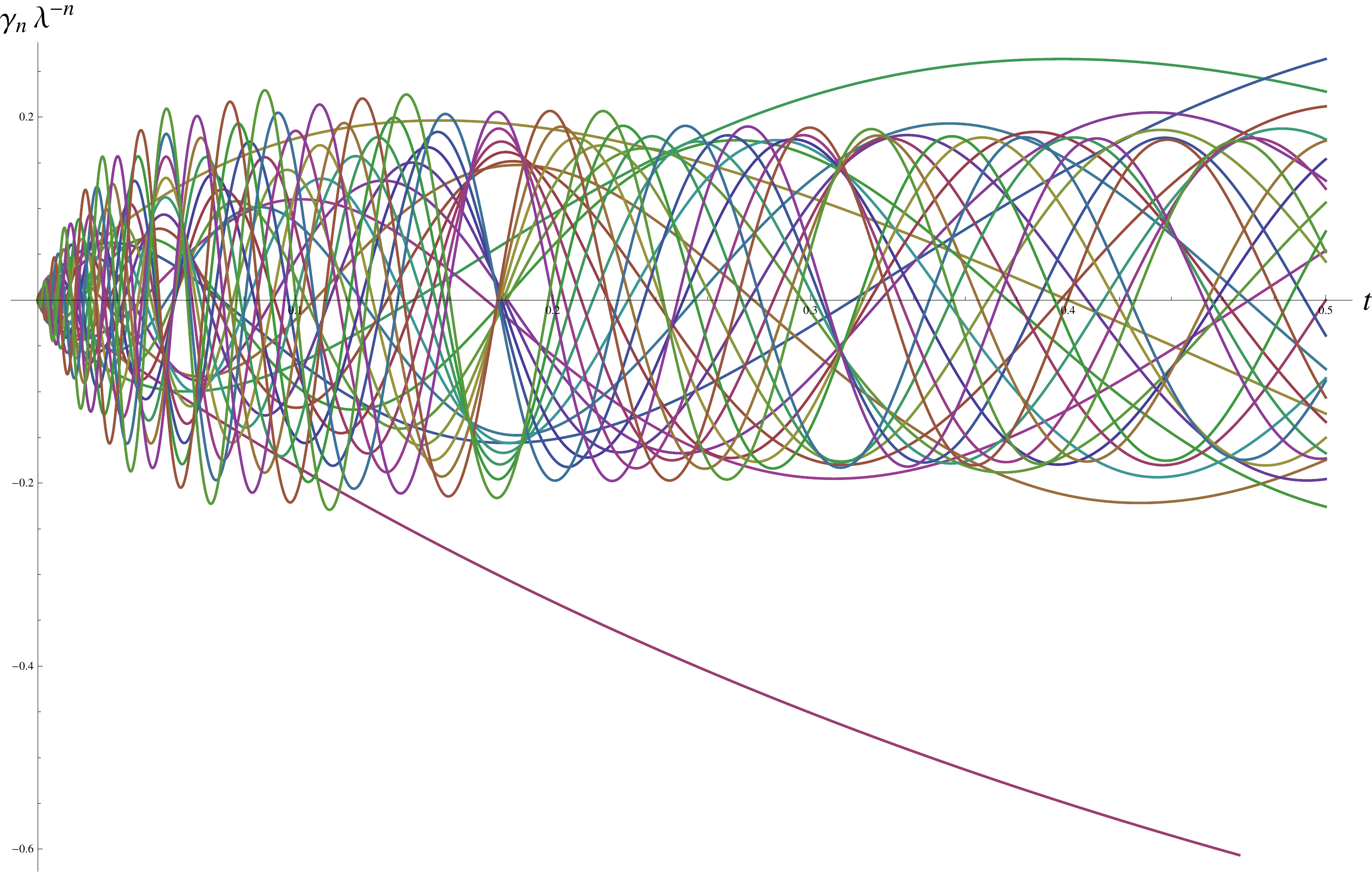} \centering %
\includegraphics[scale=0.57]{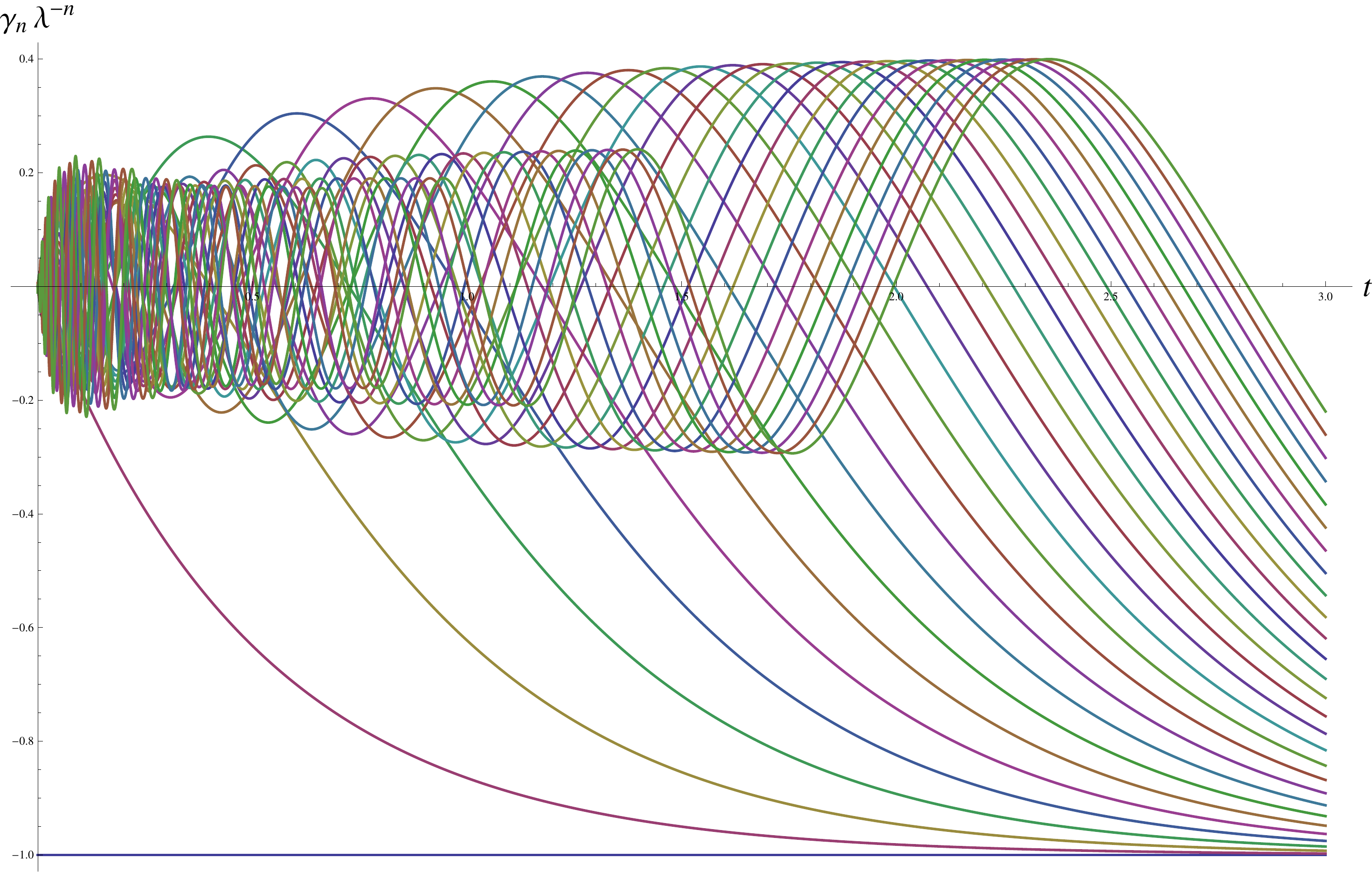}
\caption{The first 29 coefficients of the power series of $\protect\rho 
\protect\varphi _{\protect\rho }$ in $r = \protect\lambda \protect\rho $, as 
$t$ varies in $[0,0.5]$ (top) and $[0,3]$ (botom). }
\label{fig6}
\end{figure}

\begin{figure}[tbp]
\centering \includegraphics[scale=0.47]{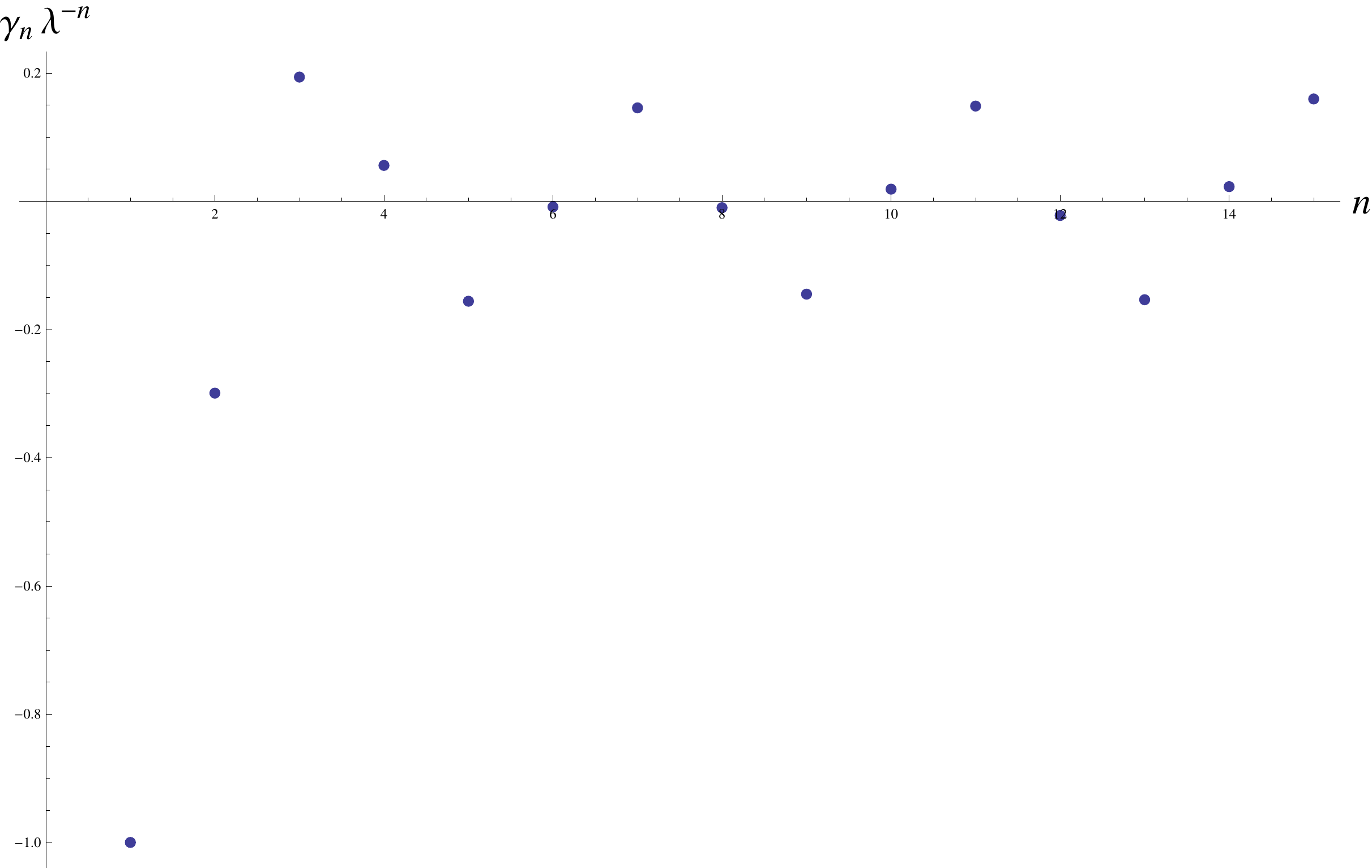} \centering %
\includegraphics[scale=0.46]{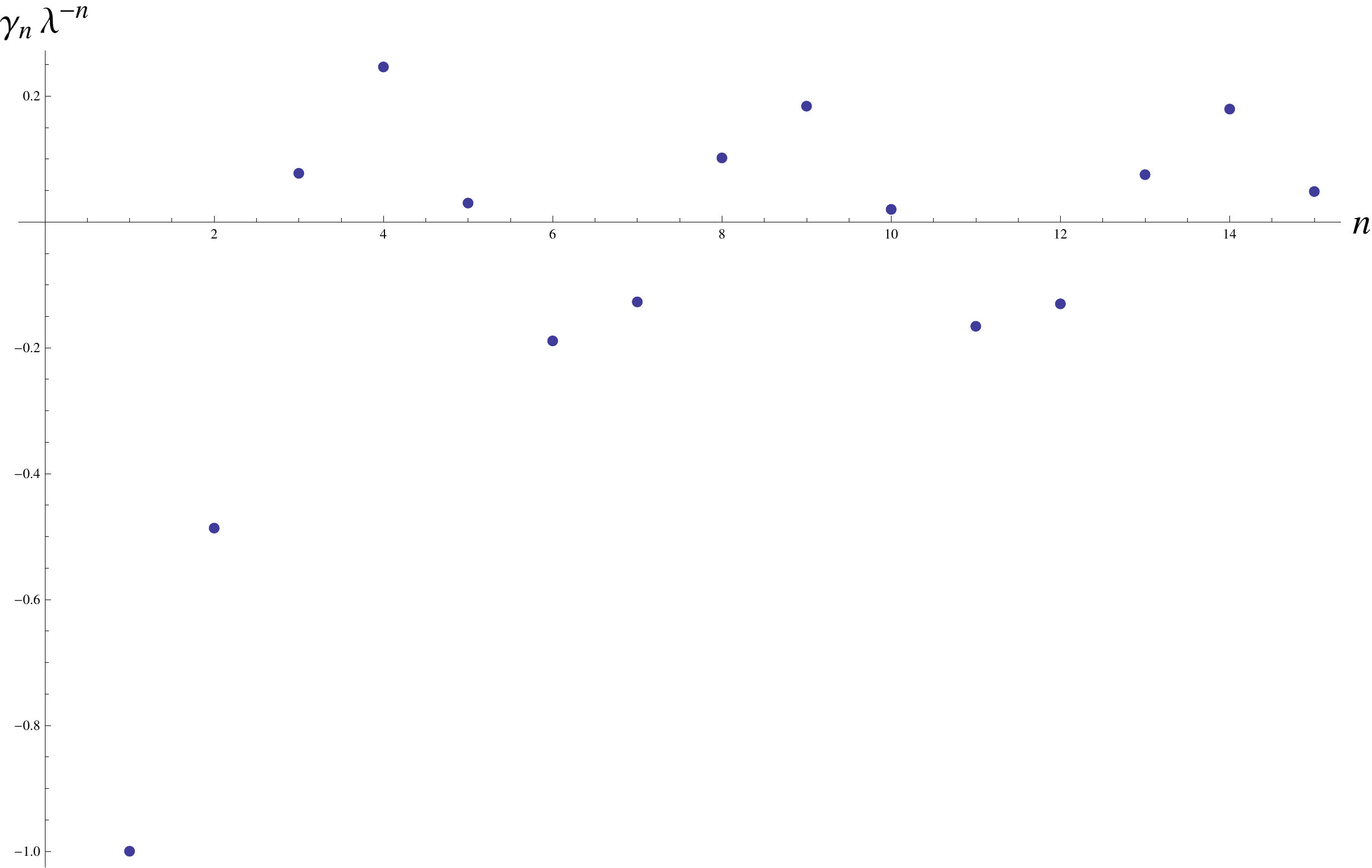} \centering %
\includegraphics[scale=0.47]{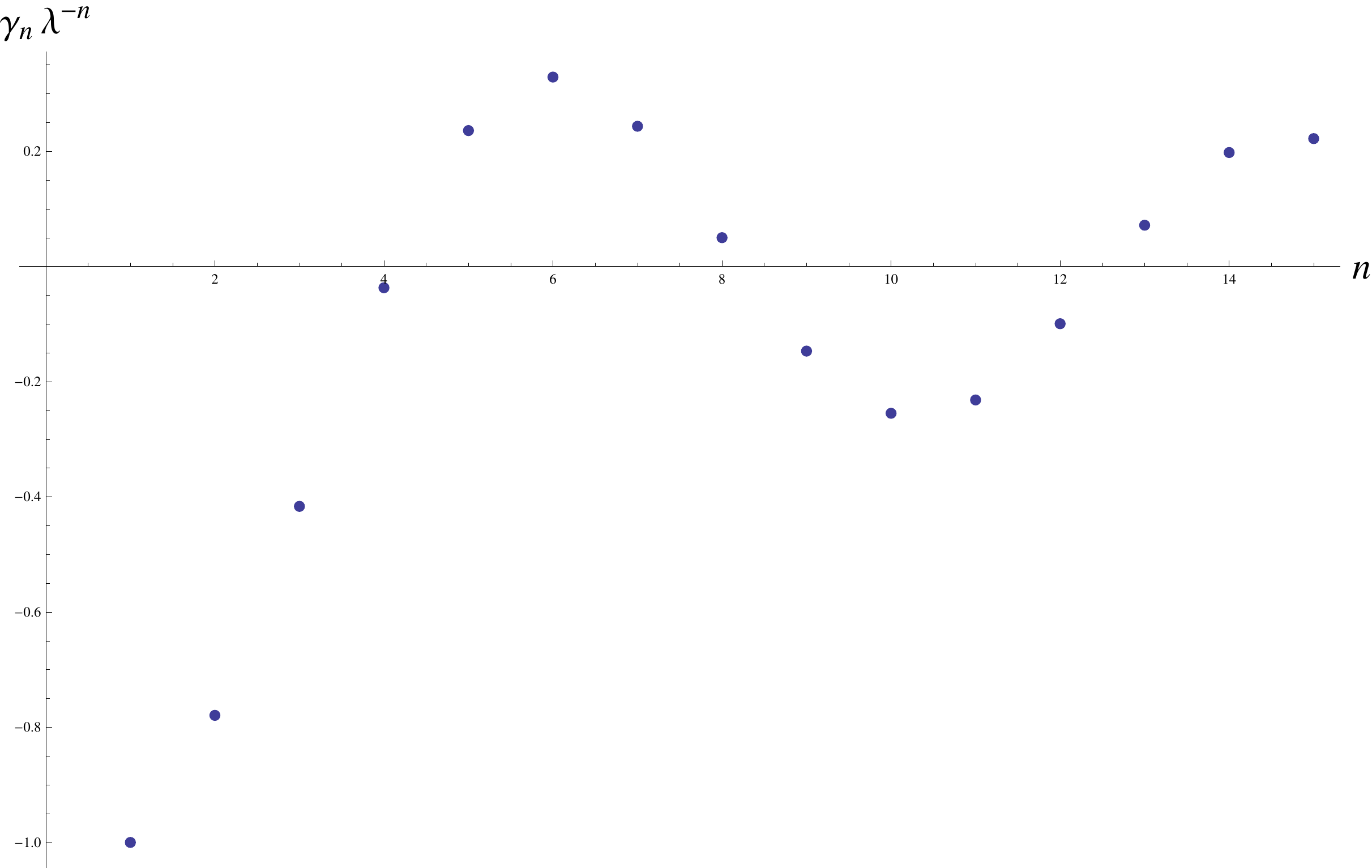}
\caption{The value of the first 15 coefficients of the power series of $%
\protect\rho \protect\varphi _{\protect\rho }$ in $r$ for $t=0.17$, $0.33$
and $0.75$. }
\label{fig7}
\end{figure}

\paragraph{Comparison with the Mayer series}

Equations (\ref{pde1}) and (\ref{gfeq}) have seen to be related to a
combinatorial problem involving the sum of weights (\ref{l}) over simply and
two-connected, respectively, Mayer graphs. Our aim is to investigate how the
reduction from simply to two-connected Mayer graphs affects the convergence
of their generating functions and characterize, as much as possible, the
nature of their leading singularities. In this investigation we are not
concerned with the Lagrange's inversion formula -- we just compare both
solutions, of (\ref{pde1}) and (\ref{gfeq}), as a power series in $z=e^{\mu
} $ and $\rho $, respectively.

Restricting our attention to (\ref{pde1}), we shall write a majorant series
whose radius of convergence attains the best known lower bound for $\mathcal{%
R}_{\wp }(t)$. We prove, in addition, that the radii of convergence of both,
the majorant and the Mayer series for equation (\ref{pde1}), are exactly the
same in the limit as $\varepsilon t$ tends to $0$.

\begin{theorem}
\label{bestmate}The (normalized) radius of convergence $\lambda (t)\mathcal{R%
}_{\wp }(t)$ of the Mayer series for the pressure (or density) of a system
of point--particles with uniformly repulsive pairwise potential is a
(strictly) monotone increasing function of $t\in \mathbb{R}_{+}$ and
satisfies%
\begin{equation}
\frac{1}{e}\leq \lambda \left( t\right) \mathcal{R}_{\wp }(t)\leq 1~
\label{e-1}
\end{equation}%
with equalities at the two extreme points $t=0$ and $\infty $. Moreover, $%
\lim_{\varepsilon \rightarrow 0}e\lambda \mathcal{R}_{p}=1$ for any $%
0<t<\infty $.
\end{theorem}

We thus conclude:

\begin{enumerate}
\item[5.] As $\varepsilon t$ tends to $0$, an expressive gain of equation (%
\ref{gfeq}) is seen by Theorem \ref{asympt} as $\allowbreak \allowbreak
0.367\,88 = e^{-1}\leq \lambda \left( t\right) \mathcal{R}_{\wp }(t)<\lambda
\left( t\right) \mathcal{R}_{\varphi }(t)=1$. In this limit, the radius of
convergence of the generator function of two-connected Mayer graphs is $2.7$
times larger then the radius of convergence of the generator function of
simply-connected ones.
\end{enumerate}

\noindent \textit{Proof of Theorem \ref{bestmate}.} Writing $q(t,z)=\wp
_{z}(t,z)=\displaystyle\sum_{k=0}^{\infty }q_{k}z^{k}$, by (\ref{ptmu}), we
have 
\begin{equation}
q_{k}=(k+1)b_{k+1}  \label{qb}
\end{equation}%
and equality of radii of convergences: $\mathcal{R}_{\wp }(t)=\mathcal{R}%
_{q}(t)$. The relation of the $q$--function to the density is $\rho
(t,z)=z\wp _{z}(t,z)=zq(t,z)$ and equation (\ref{pde1}) can be written in
the form of a conservation law%
\begin{equation}
q_{t}+\left( \varepsilon z^{2}q_{z}+\frac{1}{2}z^{2}q^{2}\right) _{z}=0
\label{qedp}
\end{equation}%
with $q(0,z)=1$. The sequence $\left( q_{k}\right) _{k\geq 0}$ of
coefficients of its power series, denoted by $\mathbf{q}$, satisfies%
\begin{eqnarray}
\dot{q}_{0} &=&0  \notag \\
\frac{1}{k+1}\dot{q}_{k}+\varepsilon kq_{k} &=&-\frac{1}{2}\left( \mathbf{q}%
\ast \mathbf{q}\right) _{k-1}~,\qquad k\geq 1  \label{q0qk}
\end{eqnarray}%
with $q_{0}=1$ and $q_{k}=0$, $k\geq 1$. Here, the convolution product $%
\left( \mathbf{q}\ast \mathbf{q}\right) _{m}=\displaystyle%
\sum_{j=0}^{m}q_{j}q_{m-j}$, in contrast with (\ref{prod}), is defined for $%
m\geq 0$. By the variation of constants formula, (\ref{q0qk}) are analogous
to%
\begin{equation*}
q_{0}(t)=1
\end{equation*}%
and%
\begin{equation}
q_{k}(t)=\frac{-1}{2}(k+1)\int_{0}^{t}e^{-\varepsilon k(k+1)(t-s)}\left( 
\mathbf{q}(s)\ast \mathbf{q}(s)\right) _{k-1}ds\ ,\qquad k\geq 1~.
\label{uint}
\end{equation}%
For $k=1$, we have%
\begin{equation*}
q_{1}(t)=\frac{-1}{2\varepsilon }\left( 1-e^{-2\varepsilon t}\right)
=-\lambda ~
\end{equation*}%
where $\lambda =\lambda \left( t,1/(2\varepsilon )\right) $ is given by (\ref%
{ft}). By convenience, we write $\lambda =(1-\eta )/(2\varepsilon )$, $\eta
(t)=e^{-2\varepsilon t}$, and introduce $\mathbf{\tilde{q}}=\left( \tilde{q}%
_{k}\right) _{k\geq 0}$ with $\tilde{q}_{k}=q_{k}/\eta ^{k}$, so (\ref{uint}%
) can be written as 
\begin{equation}
\tilde{q}_{k}(t)=\frac{-1}{2}(k+1)\int_{0}^{t}e^{-\varepsilon k(k-1)(t-s)}%
\frac{1}{\eta (s)}\left( \mathbf{\tilde{q}}(s)\ast \mathbf{\tilde{q}}%
(s)\right) _{k-1}ds\ .~  \label{qk}
\end{equation}

A sequence $\mathbf{\tilde{Q}}=\left( \tilde{Q}_{k}\right) _{k\geq 0}$ that
majorizes $\mathbf{\tilde{q}}=\left( \tilde{q}_{k}\right) _{k\geq 0}$
(putting $\tilde{Q}_{k}=Q_{k}/\eta ^{k}$, $\left( q_{k}\right) _{k\geq 0}$
is majorized by $\left( Q_{k}\right) _{k\geq 0}$), in the sense of
Definition (\ref{maj}), is obtained as follows: 
\begin{equation*}
\tilde{q}_{0}(t)=\tilde{Q}_{0}(\tau )=1\ 
\end{equation*}%
and 
\begin{eqnarray}
\left\vert \tilde{q}_{k}(t)\right\vert &\leq &\frac{1}{2}(k+1)%
\int_{0}^{t}e^{-\varepsilon k(k-1)(t-s)}\frac{1}{\eta (s)}\left( \left\vert 
\mathbf{\tilde{q}}(s)\right\vert \ast \left\vert \mathbf{\tilde{q}}%
(s)\right\vert \right) _{k-1}ds  \notag \\
&\leq &\frac{1}{2}(k+1)\int_{0}^{t}\frac{1}{\eta (s)}\left( \left\vert 
\mathbf{\tilde{q}}(s)\right\vert \ast \left\vert \mathbf{\tilde{q}}%
(s)\right\vert \right) _{k-1}ds  \notag \\
&\leq &\frac{1}{2}(k+1)\int_{0}^{\tau }\left( \mathbf{\tilde{Q}}(\tau
^{\prime })\ast \mathbf{\tilde{Q}}(\tau ^{\prime })\right) _{k-1}d\tau
^{\prime }:=\tilde{Q}_{k}(\tau )~.  \label{uPhi}
\end{eqnarray}%
where $\tau =\tau (t)=\displaystyle\int_{0}^{t}(1/\eta (s))ds=\lambda
(t)/\eta (t)$ is strictly increasing. Writing $\omega =\eta (t)z$, we have 
\begin{equation}
Q(t,z):=\sum_{k=0}^{\infty }Q_{k}(t)z^{k}=\sum_{k=0}^{\infty }\tilde{Q}%
_{k}(\tau )\omega ^{k}:=\tilde{Q}(\tau ,\omega )~  \label{majseries}
\end{equation}%
and the system of equations for $\left( \tilde{Q}_{k}\right) _{k\geq 1}$,
going backward through the steps (\ref{q0qk})-(\ref{uint}), is equivalent to
the following PDE (compare with (\ref{qedp})) 
\begin{equation}
\tilde{Q}_{\tau }-\frac{1}{2}\left( \omega ^{2}\tilde{Q}^{2}\right) _{\omega
}=0  \label{Qedp}
\end{equation}%
with $\tilde{Q}(0,\omega )=1$, whose solution can be explicitly written in
terms of the Lambert $W$--function (see Subsection 5.1 of \cite%
{Guidi-Marchetti}) 
\begin{equation}
Q(t,z)=\tilde{Q}(\tau ,\omega )=\frac{-1}{\tau \omega }W(-\tau \omega )=%
\frac{-1}{\lambda z}W(-\lambda z)  \label{Q}
\end{equation}%
(i.e., $W(x)$ is the principal branch of the inverse of $f(W)=We^{W}$,
regular at origin \cite{Corless-et-al}). As a consequence (see Appendix \ref%
{CVSO}), (\ref{majseries}) converges provided%
\begin{equation}
e\tau \left\vert \omega \right\vert =e\lambda \left\vert z\right\vert <1
\label{elz}
\end{equation}%
which, together with (\ref{uPhi}), establishes the first inequality of (\ref%
{e-1}): $e\lambda \mathcal{R}_{\wp }\geq e\lambda \mathcal{R}_{Q}\geq 1$.
Note that the majorant relation (\ref{uPhi}) (denoted by $q\ll Q$) is
preserved by derivation, anti--derivation integration w.r.t. $t$ and
composition.

To prove equality, we introduce another sequence $\mathbf{c}=\left(
c_{k}\right) _{k\geq 0}$ with $c_{k}=\left( -1\right) ^{k}q_{k}/\lambda ^{k}$%
. As one can see from (\ref{uint}), the factor $\left( -1\right) ^{n}$
compensates the alternating sign of $\left( q_{k}\right) _{k\geq 0}$ (see
Theorem 4.5.3 of \cite{Ruelle}) so $c_{k}\geq 0$ holds for every $k$. It
also follows from (\ref{uint}) that%
\begin{equation}
c_{k}(t)=\frac{k+1}{2}\frac{1}{\lambda ^{k}(t)}\int_{0}^{t}e^{-\varepsilon
k(k+1)(t-s)}\lambda ^{k-1}(s)\left( \mathbf{c}(s)\ast \mathbf{c}(s)\right)
_{k-1}ds\ ,\qquad k\geq 1  \label{ckint}
\end{equation}%
with $c_{0}(t)\equiv 1$. The asymptotic $\lambda (t)\sim t$, as $t$ tends to 
$0$, together with right continuity of $c_{k}(t)$ at $t=0$ yield%
\begin{equation}
c_{k}(0)=\frac{k+1}{2k}\left( \mathbf{c}(0)\ast \mathbf{c}(0)\right) _{k-1}\
,\qquad k\geq 1  \label{ck0}
\end{equation}%
with $c_{0}=1$, whose solution is well known (see Lemma \ref{BK} in Appendix %
\ref{PPI}): 
\begin{equation}
c_{k}(0)=\frac{(k+1)^{k}}{(k+1)!}\ .  \label{Wic}
\end{equation}%
Note that, by (\ref{qb}), $\left( -1\right) ^{n-1}c_{n-1}(0)=nb_{n}/\lambda
^{n-1}=(-n)^{n-1}/n!$ coincide with the coefficients of the Mayer series (%
\ref{nbn}) for the (conveniently normalized) density of a hard sphere gas in 
$d=\infty $ and also (in absolute value) with the coefficients of the
majorant function (\ref{Q}), proving the equality $\lim_{t\rightarrow
0}e\lambda \left( t\right) \mathcal{R}_{\wp }(t)=1$. The same is true as $%
\varepsilon $ tends to $0$ for any $t\in \mathbb{R}_{+}$, as one can see by
taking $\varepsilon =0$ in (\ref{qedp}) (or indirectly from (\ref{uint})),
whose solution is given by (\ref{Q}) with $z$ replaced by $-z$, by which the
sign of the nonlinear term in (\ref{Qedp}) changes from minus to plus.

\begin{figure}[tbp]
\centering\includegraphics[scale=0.55]{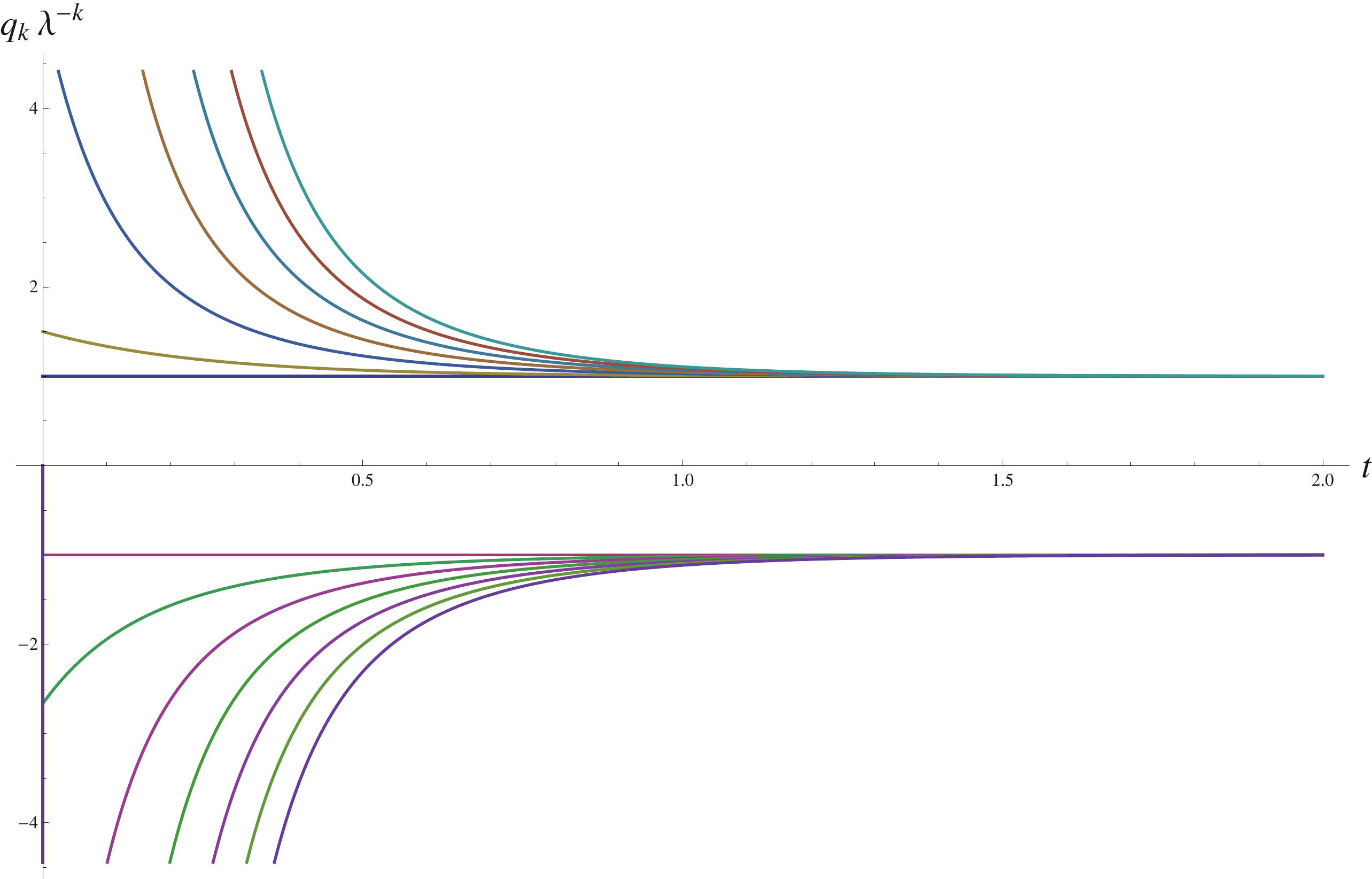}
\caption{Mayer's coefficients $q_{k}/\protect\lambda ^{k}$, $k=0,\ldots ,13$%
, as a function of $t$}
\label{figq}
\end{figure}

Differentiating (\ref{ckint}) w.r.t. $t$, gives 
\begin{equation}
\dot{c}_{k}=-\varepsilon k\left( k+1\right) c_{k}-\frac{1}{2\lambda }\left(
2\eta kc_{k}-\left( k+1\right) \left( \mathbf{c}\ast \mathbf{c}\right)
_{k-1}\right) ~,\qquad k\geq 2  \label{ckdot}
\end{equation}%
with $\dot{c}_{0}=\dot{c}_{1}\equiv 0$ and, to prove Theorem \ref{bestmate},
we need to show that $\dot{c}_{k}(t)$ remains negative for all $t\in \mathbb{%
R}_{+}$. We also need $\lim_{t\rightarrow \infty }c_{k}(t)=1$ for the upper
bound in (\ref{e-1}) -- see computer evaluation of $q_{k}(t)/\lambda ^{k}(t)$%
, $k=0,\ldots ,13$, in Fig. \ref{figq}. We formulate these statements in the
following proposition, whose proof (considerably more technical) is deferred
to Appendix \ref{PPI}.

\begin{proposition}
\label{contra}Let $\left( c_{k}(t)\right) _{k\geq 0}$ be given by $%
c_{k}=\left( -1\right) ^{k}q_{k}/\lambda ^{k}$ with $\left( q_{k}\right)
_{k\geq 0}$ the solution of (\ref{uint}) with $q_{0}(t)\equiv 1$. Then, for
every $k$, $c_{k}(t)$ is monotone decreasing in $t$, $c_{k}(0)=\lim_{t%
\searrow 0}c_{k}(t)$ satisfies (\ref{Wic}) and $\lim_{t\rightarrow \infty
}c_{k}(t)=1$.
\end{proposition}

This concludes the proof of Theorem \ref{bestmate}.

\hfill $\Box $

\begin{remark}
\label{majHJ}As $t$ tends to $\infty $, $\lambda \rightarrow 1/(2\varepsilon
)$, $q\rightarrow \displaystyle\sum_{k\geq 0}\left( -z/2\varepsilon \right)
^{k}=1/(1+z/2\varepsilon )$ and $\rho =zq\rightarrow z/(1+z/2\varepsilon )$
can be inverted $Z(\rho )=\rho /(1-\rho /2\varepsilon )$. The pressure $\wp
_{0}=2\varepsilon \displaystyle\sum_{k\geq 1}\left( -z/2\varepsilon \right)
^{k}/k$ composed with $Z(\rho )$: 
\begin{equation*}
P_{0}(\rho )=\wp _{0}\circ Z(\rho )=2\varepsilon \log \left( 1+\frac{\rho
/2\varepsilon }{1-\rho /2\varepsilon }\right) ~,
\end{equation*}%
agrees with the limit pressure (\ref{Pinfty}). In this limit, $Z(\rho )$ and
its inverse $\rho (z)$ are single--valued inside the disc $D_{2\varepsilon }$%
, the same disc for which the power series of $Z(\rho )$ and $P_{0}(\rho )$
converge. However, $D_{\varepsilon }$ is the largest disc such that $%
Z(D_{\varepsilon })\subset D_{2\varepsilon }$ and we can only conclude with
this information that $P_{0}(\rho )=\wp _{0}\circ Z(\rho )$ converges, at
least, in $D_{\varepsilon }$.
\end{remark}

\appendix

\section{Irreducible Cluster Integrals and their Relation to the Virial
Coefficients\label{ICI}}

\setcounter{equation}{0} \setcounter{theorem}{0}

The grand--canonical ensemble of interacting particles in a container $%
\Lambda $, i.e. a regular domain in $\mathbb{R}^{d}$ of volume $V=\left\vert
\Lambda \right\vert $, with activity (fugacity) $z$ at the inverse
temperature $\beta =1/(kT)$\ has a partition function%
\begin{equation}
\Xi _{\Lambda }(\beta ,z)=\sum_{n=0}^{\infty }z^{n}Q_{\Lambda ,n}(\beta )
\label{partition}
\end{equation}%
where ($Q_{\Lambda ,0}=1$)%
\begin{equation}
Q_{\Lambda ,n}(\text{$\beta $})=\frac{1}{n!}\int_{\Lambda ^{n}}e^{-\text{$%
\beta $}U(x)}d^{dn}x  \label{canonical}
\end{equation}%
is the canonical partition function and $U=U(x)$ is the pairwise interacting
energy of a configuration $x=\left( x_{1},\ldots ,x_{n}\right) \in \Lambda
^{n}=\Lambda \times \cdots \times \Lambda $ of $n$ particles.

Let $\Lambda ^{\sharp }=\bigcup_{n=0}^{\infty }\Lambda ^{n}$ be the space of
all (finite) $n$--tuple $x=\left( x_{1},\ldots ,x_{n}\right) \in \Lambda
^{n} $ ($\Lambda ^{0}$ is a set of a single point; the integral over $%
\Lambda ^{0} $ in (\ref{canonical}) is $1$, by convention) and let 
\begin{equation}
N:\Lambda ^{\sharp }\longrightarrow \left\{ 0,1,\ldots \right\} \equiv 
\mathbb{Z}_{+}  \label{N}
\end{equation}%
be a function that assigns the number of particles (components) to each
state $x=\left( x_{1},\ldots ,x_{n}\right) $ in $\Lambda ^{\sharp }$: $%
N(x)=n $. The equilibrium measure in $\Lambda ^{\sharp }$ can thus be
written as 
\begin{equation}
\mu _{\Lambda ,\text{$\beta $},z}(d^{\#}x)=\frac{1}{\Xi _{\Lambda }(\text{$%
\beta $},z)}\frac{z^{N(x)}}{N(x)!}e^{-\text{$\beta $}U(x)}d^{\sharp }x
\label{mu}
\end{equation}%
where, for each $n=0,1,\ldots $, the restriction of $d^{\sharp }x$ to $%
\Lambda ^{n}$ is the Lebesgue measure.

\begin{proposition}
\label{convex}The formal pressure $p=p_{\Lambda }(\beta ,\mu )$ and mean
density $\rho =\rho _{\Lambda }(\beta ,z)$, defined by\footnote{%
We denote by $p$ and $\wp $ the pressure as a function of the chemical
potential $\mu $ and activity $z=e^{\mu }$, respectively. The derivative of
the pressure with respect to $\mu $ and $z$ are related by $\partial
p/\partial \mu =z\partial \wp /\partial z$. The capital letter $P$ is
reserved to the pressure as a function of the density $\rho $.} 
\begin{equation}
\text{$\beta $}p_{\Lambda }(\text{$\beta $},\mu )=\frac{1}{\left\vert
\Lambda \right\vert }\log \Xi _{\Lambda }(\text{$\beta $},e^{\mu })
\label{p}
\end{equation}%
and%
\begin{equation}
\rho _{\Lambda }(\text{$\beta $},z)=\text{$\beta $}\frac{\partial p_{\Lambda
}}{\partial \mu }(\text{$\beta $},\mu )  \label{rho}
\end{equation}%
are, respectively, convex and monotone non--decreasing function of $\mu
=\log z$.
\end{proposition}

\noindent \textit{Proof.}%
\begin{eqnarray*}
\text{$\beta $}\frac{\partial ^{2}p_{\Lambda }}{\partial \mu ^{2}} &=&\frac{1%
}{\left\vert \Lambda \right\vert }\left( \mathbb{E}N^{2}-(\mathbb{E}%
N)^{2}\right) =\frac{1}{\left\vert \Lambda \right\vert }\text{Var}N\geq 0~ \\
\text{$\beta $}\frac{\partial p_{\Lambda }}{\partial \mu } &=&\frac{1}{%
\left\vert \Lambda \right\vert }\mathbb{E}N\geq 0
\end{eqnarray*}%
where $\mathbb{E}(\cdot )$ means expectation with respect to (\ref{mu}).

\hfill $\Box $

\paragraph{Ideal gas equation of state}

For an ideal gas, we set $U\equiv 0$ in (\ref{mu}). By (\ref{p}) and (\ref%
{rho}), we have 
\begin{equation}
\text{$\beta \wp $}=z\qquad \text{and}\qquad \rho =z~;  \label{pz-rhoz}
\end{equation}%
therefore, the equation of state $P(\rho )=\wp \circ Z(\rho )$ ($Z$ is the
inverse of $\rho (z)$) for an ideal gas reads%
\begin{equation*}
\text{$\beta $}P=\rho \ ,
\end{equation*}%
for all $\Lambda $.

\paragraph{Mayer series}

For a real (non--ideal) gas of particles interacting via a two--body
potential 
\begin{equation}
\phi (x_{i}-x_{j})\equiv \phi _{ij}  \label{u}
\end{equation}%
invariant under translations and rotations in $\mathbb{R}^{d}$, we have%
\begin{equation*}
U(x)=\sum_{1\leq i<j\leq N(x)}\phi _{ij}~.
\end{equation*}%
We define%
\begin{equation}
f_{ij}=e^{-\phi _{ij}}-1  \label{f}
\end{equation}%
and write the Boltzmann factor as a sum over the set $\mathcal{M}$ of Mayer
graphs $G$ (i.e., simple linear graphs) in the set $\left\{ 1,2,\ldots
,N(x)\right\} $ of labelled vertices: 
\begin{equation}
e^{-\text{$\beta $}U(x)}=\prod_{1\leq i<j\leq N(x)}\left( 1+f_{ij}\right)
=\sum_{G\in \mathcal{M}}\prod_{(ij)\in E}f_{ij}  \label{Bfactor}
\end{equation}%
where the product runs over the $(ij)$ in the set of edges $E=E(G)$ of $G$.

We define for every Mayer graph $G$ a weight 
\begin{equation}
W(G)=\int_{\Lambda ^{N}}\prod_{(ij)\in E(G)}f_{ij}(x)d^{\sharp }x~,
\label{W}
\end{equation}%
($N=N(G)$ being, by definition, the number of vertices in $G$) and write the
grand--partition function as%
\begin{equation*}
\Xi _{\Lambda }(z)=\sum_{G\in \mathcal{M}}\frac{z^{N}}{N!}W(G)~.
\end{equation*}%
The first Mayer theorem reads (see Theorem I of \cite{Uhlenbeck-Ford})

\begin{theorem}
\label{T1}%
\begin{equation}
\log \Xi _{\Lambda }(z)=\sum_{G\in \mathcal{M}:~G\text{ connected}}\frac{%
z^{N}}{N!}W(G)  \label{log}
\end{equation}
\end{theorem}

\noindent \textit{Proof. }We observe that the weight function (\ref{W}) is
independent of the labelling of the $N$ vertices and, for any Mayer graph $G$
whose connected parts are $G_{1}$, $\ldots $, $G_{k}$, we have $%
W(G)=W(G_{1})\cdots W(G_{k})$. These are the ingredients behind its proof,
which we refer to \cite{Ruelle,Brydges}.

\hfill $\Box $

By Theorem \ref{T1}, the pressure (\ref{p}) can be written as a formal power
series%
\begin{eqnarray}
\text{$\beta $}\wp _{\Lambda }(\text{$\beta $},z) &=&\frac{1}{\left\vert
\Lambda \right\vert }\sum_{G\in \mathcal{M}:~G~\text{connected}}\frac{z^{N}}{%
N!}W(G)  \notag \\
&=&\sum_{n=1}^{\infty }b_{\Lambda ,n}z^{n}:=\chi _{\Lambda }(z)  \label{b}
\end{eqnarray}%
where, for $n\geq 1$, 
\begin{equation}
b_{\Lambda ,n}=\frac{1}{n!}\sum_{\substack{ G\in \mathcal{M}:G\text{
connected,}  \\ N(G)=n}}\frac{1}{\left\vert \Lambda \right\vert }%
\int_{\Lambda ^{n}}\prod_{(ij)\in E(G)}f_{ij}(x)d^{\sharp }x  \label{bb}
\end{equation}%
are the Mayer coefficients. From now on we omit the dependence on the
inverse temperature $\beta $ to avoid notational conflict.

\paragraph{Irreducible cluster integrals}

Referring to (\ref{W}) with $G$ connected, we define $w(G)$ by holding one
vertex, let us say $x_{1}$, fixed at origin while $x=(x_{1},\ldots ,x_{N})$
is integrated over $\Lambda ^{N-1}$%
\begin{equation}
w(G)=\int_{\Lambda ^{N-1}}\prod_{(ij)\in E(G)}f_{ij}(x)d^{\sharp }x~.
\label{w}
\end{equation}%
By translational invariance of $\phi _{ij}$, in the thermodynamic limit, $%
w(G)$ is independent of the vertex to be fixed. Assuming $f_{ij}\in L^{1}(%
\mathbb{R}^{d})$, the limit of $W(G)/\left\vert \Lambda \right\vert $ and $%
w(G)$ exist along any sequence $\left( \Lambda _{m}\right) _{m\geq 1}$ of
regular domains tending to $\mathbb{R}^{N}$ (see e.g. \cite{Ruelle}), and
for any connected graph $G$, we have%
\begin{equation*}
\lim_{m\rightarrow \infty }\frac{1}{\left\vert \Lambda _{m}\right\vert }%
W_{m}(G)=\lim_{m\rightarrow \infty }w_{m}(G)=w_{\infty }(G)~.
\end{equation*}

\begin{definition}
A vertex $i_{0}$ is said to be an \textbf{articulation point} of a connected
Mayer graph $G$ if $G$ becomes disconnected after its removal. A graph $G$
with no articulation points is called a \textbf{block} or a \textbf{%
irreducible graph}.

A weight function $w$ is said to be \textbf{block--multiplicative} if for
any connected Mayer graph $G$, whose blocks are $G_{1}$, $\ldots $, $G_{k}$,
we have%
\begin{equation}
w(G)=w(G_{1})\cdots w(G_{k})~.  \label{b-m}
\end{equation}
\end{definition}

We define the Mayer graph consisting of a single vertex to be not a block.
The simplest block ($2$--block) consists of a single edge together with two
end points. The next simplest one ($3$--block) has three vertices and three
edges cyclically connected.

Referring to (\ref{log}), with connected Mayer graphs $G$ replaced by
blocks, we define 
\begin{equation*}
\mathcal{B}_{\Lambda }(\rho )=\sum_{G\in \mathcal{M}:G~\text{is a block}}%
\frac{\rho ^{N}}{N!}W(G)
\end{equation*}%
and, analogously,%
\begin{eqnarray}
\mathfrak{\beta }_{\Lambda }(\rho ) &=&\sum_{G\in \mathcal{M}:G~\text{is a
block}}\frac{\rho ^{N}}{N!}w(G)  \notag \\
&=&\sum_{n=1}^{\infty }\frac{1}{n+1}\beta _{n}\rho ^{n+1}  \label{beta}
\end{eqnarray}%
where (\ref{beta}) defines the $\beta _{n-1}$, which are called (\textbf{%
irreducible})\textbf{\ cluster integrals} of order $n$. The second Mayer
theorem may be stated as (see Theorem II of \cite{Uhlenbeck-Ford} for a
proof)

\begin{theorem}
\label{T2}The thermodynamic limits $\rho (z)=\lim_{n\rightarrow \infty }\rho
_{\Lambda _{n}}(z)$ and $\mathfrak{\beta }(\rho )=\lim_{n\rightarrow \infty }%
\mathfrak{\beta }_{\Lambda _{n}}(\rho )$, satisfy a functional equation 
\begin{equation}
\rho (z)=ze^{\mathfrak{\beta }^{\prime }\circ \rho (z)}~.  \label{rho-z}
\end{equation}
\end{theorem}

A simpler proof which holds for formal power series is provided in Leroux's
article on combinatorial species \cite{Leroux}, Theorem 1.3. The key
ingredient is the block--multiplicative property of the weight function (\ref%
{w}) in the thermodynamic limit (see Proposition 2.2 of \cite{Leroux}).

Theorem \ref{T2} also holds for finite $\Lambda $ provided the limit is
taken over $\Lambda _{m}=\left[ k_{m},k_{m}\right] ^{d}$, for an strictly
increasing sequence of positive numbers $\left( k_{m}\right) _{m\geq 1}$,
with $\phi =\phi (x)$ satisfying periodic boundary conditions: $\phi
(x)=\phi (x+k_{m}e_{j})$ for each direction $e_{j}$ of $\mathbb{R}^{d}$. In
this case $w_{m}(G)$ for any finite volume $\Lambda _{m}$ is
block--multiplicative (see Lemma 2 of \cite{Pulvirenti-Tsagkarogiannis}). We
gave in (\ref{ip}) another example of pair potential in which
block--multiplicativity (\ref{b-m}) holds for any $\Lambda \subset \mathbb{R}%
^{d}$, with $w(G)$ depending only on its volume $\left\vert \Lambda
\right\vert $. From here on, we deal with the formal power series (\ref{b})
and (\ref{beta}). For notational simplicity, we drop the index $\Lambda $
everywhere, independently of whether the thermodynamic limit has already
been taken.

\paragraph{Reduction of cluster integrals}

The relation between the cluster integrals $\left( b_{l}\right) _{l\geq 1}$
(Mayer coefficients) and the irreducible cluster integrals $\left( \beta
_{l}\right) _{l\geq 1}$ (related to virial coefficients) is obtained as
follows. By definitions (\ref{beta}),%
\begin{equation}
\mathfrak{\beta }^{\prime }(\rho )=\sum_{n=1}^{\infty }\beta _{n}\rho
^{n}:=\varphi (\rho )  \label{phi}
\end{equation}%
and the equation (\ref{rho-z}) can be written as%
\begin{equation}
z=\rho (z)e^{-\varphi \circ \rho (z)}~.  \label{z-rho}
\end{equation}%
Taking the derivative in both sides 
\begin{eqnarray*}
1 &=&\left( 1-\rho (z)(\varphi ^{\prime }\circ \rho )(z)\right) e^{-\varphi
\circ \rho (z)}\rho ^{\prime }(z) \\
&=&\left( 1-\rho (z)(\varphi ^{\prime }\circ \rho )(z)\right) \frac{z\rho
^{\prime }(z)}{\rho (z)}
\end{eqnarray*}%
yields, assuming provisionally that the series (\ref{phi}) converges
absolutely in a disc $D$ centered at $\rho =0$,%
\begin{eqnarray}
z\rho ^{\prime }(z) &=&\frac{\rho (z)}{1-\rho (z)(\varphi ^{\prime }\circ
\rho )(z)}  \label{z} \\
&=&\frac{1}{2\pi i}\oint_{C}\frac{ze^{\varphi (\rho )}}{\rho -ze^{\varphi
(\rho )}}d\rho  \label{zrho}
\end{eqnarray}%
by the residue theorem (see Theorem 9.1.1 of \cite{Hille} et seq.), with the
integral over a contour $C$ in $D$ containing the origin in its interior.
Note that, by the implicit function theorem (see e.g. Theorem 9.4.4 of \cite%
{Hille}) 
\begin{equation}
\rho -ze^{\varphi (\rho )}=0  \label{implicity}
\end{equation}%
has a unique holomorphic solution $\rho ^{\ast }=\rho ^{\ast }(z)$ for $z$
in a disc $D^{\prime }$ centered at $z=0$ with $\rho ^{\ast }(0)=0$ and $%
\rho ^{\ast \prime }(0)=1$ and $C$ can be chosen so that $\rho ^{\ast }(z)$
remains inside $C$ for every $z\in D^{\prime }$. By Rouch\'{e}'s theorem, 
\cite{Hille} that condition is expressed by%
\begin{equation}
\left\vert \frac{z}{\rho }e^{\varphi (\rho )}\right\vert <1\ ,\qquad \rho
\in C,  \label{rouche}
\end{equation}%
and (\ref{zrho}) can be written as%
\begin{eqnarray*}
z\rho ^{\prime }(z) &=&\frac{1}{2\pi i}\oint_{C}\sum_{l=1}^{\infty }\frac{%
z^{l}e^{l\varphi (\rho )}}{\rho ^{l}}d\rho \\
&=&\sum_{l=1}^{\infty }\left( \frac{1}{2\pi i}\oint_{C}\frac{e^{l\varphi
(\rho )}}{\rho ^{l}}d\rho \right) z^{l}
\end{eqnarray*}%
which, when compared to the series (the second derivative of (\ref{b}) times 
$z^{2}$):%
\begin{equation*}
z\rho ^{\prime }(z)=\sum_{l=1}^{\infty }l^{2}b_{l}z^{l}~,
\end{equation*}%
together with Cauchy theorem, gives $b_{1}=1$ and 
\begin{equation*}
b_{l}=\frac{1}{l^{2}}\left( e^{l\varphi (\rho )}\right) ^{[l-1]}(0)~
\end{equation*}%
for every $l\geq 2$ (from here on, $h^{[l]}(\rho _{0})=h^{(l)}(\rho _{0})/l!$
denotes $l$--th derivative of $h(\rho )$, divided by $l!$, at the point $%
\rho _{0}$).

Applying Fa\`{a} di Bruno formula (see e.g. \cite{Floater-Lyche}) 
\begin{equation}
(f\circ g)^{[l]}(\rho )=\sum_{\substack{ n_{1},\ldots ,n_{l}\geq 0:  \\ %
n_{1}+2n_{2}+\cdots +ln_{l}=l}}\frac{1}{n_{1}!\cdots n_{n}!}g^{[1]}(\rho
)^{n_{1}}\cdots g^{[l]}(\rho )^{n_{l}}\left( f^{(n_{1}+\cdots +n_{l})}\circ
g\right) (\rho )  \label{Faa}
\end{equation}%
for high order chain with $f(y)=e^{y}$ and $g(\rho )=l\varphi (\rho )=l\beta
^{\prime }(\rho )$ at $\rho =0$, yields the desired reduction of cluster
integrals in terms of the irreducible ones:%
\begin{equation}
b_{l}=\frac{1}{l^{2}}\sum_{\substack{ n_{1},\ldots ,n_{l-1}\geq 0:  \\ %
n_{1}+2n_{2}+\cdots +(l-1)n_{l-1}=l-1}}\prod_{i=1}^{l-1}\frac{\left( l\beta
_{i}\right) ^{n_{i}}}{n_{i}!}~.  \label{bbeta}
\end{equation}

\hfill $\Box $

\begin{remark}
We have applied the analytic function method (Lagrange's inversion formula)
proposed in \cite{Uhlenbeck-Khan} and developed in \cite{Born-Fuchs}
backward, i.e., starting from the functional equation (\ref{rho-z}). A
combinatorial interpretation of (\ref{bbeta}) is possible in terms of Husimi
graphs $G$ with $N(G)=l$, labelled by $\left\{ 1,\ldots ,l\right\} $, all of
whose blocks $K_{n_{j}}$, $j=1,\ldots ,k$, are complete graphs of size $%
n_{j} $ (see eq. (36) and Theorem 3.2 of \cite{Leroux}).
\end{remark}

\begin{remark}
\label{formal}The very same equations hold as a formal power series if,
instead of (\ref{zrho}), Lagrange --B\"{u}rmann formula \cite{Henrici} 
\begin{equation}
R\circ \rho (z)=\sum_{n=1}^{\infty }\frac{1}{n}\text{Res}\left( R^{\prime
}Z^{-n}\right) z^{n}  \label{R}
\end{equation}%
is applied to (\ref{z}). Here $Z\circ \rho (z)=z$, i.e., $Z(\rho )=%
\displaystyle\sum_{n\geq 1}c_{n}\rho ^{n}$ is the formal inverse of $\rho
(z) $%
\begin{equation*}
R(\rho )=\frac{\rho }{1-\rho \varphi ^{\prime }(\rho )}=\sum_{n=1}^{\infty
}a_{n}\rho ^{n}
\end{equation*}%
and, for $m\in \mathbb{Z}$, 
\begin{equation*}
Z^{m}(\rho )=\sum_{n=-\infty }^{\infty }c_{n}^{(m)}\rho ^{n}
\end{equation*}%
is the $m$--th power (not necessarily positive) of $Z$, defined on the ring
of formal Laurent series with finitely many negative subscripts different
from zero; $\text{Res}(F)$ in (\ref{R}) means the $a_{-1}$ coefficient of $F$%
.
\end{remark}

The inverse of (\ref{bbeta}),%
\begin{equation}
\beta _{k}=\sum_{\substack{ n_{2},\ldots ,n_{k+1}\geq 0:  \\ %
n_{2}+2n_{3}+\cdots +kn_{k+1}=k}}\left( -1\right) ^{n_{2}+\cdots +n_{k+1}-1}%
\frac{(k+n_{2}+\cdots +n_{k+1}-1)!}{k!}\prod_{i=2}^{k+1}\frac{\left(
ib_{i}\right) ^{n_{i}}}{n_{i}!}  \label{inverse}
\end{equation}%
can be obtained analogously (see eq. (49) of \cite{Mayer} and eqs.
(2.1)-(2.5) of \cite{Lebowitz-Penrose}). Let (\ref{rho-z}) be written as%
\begin{equation*}
\rho =Z(\rho )e^{\varphi (\rho )}~,
\end{equation*}%
where $Z(\rho )$ is the inverse of $\rho (z)$. Differentiating both sides
w.r.t. $\rho $ and using the fact that the condition (\ref{rouche}),
expressed by Rouch\'{e}'s theorem, is equivalent to $\left\vert \rho /\rho
(z)\right\vert <1$ for $z\in C^{\prime }=Z(C)$ where $C$ is the contour
chosen in (\ref{zrho}), we have%
\begin{eqnarray*}
\rho \varphi ^{\prime }(\rho ) &=&1-\frac{\rho }{Z(\rho )}\frac{1}{\rho
^{\prime }\circ Z(\rho )} \\
&=&\frac{1}{2\pi i}\oint_{C^{\prime }}\frac{1}{z}\left( 1-\frac{\rho (z)}{%
\rho (z)-\rho }\right) dz \\
&=&\frac{-1}{2\pi i}\oint_{C^{\prime }}\frac{1}{z}\frac{\rho /\rho (z)}{%
1-\rho /\rho (z)}dz \\
&=&\sum_{k=1}^{\infty }\left( \frac{-1}{2\pi i}\oint_{C^{\prime }}\frac{1}{%
\left( \rho (z)/z\right) ^{k}}\frac{1}{z^{k+1}}dz\right) \rho ^{k}
\end{eqnarray*}%
which implies%
\begin{equation*}
k\beta _{k}=-\left( (\rho (z)/z)^{-k}\right) ^{[k]}(0)~,
\end{equation*}%
by Cauchy formula, and (\ref{inverse}) by applying Fa\`{a} di Bruno formula (%
\ref{Faa}) -- the argument $\rho $ changed to $z$ -- with $f(y)=y^{-k}$ and $%
g(z)=\rho (z)/z=1+\displaystyle\sum_{n\geq 2}nb_{n}z^{n-1}$ at $z=0$.

\paragraph{Kamerlingh Onnes virial series\label{KO}}

Equation (\ref{z}) together with (\ref{rho}) yields (with $P(\rho )=\wp
\circ Z(\rho )$)%
\begin{eqnarray}
\beta P^{\prime }(\rho ) &=&\frac{\beta \wp ^{\prime }}{\rho ^{\prime }} 
\notag \\
&=&1-\rho \varphi ^{\prime }(\rho )  \label{phi-prime} \\
&=&1-\sum_{l=1}^{\infty }l\beta _{l}\rho ^{l}~.  \label{minus}
\end{eqnarray}

The Kamerlingh Onnes virial series%
\begin{equation}
\beta P(\rho )=\sum_{n=1}^{\infty }B_{n}\rho ^{n}  \label{vseries}
\end{equation}%
is thus obtained by integrating term--by--term (\ref{minus}):%
\begin{equation}
\beta P(\rho )=\rho \left( 1-\sum_{l=1}^{\infty }\frac{l}{l+1}\beta _{l}\rho
^{l}\right)  \label{P}
\end{equation}%
which gives, by (\ref{beta}), 
\begin{equation}
B_{n}=-\frac{n-1}{n}\beta _{n-1}  \label{Bbeta}
\end{equation}%
with%
\begin{equation}
\beta _{n-1}=\frac{1}{(n-1)!}\sum_{\substack{ G:G~\text{is a block}  \\ 
\text{of order }n}}\frac{1}{\left\vert \Lambda \right\vert }\int_{\Lambda
^{N}}\prod_{(ij)\in E(G)}f_{ij}(x)d^{\sharp }x~.  \label{betan-1}
\end{equation}

\paragraph{The Helmholtz Free--Energy}

Alternatively, we may follow another direction already known by Mayer (see
e.g. \cite{Mayer-Mayer}, eqs. (13.47)-(13.50))

\begin{proposition}
\label{legendre}The Helmholtz free energy density $F(\rho )=-(1/\left\vert
\Lambda \right\vert )\log Q_{\Lambda ,N}$, with $\rho =N/\left\vert \Lambda
\right\vert $ finite, in the thermodynamic limit, is formally given by the
Legendre transform%
\begin{equation}
F(\rho )=\sup_{\mu }\left( \mu \rho -\text{$\beta $}\wp (e^{\mu })\right)
=\rho \log \rho -\rho -\mathfrak{\beta }(\rho )  \label{F}
\end{equation}%
where $\mathfrak{\beta }(\rho )$ is defined by (\ref{beta}). Hence, by (\ref%
{phi}), $\varphi (\rho )=\mathfrak{\beta }^{\prime }(\rho )=-F^{\prime
}(\rho )+\log \rho $ generates the irreducible cluster integrals $\beta _{n}$%
, $n=1,2,\ldots $.
\end{proposition}

\noindent \textit{Proof.} As a formal power series, $\mu ^{\ast }=\mu ^{\ast
}(\rho )$ solves the equation $\rho =\beta \wp ^{\prime }(e^{\mu })\cdot
e^{\mu }$ for $\mu $, and $P(\rho )=\wp \circ e^{\mu ^{\ast }}(\rho )$ is,
by definition, the pressure (as a function of $\rho $) where $p(\mu )=\wp
(e^{\mu })$ is a convex function of $\mu \in \mathbb{R}$, by Proposition \ref%
{convex}. This implies that $F$ is a convex function of $\rho \in \mathbb{R}%
_{+}$ and 
\begin{equation*}
\text{$\beta $}p(\mu )=\sup_{\rho }\left( \mu \rho -F(\rho )\right) ~.
\end{equation*}%
The two first terms in the r.h.s. of (\ref{F}) are ideal gas contributions
to the Legendre transform: by (\ref{pz-rhoz}), $F^{\mathrm{ideal}}(\rho
)=\sup_{\mu }\left( \mu \rho -e^{\mu }\right) =$ $\mu ^{\ast }\rho -$$e^{\mu
^{\ast }}=\rho \log \rho -\rho $. The last term of (\ref{F}) is obtained by
solving (\ref{z-rho}) with $z=e^{\mu }$ and $\varphi $ given by (\ref{phi})
for $\mu $:%
\begin{equation}
\mu ^{\ast }=\mu ^{\ast }(\rho )=\log \rho -\sum_{n=1}^{\infty }\beta
_{n}\rho ^{n}  \label{mustar}
\end{equation}%
Replacing $\mu ^{\ast }$ into $F(\rho )=\mu ^{\ast }\rho -\beta \wp (e^{\mu
^{\ast }})=\mu ^{\ast }\rho -\beta P(\rho )$, together with (\ref{P}) and (%
\ref{beta}), yields (\ref{F}).

\hfill $\Box $

\section{Convergence of Virial Series: Overview and Previous Results\label%
{CVSO}}

\setcounter{equation}{0} \setcounter{theorem}{0}

The first convergence proof of virial series by Lebowitz--Penrose\cite%
{Lebowitz-Penrose} (see Ruelle \cite{Ruelle}, Theorem 4.3.2, commentary on
p. 86 for non--negative potentials and review MR0226924 by O. Penrose)
estimates $\mathcal{R}_{P}$ from an estimate on the convergence of the Mayer
series (\ref{b}). Lebowitz--Penrose's method provides a lower bound for $%
\mathcal{R}_{P}$ which inherits a limitation coming from a (nonphysical)
singularity, of combinatorial origin, that prevents the Mayer series to be
convergent beyond that point. We shall explain how this limitation has been
circumvented by our method.

\paragraph{ Gas of hard spheres in $d=\infty $: virial series}

\begin{figure}[tbp]
\centering\includegraphics[scale=0.50]{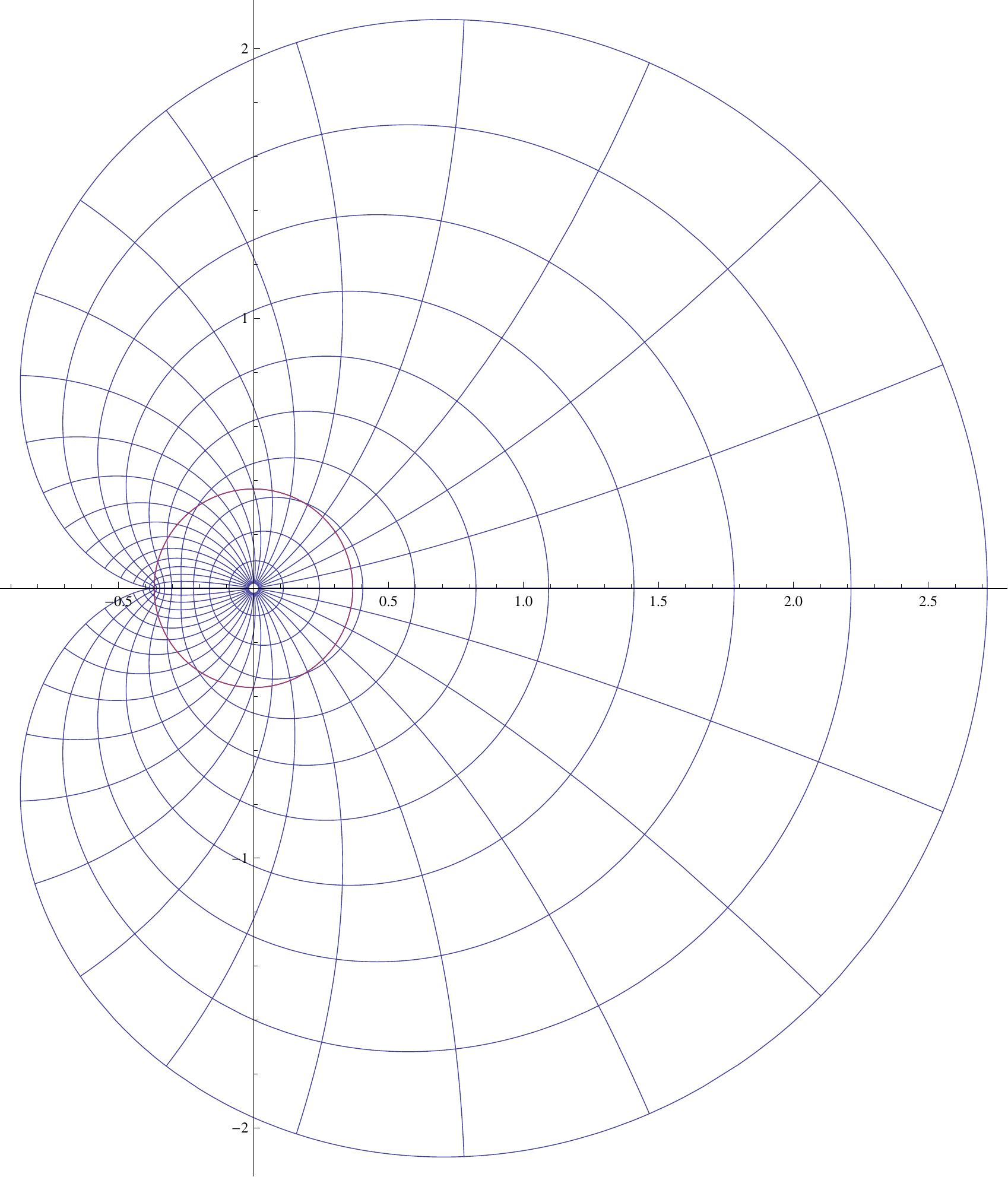}
\caption{Image of circles under $\protect\rho \longmapsto Z(\protect\rho )$}
\label{fig1}
\end{figure}

We begin by examining (\ref{implicity}) for a gas of hard spheres in
infinitely many dimensions. In this limit, since all $n$--blocks with $n>2$
give no contributions, we have (see eq. (6) of \cite{Frisch-Rivier-Wyler}) 
\footnote{%
We need to multiply $p$, $\rho $ and $z$ by its proper natural scale $%
\upsilon _{0}$, the volume of a $d$--dimensional sphere with radius $a$,
prior the limit of $d$ to infinity.} 
\begin{equation}
\varphi (\rho )=\rho ~.  \label{phi-rho}
\end{equation}%
We look at (\ref{z-rho}) as a map from the complex $\rho $--plane to the
complex $z$--plane: $\rho \in \mathbb{C}\longmapsto Z(\rho )\in \mathbb{C}$
where 
\begin{equation}
Z(\rho )=\rho e^{\rho }~.  \label{Z}
\end{equation}%
Figure \ref{fig1} depicts circles images $Z(\rho )$, $\left\vert \rho
\right\vert =\mathrm{const.}$, in the complex $\rho $--plane. Note the
formation of a cusp at $-e^{-1}$ as a consequence of the fact that $%
Z^{\prime }(\rho )=(1+\rho )e^{\rho }$ vanishes at $\rho =-1$. Recall that
an analytic function $f:D\longrightarrow \mathbb{C}$ is \textbf{univalent}
in an open domain $D$ if $f(\rho _{1})\neq f(\rho _{2})$ for all $\rho
_{1},\rho _{2}\in D$ with $\rho _{1}\neq \rho _{2}$. Hence, $Z(\rho )$ is
univalent in any disc $\mathbb{D}_{\tau }$ centered at origin with radius $%
\tau \leq 1$.

The inverse $\rho (z)$ of $Z(\rho )$ is the Lambert $W$--function $W(z)$, a
multivalued function whose principal branch is defined in the slit domain $%
\mathbb{C}\backslash (-\infty ,e^{-1}]$ (see e.g. \cite{Corless-et-al}).

The Mayer series $\rho (z)=\displaystyle\sum_{n\geq 1}nb_{n}z^{n}$ for the
hard-sphere density function is, in this limit, a sum of free diagrams which
may be evaluated from $\rho (z)=W(z)$ by the Lagrange method:%
\begin{equation}
nb_{n}=\frac{\left( -n\right) ^{n-1}}{n!}~;  \label{nbn}
\end{equation}%
its radius of convergence is thus $r=e^{-1}$. To recover the virial series, $%
P(\rho )=\wp \circ Z(\rho )$ is defined in a disc $\mathbb{D}_{r^{\prime }}$
of radius $r^{\prime }$ in the $\rho $--plane such for $\left\vert Z(\rho
)\right\vert <e^{-1}$ for every $\rho \in \mathbb{D}_{r^{\prime }}$, i.e., $%
Z(\mathbb{D}_{r^{\prime }})\subset \mathbb{D}_{e^{-1}}$. Since $\left\vert
Z(\rho )\right\vert \leq \left\vert \rho \right\vert e^{\left\vert \rho
\right\vert }$ holds as equality for $\rho \geq 0$ (with no absolute
values), by inverting $Z$ on the semi-line, we have 
\begin{equation}
r^{\prime }=W(e^{-1})=0.278465~  \label{RW}
\end{equation}%
(in Figure \ref{fig1}, the image of two (almost three) circles of radius $%
i/10$, $i=1,2,\ldots ,10$, are inside the disc $\mathbb{D}_{e^{-1}}$)
despite of $Z(\rho )$ has a radius of univalence $\tau =1$.

We can, however, obtain the virial series directly from (\ref{Z}). By (\ref%
{phi-prime}), (\ref{phi-rho}) and $P(0)=0$%
\begin{equation*}
\beta P(\rho )=\int_{0}^{\rho }\left( 1-\tilde{\rho}\varphi ^{\prime }(%
\tilde{\rho})\right) d\tilde{\rho}=\rho +\frac{1}{2}\rho ^{2}
\end{equation*}%
which is the equation of state (\ref{eq-state}). We observe that the image $%
P(\mathbb{D}_{\tau })$ of a disc $\mathbb{D}_{\tau }$ of radius $\tau $, is
a cardioid domain and the radius of univalence of $P(\rho )$ is also $\tau
=1 $. If, on the other hand, we have applied Fa\`{a} di Bruno (or Scott's)
formula (\ref{Faa}), together with the Lagrange--B\"{u}rmann formula (\ref{R}%
), to the formal power series $\wp (z)$ and $\rho (z)=\beta z\wp ^{\prime
}(z)$, we would see the cancellation of all terms of order larger than $2$: $%
\wp \circ Z(\rho )=\rho +\rho ^{2}/2$.

\paragraph{Lebowitz--Penrose lower bound}

The Lambert $W$--function plays an important rule for a system of particles
interacting via a pair potential $\phi $ satisfying \textbf{(i)} (\textit{%
stability) there exists }$\Phi \geq 0$ such that 
\begin{equation}
\exists ~\Phi \geq 0:\sum_{1\leq i<j\leq n}\phi (x_{i}-x_{j})\geq -n\Phi
\label{i}
\end{equation}%
for every $n\geq 2$ and $\left( x_{1},\ldots ,x_{n}\right) \in \mathbb{R}%
^{nd}$; and \textbf{(ii)} 
\begin{equation}
\left\Vert e^{-\text{$\beta $}\phi }-1\right\Vert _{1}=B(\beta )<\infty ~,
\label{ii}
\end{equation}%
with $\left\Vert f\right\Vert _{1}=\int_{\mathbb{R}^{d}}\left\vert
f(x)\right\vert dx$ the $L^{1}$--norm in $\mathbb{R}^{d}$. A $\phi $
satisfying \textbf{(i)} and \textbf{(ii)} is called regular potential; note
that \textbf{(i)} implies that $\phi $ is bounded from below. For systems
with regular potentials, Penrose's estimate\cite{Penrose} yields%
\begin{equation}
\left\vert \rho _{\Lambda }(z)-z\right\vert \leq \frac{-1}{\kappa ^{2}B}%
W(-\kappa B\left\vert z\right\vert )-\frac{\left\vert z\right\vert }{\kappa }
\label{upper}
\end{equation}%
where $\kappa =\exp (2\beta \Phi )\geq 1$, uniformly in $\Lambda $, provided%
\begin{equation}
we^{-w}\equiv \kappa B\left\vert z\right\vert <e^{-1}~.  \label{e}
\end{equation}%
Note that $0\leq w<1$ preserves this inequality. (\ref{upper}) together with 
$W(-we^{-w})=-w$, by definition of $W(z)$, imply 
\begin{eqnarray}
\left\vert \rho _{\Lambda }(z)\right\vert &\geq &\left( 1+\frac{1}{\kappa }%
\right) \left\vert z\right\vert +\frac{1}{\kappa ^{2}B}W(-\kappa B\left\vert
z\right\vert )  \notag \\
&=&((1+\kappa )e^{-w}-1)\frac{w}{\kappa ^{2}B}  \label{rho-lb}
\end{eqnarray}%
which, by maximizing in $w$ satisfying (\ref{e}), yields%
\begin{equation}
\left\vert \rho _{\Lambda }(z)\right\vert \geq 0.28952\frac{1}{(1+\kappa )B}%
\equiv R_{0}\ .  \label{R0}
\end{equation}

Applying Lagrange--B\"{u}rmann formula (\ref{R}) on the other way around,
i.e., replacing $R$ by $\beta \wp (z)$ and $\rho (z)$ by its inverse $%
Z=Z(\rho )$, together with $\beta z\wp ^{\prime }(z)=\rho (z)$, the
coefficients of virial series (\ref{vseries}) reads 
\begin{subequations}
\begin{equation}
B_{n}=\frac{1}{n}\text{Res}\left( \beta \wp ^{\prime }\rho ^{-n}\right) =%
\frac{1}{n}\text{Res}\left( z^{-1}\rho ^{-n+1}\right) =\frac{1}{n}\frac{1}{%
2\pi i}\oint_{\left\vert z\right\vert =\delta _{0}}\rho ^{-n+1}(z)\frac{dz}{z%
}\leq n^{-1}R_{0}^{-n+1}~,  \label{Bn}
\end{equation}%
or, equivalently, 
\end{subequations}
\begin{equation*}
\beta _{n}\leq n^{-1}R_{0}^{-n}
\end{equation*}%
by (\ref{Bbeta}).

The integral representation of $B_{n}$ (third equality in (\ref{Bn})) has
been derived by Lebowitz--Penrose (see eq. (2.5) in \cite{Lebowitz-Penrose})
using Lagrange's theorem. By the ratio test, $R_{0}$ is thus a lower bound
for the radius of convergence $\mathcal{R}_{P}$ of the virial series.
Recently, Morais--Procacci\cite{Morais-Procacci}, using the cluster
expansion proposed by \cite{Pulvirenti-Tsagkarogiannis} for dealing with the
canonical partition function, together with Penrose's estimate on the Mayer
coefficients, have (surprisingly) produced the same lower bound on the
radius of convergence of the series of Helmholtz free energy in powers of
the density $\rho =N/\left\vert \Lambda \right\vert $ (see Theorem 1 and
Remark 2 and 4 therein. Their expansion also yields (\ref{inverse}) in the
limit as $\left\vert \Lambda \right\vert $ goes to $\infty $).

\begin{figure}[tbp]
\centering\includegraphics[scale=0.50]{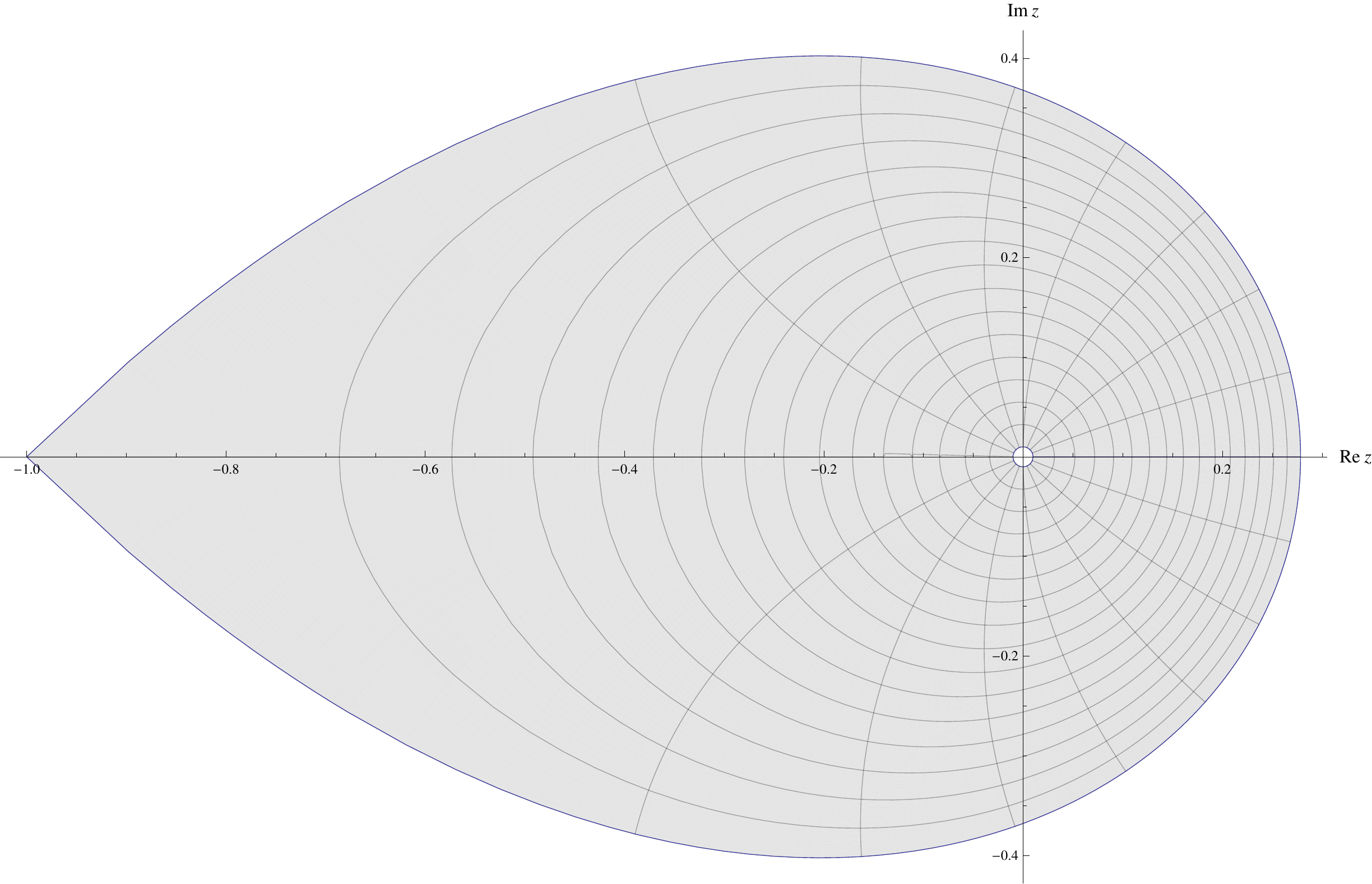}
\caption{Image of circles under $z \longmapsto \protect\rho (z)$}
\label{fig2}
\end{figure}

Apart from the factor $1+\kappa \geq 2$, which came into (\ref{rho-lb}) in
view of the estimation (\ref{upper}) by the majorant sum, the constant $%
0.28952$ appearing in both estimates is very near to $r^{\prime }$ defined
by (\ref{RW}). If Lagrange's theorem is applied to hard spheres in
infinitely dimensions, whose density is exactly given by $\rho (z)=W(z)$,
the same estimate $r^{\prime }$ on radius of convergence of its virial
series is obtained (see eq. (3.1) of \cite{Lebowitz-Penrose}):%
\begin{equation*}
\mathcal{R}_{P}^{\mathrm{h.s.}}\geq \mu \equiv \min_{z~\mathrm{on}\
C}\left\vert \rho (z)\right\vert =W(e^{-1})\equiv r^{\prime }
\end{equation*}%
where the contour $C=\left\{ e^{it-1},\ 0\leq t<2\pi \right\} $ has been
chosen on the domain $e\left\vert z\right\vert \leq 1$ such that the minimum
value on $C$ is the largest possible (see Fig. \ref{fig2}). However,
repeating the same steps (\ref{upper})-(\ref{R0}), 
\begin{equation*}
\left\vert \rho _{\Lambda }(z)-z\right\vert \leq -W(-\left\vert z\right\vert
)-\left\vert z\right\vert
\end{equation*}%
with $C=\left\{ z=re^{it},\ 0\leq t<2\pi \right\} $ and $we^{-w}\equiv
r<e^{-1}$, yields an estimate $0.52\times r^{\prime }$ (instead $0.5\times
r^{\prime }$)%
\begin{equation*}
\left\vert \rho _{\Lambda }(z)\right\vert \geq \max_{0\,\leq r<e^{-1}}\left(
2r+W(-r)\right) =\max_{w\geq 0}(2e^{-w}-1)w\simeq 0.144\,76=\frac{0.28952}{2}
~
\end{equation*}

The slightly increasing on the numerical factor, in comparison to $r^{\prime
}$, is intrinsic to the majorant method employed: the curve $C$ in the
Lagrange's theorem is chosen to minimize the amount lost by replacing the
coefficients of $\rho _{\Lambda }(z)-z$ by their absolute values. We have
done something similar in our majorant method in Section \ref{CMM} when (\ref%
{Phik}) is replaced by (\ref{bPhik}) in order the estimate on $\mathcal{R}%
_{\varphi }(t)$ exceeds the threshold $r^{\prime }$. This and the fact that
our lower bound on $\mathcal{R}_{P}(t)$ goes beyond Lebowitz-Penrose's for
nonnegative potentials are manifestations that our approach circumvents the
(nonphysical) singularity of the Lambert $W$--function.

\paragraph{Example of an equation of state presenting a plateau}

The purpose here is review an explicit example in which the condensation
phenomenon is not determined by the singularities present on $P(\rho )$.

Lee--Yang's theory\cite{Yang-Lee} is capable to explain condensation
directly from the partition function (\ref{partition}). To illustrate how
this phenomenon takes place, Uhlenbeck and Ford (see Section III.4 of \cite%
{Uhlenbeck-Ford}) have devised an artificial example in which the
grand--canonical partition function is given by\footnote{%
Although no potential system has been assigned, so far, to this model, there
is, beside the hard-core hypothesis (to make $\Xi _{\Lambda }$ a polynomial
in $z$), an \textquotedblleft attractive sign\textquotedblright\ behind the
assumption on the distribution of zeros over the unit circle.}%
\begin{eqnarray*}
\Xi _{\Lambda }(z) &=&\left( 1+z\right) ^{\left\vert \Lambda \right\vert }%
\frac{1-z^{\left\vert \Lambda \right\vert }}{1-z}~ \\
&=&\exp \left( \left\vert \Lambda \right\vert \log \left( 1+z\right)
+\sum_{n=0}^{\left\vert \Lambda \right\vert -1}\log \left( 1-ze^{-2\pi
in/\left\vert \Lambda \right\vert }\right) -\log \left( 1-z\right) \right) .
\end{eqnarray*}%
One sees that (see e.g. Example 5.7 in Chap. 0 of \cite{Saff-Totik}) 
\begin{equation}
\wp (z)=\lim_{\left\vert \Lambda \right\vert \rightarrow \infty }\frac{1}{%
\left\vert \Lambda \right\vert }\log \Xi _{\Lambda }(z)=\left\{ 
\begin{array}{ll}
\log \left( 1+z\right) & \text{if\ }\left\vert z\right\vert \leq 1 \\ 
\log \left( 1+z\right) +\log z & \text{if\ }\left\vert z\right\vert >1%
\end{array}%
\right.  \label{chi1}
\end{equation}%
is the logarithmic potential due to one unit of charge at $z=-1$ and one
unit of charge uniformly distributed on the unit circle. The pressure $%
P(\rho )=\wp \circ (z\wp ^{\prime })^{-1}(\rho )$ in the so called Ford
model does present a plateau%
\begin{equation}
P(\rho )=\left\{ 
\begin{array}{ll}
\log (1/(1-\rho )) & \text{if~}0\leq \rho <1/2 \\ 
\log 2 & \text{if~}1/2\leq \rho <3/2 \\ 
\log (\rho -1)/(2-\rho )^{2} & \text{if~}3/2\leq \rho <2%
\end{array}%
\right.  \label{Ford}
\end{equation}%
although Pad\'{e} approximation is unable to detect the singularity on its
virial expansion because is not of algebraic type and it is beyond the
critical saturation point $\rho _{\mathrm{sat.}}=1/2$.

Although the power series of $P_{1}(\rho )=\log (1/(1-\rho ))$ converges for 
$\left\vert \rho \right\vert <1$, the image of $\Re \mathrm{e}\rho >1/2$ by $%
Z_{1}(\rho )=(z\wp _{1}^{\prime })^{-1}(\rho )=\rho /(1-\rho )$ is on the
complement $\mathbb{C}\backslash \mathbb{D}_{1}$ of the unit disk $\mathbb{D}%
_{1}$, where $\wp _{1}(z)=\log \left( 1+z\right) $ is defined ($P(\rho )$
is, indeed, not defined at any point $\rho $ of the forbidden domain $%
1/2<\Re \mathrm{e}\rho <3/2$). The presence of plateau results from the
convex envelop of the Helmholtz free energy, defined by (\ref{F}) in each of
the two branches:%
\begin{equation*}
F(\rho )=\left\{ 
\begin{array}{ll}
\rho \log \rho +(1-\rho )\log (1-\rho ) & \text{if~}0\leq \rho <1/2 \\ 
-\log 2 & \text{if~}1/2\leq \rho <3/2 \\ 
(\rho -1)\log (\rho -1)+(2-\rho )\log (2-\rho ) & \text{if~}3/2\leq \rho <2%
\end{array}%
\right. \ 
\end{equation*}%
with the horizontal line being tangent to both curves $\left( \rho
,F_{i}(\rho )\right) $, $i=1,2$, the first at $\rho =1/2$ and the second at $%
\rho =3/2$.

\section{ Hamilton-Jacobi Equation\label{HJE}}

\setcounter{equation}{0} \setcounter{theorem}{0}

We shall show that the pressure $p=p(t,z)$, for a gas of interacting
point--particles with uniformly repulsive pair potential, satisfies
(exactly) a \textquotedblleft viscous\textquotedblright\ Hamilton--Jacobi
equation (\ref{pde1}) with $\mu =\log z$ the chemical potential, $%
\varepsilon =1/(2\left\vert \Lambda \right\vert )$ the inverse volume,
proving Proposition \ref{PDEs}, part \textbf{(a)}; $t$ represents both, the
\textquotedblleft inverse temperature\textquotedblright\ (repulsive
intensity) and the \textquotedblleft time evolution\textquotedblright\
(interpolating) parameter starting at $t=0$ from the ideal gas $p(0,z)=z$.
The repulsive (positive) interaction, expressed by the \textquotedblleft
wrong\textquotedblleft\ sign in front of the Laplacian, is responsible for
the equilibrium stability (i.e., avoids the collapse of a large number of
particles into a point).

\paragraph{Brydges--Kennedy's system of equations for the Ursell functions}

We modify slightly and extend some of the notions introduced in Appendix \ref%
{ICI}. For a $n$-tuple $x=\left( x_{1},\ldots ,x_{n}\right) $ of points in $%
\mathbb{R}^{d}$, let $x_{I}=\left( x_{i_{1}},\ldots ,x_{i_{s}}\right) $
denote the $s$-tuple, $s\leq n$, of points indexed by $I=\left\{
i_{1},\ldots ,i_{s}\right\} $ $\subseteq \left\{ 1,\ldots ,n\right\} $. We
denote by $\left\vert I\right\vert $ the cardinality of $I$.

Boltzmann $\psi (t,x)$ and Ursell $\psi ^{c}(t,x)$ functions are assigned to
each $t\in \mathbb{R}_{+}$ and $x\in \mathbb{R}^{\#}$ as follows. Given the
total energy of a configuration $x$ at $t$: $U(t,x)=\displaystyle\sum_{1\leq
i<j\leq N(x)}\phi \left( t;x_{i}-x_{j}\right) $, with $\phi \left(
t;x\right) $ satisfying the assumptions stated in Appendix \ref{ICI}, we
write (with $\beta =1$) 
\begin{equation}
\psi (t,x)=e^{-U(t,x)}~.  \label{psi}
\end{equation}

\begin{definition}
\label{Prod}Let $\circ $ denote the (algebraic) convolution product: 
\begin{equation*}
f\circ g(x)=\sum_{\substack{ I,J:I\cap J=\emptyset ,  \\ I\cup J=\{1,\ldots
,n\}}}f(x_{I})g(x_{J})
\end{equation*}%
for any pair of state functions $f$ and $g$ on $\mathbb{R}^{\#}$. The
(algebraic) exponential function $\text{Exp}\left( f\right) $ of $f$ is
(formally) given by%
\begin{equation}
\text{Exp}\left( f\right) (x)=\mathbf{1}(x)+f(x)+\frac{1}{2}\left( f\circ
f\right) (x)+\cdots  \label{Exp}
\end{equation}%
where $\mathbf{1}(x)=1$ if $x=0$ (i.e., $x\in \mathbb{R}^{0}=\left\{
0\right\} $) and $=0$ otherwise.

We define a inner product $\langle \cdot ,\cdot \rangle (z)$ as a (positive)
bilinear form 
\begin{equation}
\langle f,g\rangle (z)=\sum_{n=0}^{\infty }\frac{z^{n}}{n!}\int_{\mathbb{R}%
^{dn}}f(x)g(x)d^{\#}x~  \label{bilinear}
\end{equation}%
over any two state functions $f$ and $g$ on $\mathbb{R}^{\#}$ and $z\in 
\mathbb{R}_{+}$.
\end{definition}

We define the Ursell function recursively (in $n=N(x)$, $n=1,2,\ldots $) by
the equation%
\begin{equation}
\psi (t,x)=\text{Exp}\left( \psi ^{c}\right) (t,x)  \label{psic}
\end{equation}%
and write the indicator function of a state in $\Lambda ^{\sharp }$ as%
\begin{equation}
\chi _{\Lambda }(x)=\left\{ 
\begin{array}{ll}
1 & \text{if\ }x\in \Lambda ^{\sharp } \\ 
0 & \text{otherwise}%
\end{array}%
\right. ~.  \label{chi}
\end{equation}%
With these notations and definitions, we have

\begin{proposition}
\label{fpressure}The formal pressure (\ref{p}) may be written as%
\begin{equation}
p_{\Lambda }(t,\mu )=\frac{1}{\left\vert \Lambda \right\vert }\langle \chi
_{\Lambda },\psi ^{c}\rangle (e^{\mu })~.  \label{press}
\end{equation}
\end{proposition}

We need the following

\begin{lemma}
\label{deconv}%
\begin{equation*}
\langle \chi _{\Lambda },f\circ g\rangle =\langle \chi _{\Lambda },f\rangle
\langle \chi _{\Lambda },g\rangle
\end{equation*}
\end{lemma}

\noindent \textit{Proof.} Write $x=\left( x_{1},\ldots ,x_{n}\right) =\left(
x_{I},x_{J}\right) $ for any $I,J$ such that $I\cap J=\emptyset $ and $I\cup
J=\{1,\ldots ,n\}$. Then, by (\ref{chi}) and Definitions \ref{Prod}, we have 
\begin{equation*}
\chi _{\Lambda }(x)=\chi _{\Lambda }(x_{I})\chi _{\Lambda }(x_{J})~
\end{equation*}%
and%
\begin{eqnarray*}
\langle \chi _{\Lambda },f\circ g\rangle (z) &=&\sum_{n=0}^{\infty }\frac{%
z^{n}}{n!}\sum_{\substack{ I,J:I\cap J=\emptyset ,  \\ I\cup J=\{1,\ldots
,n\} }}\int_{\Lambda ^{I}}f(x_{I})d^{\#}x\int_{\Lambda ^{J}}g(x_{J})d^{\#}x
\\
&=&\sum_{n=0}^{\infty }\frac{z^{n}}{n!}\sum_{s=0}^{n}\dbinom{n}{s}%
\int_{\Lambda ^{s}}f(x)d^{\#}x\int_{\Lambda ^{n-s}}g(y)d^{\#}y \\
&=&\sum_{s=0}^{\infty }\frac{z^{s}}{s!}\int_{\Lambda
^{s}}f(x)d^{\#}x\sum_{n=s}^{\infty }\frac{z^{n-s}}{(n-s)!}\int_{\Lambda
^{n-s}}g(y)d^{\#}y \\
&=&\langle \chi _{\Lambda },f\rangle (z)\cdot \langle \chi _{\Lambda
},g\rangle (z)~.
\end{eqnarray*}

\hfill $\Box $

\noindent \textit{Proof of Proposition \ref{fpressure}.} By (\ref{p}), (\ref%
{partition}), (\ref{bilinear}) and (\ref{chi}), we have%
\begin{eqnarray*}
p_{\Lambda }(t,\mu ) &=&\frac{1}{\left\vert \Lambda \right\vert }\log \left(
\langle \chi _{\Lambda },\psi \rangle (e^{\mu })\right) \\
&=&\frac{1}{\left\vert \Lambda \right\vert }\log \left( \langle \chi
_{\Lambda },\text{Exp}(\psi ^{c})\rangle (e^{\mu })\right) \\
&=&\frac{1}{\left\vert \Lambda \right\vert }\langle \chi _{\Lambda },\psi
^{c}\rangle (e^{\mu })~.
\end{eqnarray*}%
in view of (\ref{Exp}) and Lemma \ref{deconv}.

\hfill $\Box $

The following is our main result in this paragraph. We refer to Lemma 3.3 of 
\cite{Brydges-Kennedy} for a proof.

\begin{proposition}
\label{BKeqs}If $\phi \left( t;x\right) $ is differentiable w.r.t. $t$ and
satisfies $\phi \left( 0;x\right) \equiv 0$, then the Brydges--Kennedy
system of equations 
\begin{equation}
f_{t}(t,x)+U_{t}(t,x)f(t,x)+\frac{1}{2}\sum_{\substack{ I,J:I\cap
J=\emptyset ,  \\ I\cup J=\left\{ 1,\ldots ,n\right\} }}%
U_{t}(t,x_{I},x_{J})f(t,x_{I})f(t,x_{J})=0\ ,\qquad t>0  \label{B-K}
\end{equation}%
for all $x\in \mathbb{R}^{\#}$, $N(x)=n$ and $n=1,2,\ldots $, together with
the initial condition%
\begin{equation*}
f(0,x)=\left\{ 
\begin{array}{lll}
1 &  & \text{if\ }N(x)=1 \\ 
0 &  & \text{if\ }N(x)>1%
\end{array}%
\right.
\end{equation*}%
has a unique solution given by the Ursell functions: $f=\psi ^{c}$. Here $%
U_{t}$ means derivative of $U$ w.r.t. $t$ and%
\begin{equation*}
U_{t}(t,x_{I},x_{J})=\sum_{i\in I,j\in J}\phi _{t}\left( t,\left\vert
x_{i}-x_{j}\right\vert \right) ~.
\end{equation*}
\end{proposition}

We apply Proposition \ref{BKeqs} to our system of point--particles with
uniformly repulsive pair--interacting potential (\ref{ip}).

\paragraph{Proof of Proposition \protect\ref{PDEs}, part (a)}

For $\phi _{\Lambda }(t;x)$ given by (\ref{ip}), write 
\begin{equation*}
U(t,x)=\displaystyle\sum_{1\leq i<j\leq N(x)}\phi _{\Lambda }(t;x_{i}-x_{j})
\end{equation*}%
and let $\psi (t,x)$ and $\psi ^{c}(t,x)$ be given by (\ref{psi}) and (\ref%
{psic}). Then, 
\begin{equation*}
U(t,x)=\frac{N(x)(N(x)-1)}{2}\frac{t}{\left\vert \Lambda \right\vert }
\end{equation*}%
holds for any state $x\in \Lambda ^{\#}$ and Brydges-Kennedy's system (\ref%
{B-K}) reads%
\begin{equation*}
\psi _{t}^{c}(t,x)+\frac{1}{2\left\vert \Lambda \right\vert }(n^{2}-n)\psi
^{c}(t,x)+\frac{1}{2\left\vert \Lambda \right\vert }\sum_{\substack{ %
I,J:I\cap J=\emptyset ,  \\ I\cup J=\left\{ 1,\ldots ,n\right\} }}\left\vert
I\right\vert \psi ^{c}(t,x_{I})\left\vert J\right\vert \psi ^{c}(t,x_{J})=0~.
\end{equation*}%
By (\ref{N}) and Definition \ref{Prod}, they can be written, more compactly,
as%
\begin{equation}
\psi _{t}^{c}+\frac{1}{2\left\vert \Lambda \right\vert }(N^{2}-N)\psi ^{c}+%
\frac{1}{2\left\vert \Lambda \right\vert }\left( N\psi ^{c}\circ N\psi
^{c}\right) =0~.  \label{PPPP}
\end{equation}%
Applying $\dfrac{1}{\left\vert \Lambda \right\vert }\langle \chi _{\Lambda
},\cdot \rangle (e^{\mu })$ to equation (\ref{PPPP}) together with the
following

\begin{lemma}
\label{two}%
\begin{eqnarray}
\frac{1}{\left\vert \Lambda \right\vert }\langle \chi _{\Lambda },N^{k}\psi
^{c}\rangle (e^{\mu }) &=&\frac{\partial ^{k}p_{\Lambda }}{\partial \mu ^{k}}%
(t,\mu )\ ,\qquad k=1,2  \label{Nk} \\
\frac{1}{\left\vert \Lambda \right\vert ^{2}}\langle \chi _{\Lambda },N\psi
^{c}\circ N\psi ^{c}\rangle (e^{\mu }) &=&\left( \frac{\partial p_{\Lambda }%
}{\partial \mu }\right) ^{2}(t,\mu )~;  \label{NN}
\end{eqnarray}
\end{lemma}

\noindent we arrive to desired PDE equation:%
\begin{equation*}
p_{t}+\varepsilon (p_{\mu \mu }-p_{\mu })+\frac{1}{2}\left( p_{\mu }\right)
^{2}=0~
\end{equation*}%
with $\varepsilon =1/(2\left\vert \Lambda \right\vert )$, reducing the proof
of Proposition \ref{PDEs}, part \textbf{(a)}, to the proof of Lemma \ref{two}%
.

\medskip

\noindent \textit{Proof of Lemma \ref{two}.} By (\ref{chi}) and Definitions %
\ref{Prod}, we have%
\begin{eqnarray*}
\frac{1}{\left\vert \Lambda \right\vert }\langle \chi _{\Lambda },N^{k}\psi
^{c}\rangle (e^{\mu }) &=&\sum_{n=0}^{\infty }\frac{n^{k}e^{n\mu }}{n!}\frac{%
1}{\left\vert \Lambda \right\vert }\int_{\Lambda ^{n}}\psi ^{c}(t,x)d^{\#}x
\\
&=&\frac{\partial ^{k}}{\partial \mu ^{k}}\sum_{n=0}^{\infty }\frac{e^{n\mu }%
}{n!}\frac{1}{\left\vert \Lambda \right\vert }\int_{\Lambda ^{n}}\psi
^{c}(t,x)d^{\#}x \\
&=&\frac{\partial ^{k}}{\partial \mu ^{k}}\langle \chi _{\Lambda },\psi
^{c}\rangle (e^{\mu })~,
\end{eqnarray*}%
by formal manipulation of derivatives, and (\ref{Nk}) follows by Proposition %
\ref{fpressure}. The proof of (\ref{NN}) is analogous. By Proposition \ref%
{deconv},%
\begin{equation*}
\frac{1}{\left\vert \Lambda \right\vert ^{2}}\langle \chi _{\Lambda },N\psi
^{c}\circ N\psi ^{c}\rangle (e^{\mu })=\frac{1}{\left\vert \Lambda
\right\vert }\langle \chi _{\Lambda },N\psi ^{c}\rangle (e^{\mu })\cdot 
\frac{1}{\left\vert \Lambda \right\vert }\langle \chi _{\Lambda },N\psi
^{c}\rangle (e^{\mu })
\end{equation*}%
which, in view of (\ref{Nk}), proves (\ref{NN}).

\hfill $\Box $

\paragraph{Hopf-Lax-Oleinik formula}

We have devised also a way of deriving, from (\ref{pde}), another equation
directly related to the virial series. Assuming that (\ref{pde}) can be
solved by Hopf-Lax-Oleinik (HLO) formula\footnote{%
With the Inf-convolution transformation in the formal sense: $%
p(t,x)=g_{\varepsilon }(t,y^{\ast })+\frac{1}{2t}\left( x-y^{\ast }\right)
^{2}$ where $y^{\ast }=y^{\ast }(t,x)$ solves (formally) the equation $%
g_{\varepsilon }^{\prime }(y)-(x-y)/t=0$ for $y$.}%
\begin{equation}
p(t,x)=\min_{y}\left( g_{\varepsilon }(t,y)+\frac{1}{2t}\left( x-y\right)
^{2}\right) ~,~  \label{pg}
\end{equation}%
$g_{\varepsilon }(t,x)$ satisfies the following initial value problem%
\begin{equation}
g_{t}+\varepsilon \left( \frac{g_{xx}}{1+tg_{xx}}-g_{x}\right) =0  \label{g}
\end{equation}%
with $g(0,x)=e^{x}$. The formal series solution of (\ref{g}), in powers of $%
z=e^{x}$, is related with edge--irreducible graphs, i.e., connected graphs $%
G $ whose removal of an edge $e\in E(G)$ remain connected. We prove in
Appendix \ref{CVS} a general convergence theorem which implies, in
particular, convergence of the (Mayer) power series solution of (\ref{g}).

\section{Global existence and uniqueness of Mayer type solution\label{CVS}}

\setcounter{equation}{0} \setcounter{theorem}{0}

\paragraph{ Power series solutions: General result}

This section is devoted to the investigation of partial differential
equations of the form 
\begin{equation}
u_{t}+\mathcal{A}(t,u_{x},u_{xx})=0  \label{PDE}
\end{equation}%
for some function $\mathcal{A}:\mathbb{R}_{+}\times U\times V\longrightarrow 
\mathbb{C}$ smooth in the variable $t$ and holomorphic in both domains $%
U,~V\subset \mathbb{C}$ containing the origin.

\begin{definition}
A solution of (\ref{PDE}) is said to be of \textbf{Mayer type} if may be
represented by a (formal) power series of $z=e^{x}$:%
\begin{equation}
u(t,x)=\sum_{n=1}^{\infty }u_{n}(t)e^{nx}~~.  \label{sol}
\end{equation}
\end{definition}

We are interested in the (ideal gas) initial condition 
\begin{equation}
u(0,x)=e^{x}\ .  \label{ic}
\end{equation}
The solution of (\ref{PDE}) starting from (\ref{ic}), or any other initial
conditions of (Mayer) power series type, preserve the form (\ref{sol}). For
simplicity, we restrict ourselves to (\ref{ic}).

Equations (\ref{pde1}) and (\ref{g}) are examples of (\ref{PDE}) with $%
\mathcal{A}(t,a,b)=\dfrac{1}{2}a^{2}+\varepsilon (b-a)$ and $\mathcal{A}%
(t,a,b)=\varepsilon \left( \dfrac{b}{1+tb}-a\right) $, respectively,
satisfying the initial condition (\ref{ic}). The other equation (\ref{Feq})
for the free energy $F$ cannot be handled directly but equation (\ref{eq})
for $\varphi (t,\rho )=F_{\rho }(t,\rho )-\log \rho $, can also be dealt by
the following procedure (the initial condition is trivial but $A_{0,0}$ in (%
\ref{Fab}) depends on $\rho $).

Our aim is to prove that (\ref{sol}) converges uniformly in $t\in \left[ 0,T%
\right] $, for all $T>0$, and $z=e^{x}$ in a domain $D_{t}=\mathbb{D}_{r(t)}$
where $\mathbb{D}_{r}$ is an open disc in $\mathbb{C}$ centered at origin
with radius $r$, provided it solves (\ref{PDE}) with 
\begin{equation}
\mathcal{A}(t,a,b)=\sum_{\substack{ n,m=0:  \\ n+m\neq 0}}^{\infty
}A_{n,m}(t)a^{n}b^{m}~  \label{Fab}
\end{equation}%
satisfying the assumptions:

\begin{enumerate}
\item $A_{1,0}=\varepsilon \alpha $, $A_{0,1}=\varepsilon $ and $%
A_{2,0}=\gamma $, are constant in $t$, for some $\varepsilon ,\gamma >0$ and 
$\alpha \geq -1$;

\item if $n+m>1$ and $(n,m)\neq (2,0)$, 
\begin{equation}
\left\vert A_{n,m}(t)\right\vert \leq C\left( t\eta \right) ^{n+m-1}
\label{eta}
\end{equation}%
holds for some positive constants $C$ and $\eta $.
\end{enumerate}

Equations (\ref{pde1}) and (\ref{g}) satisfy the assumptions: $\alpha =-1$, $%
\gamma =1/2$ and $C=0$ for the former; $\alpha =-1$, $\gamma =0$, $A_{n,m}=0$
if $n\geq 2$, $C=\varepsilon $ and $\eta =1$ for the latter.

\begin{definition}
\label{Conv}For any two sequences $\mathbf{\alpha }=\left( \alpha
_{n}\right) _{n\geq 1}$ and $\mathbf{\beta }=\left( \beta _{n}\right)
_{n\geq 1}$, their product $\mathbf{\alpha }\cdot \mathbf{\beta }=\left( (%
\mathbf{\alpha }\cdot \mathbf{\beta })_{k}\right) _{k\geq 1}$ and
convolution product $\alpha \ast \beta =\left( (\alpha \ast \beta
)_{k}\right) _{k\geq 1}$ are sequences defined, respectively, by%
\begin{equation}
(\mathbf{\alpha }\cdot \mathbf{\beta })_{k}=\alpha _{k}\beta _{k}
\label{prod0}
\end{equation}%
and by $\left( \alpha \ast \beta \right) _{1}=0$ and for any $k\geq 2$ 
\begin{equation}
\left( \mathbf{\alpha }\ast \mathbf{\beta }\right)
_{k}=\sum_{l=1}^{k-1}\alpha _{l}\beta _{k-l}~.  \label{prod}
\end{equation}
\end{definition}

\begin{proposition}
\label{fg}If $f(z)=\displaystyle\sum_{n\geq 1}\alpha _{n}z^{n}$ and $g(z)=%
\displaystyle\sum_{n\geq 1}\beta _{n}z^{n}$ are two formal series, then%
\begin{equation*}
\left( fg\right) (z)=\sum_{k\geq 1}\left( \mathbf{\alpha }\ast \mathbf{\beta 
}\right) _{k}z^{k}~.
\end{equation*}
\end{proposition}

\noindent \textit{Proof.} Proposition \ref{fg} is proved by rearranging the
double sum%
\begin{equation*}
\left( fg\right) (z)=\sum_{n\geq 1}\sum_{m\geq 1}\alpha _{n}\beta
_{m}z^{n+m}=\sum_{k\geq 1}\left( \sum_{\substack{ n,m\geq 1:  \\ n+m=k}}%
\alpha _{n}\beta _{m}\right) z^{k}
\end{equation*}%
with the sum between parenthesis equivalent to the r.h.s. of (\ref{prod}).

\hfill $\Box $

Formal derivatives of (\ref{sol}) with respect to $t$ and $x$, i.e., term by
term, lead to a formal power series of the same type with coefficients of $%
u_{t}$, $u_{x}$ and $u_{xx}$ given by $\dot{u}_{n}$, $nu_{n}\equiv v_{n}$
and $n^{2}u_{n}\equiv w_{n}$, respectively. Plugging the expansion of these
functions into (\ref{PDE}) yields for the $k$--th coefficient of the
equation, $k\geq 1$,%
\begin{equation}
\dot{u}_{k}+\sum_{\substack{ n,m:  \\ 1\leq n+m\leq k}}A_{n,m}(t)\left( 
\underset{n}{\underbrace{\mathbf{v}\ast \cdots \ast \mathbf{v}}}\ast 
\underset{m}{\underbrace{\mathbf{w}\ast \cdots \ast \mathbf{w}}}\right)
_{k}=0~.  \label{sum}
\end{equation}%
The restriction $n+m\leq k$ in (\ref{sum}) results from the fact that our
sequence $\mathbf{u}=\left( u_{n}\right) _{n\geq 1}$ starts with $n=1$ and a
convolution involving $n+m$ sequences cannot have nonvanishing component $k$
if $k>n+m$. Consequently, for any $K\in \mathbb{N}$ (\ref{sum}) with $1\leq
k\leq K$, form a \textbf{closed} system of $K$ (first order) differential
equations, involving $K$ unknown functions: $u_{1}(t),\ldots ,u_{K}(t)$
satisfying the initial condition%
\begin{equation}
u_{k}(0)=\left\{ 
\begin{array}{ll}
1 & \text{if~}k=1 \\ 
0 & \text{otherwise}%
\end{array}%
\right. \ .  \label{un0}
\end{equation}

\begin{theorem}
\label{convergence}There exists an unique formal series of the form (\ref%
{sol}) that solves (\ref{PDE}) with the initial condition (\ref{ic}). Under
assumptions $1.$ and $2.$ on the coefficients $A_{n,m}$ of $\mathcal{A}$ one
can find a power series $U(t,x)$ of the form%
\begin{equation}
U(t,x)=\sum_{n=1}^{\infty }P_{n}(t)e^{-\varepsilon (1+a)nt}e^{nx}  \label{U}
\end{equation}%
where $P_{n}(t)=\displaystyle\sum_{l=1}^{n}C_{n,l}t^{l-1}$ is a polynomial
of order $n-1$ with (positive) coefficients such that $U(t,x)$ majorizes (%
\ref{sol}): $u(t,x)\ll U(t,x)$ in the sense that%
\begin{equation*}
\left\vert u_{n}(t)\right\vert \leq P_{n}(t)e^{-\varepsilon (1+a)nt}
\end{equation*}%
holds for any $n\geq 1$ and $t\in \mathbb{R}_{+}$. Moreover, 
\begin{equation}
C_{n,l}\leq \delta A^{2}\frac{B^{n}}{n^{4}}\frac{1}{l^{2}}  \label{Cnk}
\end{equation}%
with $B=\left( \delta A^{2}\right) ^{-1}$, $A=3/(2\pi ^{2})$ and $\delta
=1/\eta +C/\varepsilon -\sqrt{2/\eta +C^{2}/\varepsilon ^{2}}$, and the
series (\ref{sol}) converges uniformly for $t$ and $z=e^{x}$ in the domain $%
\Omega \subset \mathbb{R}_{+}\times \mathbb{C}$ defined by 
\begin{equation}
B\max (1,t)e^{-\varepsilon (1+a)t}\left\vert z\right\vert <1~,
\label{domain}
\end{equation}%
representing therein the unique solution of the initial value problem.
\end{theorem}

\paragraph{Proof of Theorem \protect\ref{convergence}}

The formal series solution of (\ref{PDE}) is obtained by integrating the
system of first order differential equations (\ref{sum}). Isolating its
linear term, equation (\ref{sum}) reads%
\begin{equation}
\dot{u}_{k}+\varepsilon (k^{2}+ak)u_{k}=f_{k}\left( u_{1},\ldots
,u_{k-1};t\right)  \label{uk}
\end{equation}%
where, by definition (\ref{prod}),%
\begin{equation}
f_{k}\left( u_{1},\ldots ,u_{k-1};t\right) =-\sum_{\substack{ n,m:  \\ 2\leq
n+m\leq k}}A_{n,m}(t)\left( \underset{n}{\underbrace{\mathbf{v}\ast \cdots
\ast \mathbf{v}}}\ast \underset{m}{\underbrace{\mathbf{w}\ast \cdots \ast 
\mathbf{w}}}\right) _{k}~  \label{fut}
\end{equation}%
does not depend on $u_{n}$ with $n\geq k$.

The solution of (\ref{uk}) for $k=1$: 
\begin{equation*}
\dot{u}_{1}+\varepsilon \left( 1+a\right) u_{1}=0
\end{equation*}%
with $u_{1}(0)=1$ is%
\begin{equation}
u_{1}(t)=e^{-\varepsilon (1+a)t}~.  \label{u1t}
\end{equation}%
Given that we have already solved for $u_{1}(t),\ldots ,u_{k-1}(t)$, we
solve equation (\ref{uk}) with $u_{k}(0)=0$:

\begin{equation}
u_{k}(t)=e^{-\varepsilon (k^{2}+ak)t}\int_{0}^{t}e^{\varepsilon
(k^{2}+ak)s}f_{k}\left( u_{1},\ldots ,u_{k-1};s\right) ds~  \label{ukt}
\end{equation}
uniquely defines the coefficient of the formal series, proving the first
statement of Theorem \ref{convergence}.

To prove the existence of a majorant series of the form (\ref{U}), we show
that the sequence $\left( U_{n}(t)\right) _{n\geq 1}$ with $%
U_{n}(t)=P_{n}(t)e^{-\varepsilon (1+a)nt}$ satisfies (\ref{ukt}), with $%
-A_{n,m}$ in (\ref{fut}) replaced by the upper bound of $\left\vert
A_{n,m}\right\vert $, as an equality. We then find $\delta >0$ and $B>1$
such that (\ref{Cnk}) holds for every $n\geq 1$ and $1\leq l\leq n$, by
induction. For this, we need

\begin{lemma}
\label{conv}Let $\mathbf{c}=\left( c_{i}\right) _{i\geq 1}$ be a sequence of
positive numbers defined by%
\begin{equation}
c_{i}=A\frac{1}{i^{2}}d^{i}  \label{ci}
\end{equation}%
for some $d>0$ and $A=3/(2\pi ^{2})=0.1519...$ as in Theorem \ref%
{convergence}. Then the sequence formed by convolution of $\mathbf{c}$ with
itself is dominated by the sequence $\mathbf{c}$, i.e.,%
\begin{equation*}
\left( \mathbf{c}\ast \mathbf{c}\right) _{n}\leq c_{n}
\end{equation*}%
holds for any $n\geq 1$.
\end{lemma}

\noindent \textit{Proof. By definition (\ref{prod}) together with \ref{ci},
we have}%
\begin{eqnarray*}
\frac{1}{c_{n}}\sum_{i=1}^{n-1}c_{i}c_{n-1} &=&A\sum_{i=1}^{n-1}\frac{n^{2}}{%
i^{2}(n-i)^{2}} \\
&=&A\sum_{i=1}^{n-1}\left( \frac{1}{i}+\frac{1}{n-i}\right) ^{2} \\
&\leq &2A\sum_{i=1}^{n-1}\left( \frac{1}{i^{2}}+\frac{1}{(n-i)^{2}}\right)
\leq \frac{2\pi ^{2}}{3}A=1~.
\end{eqnarray*}%
For the two inequalities, we have used%
\begin{equation*}
0\leq \left( a-b\right) ^{2}=2(a^{2}+b^{2})-(a+b)^{2}\Longrightarrow
(a+b)^{2}\leq 2(a^{2}+b^{2})
\end{equation*}%
with $a=1/i$ and $b=1/(n-i)$ and $\displaystyle\sum_{i\geq 1}1/i^{2}=\pi
^{2}/6$. This concludes the proof of Lemma \ref{conv}.

\hfill $\Box $

We also need

\begin{lemma}
\begin{equation}
\int_{0}^{T}s^{r-1}e^{s}ds\leq T^{r-1}e^{T}  \label{parts}
\end{equation}%
holds for any integer $r\geq 1$ and $0\leq T<\infty $.
\end{lemma}

\noindent \textit{Proof.} For $r=1$, (\ref{parts}) is easily true. For $%
r\geq 2$, by integration by parts, we have%
\begin{equation*}
\int_{0}^{T}s^{r-1}e^{s}ds=T^{r-1}e^{T}-(r-1)\int_{0}^{T}s^{r-2}e^{s}ds
\end{equation*}%
from which the proof is concluded.

\hfill $\Box $

Convolutions appears not only in the product of power series but also due to
products of polynomials when two or more majorant solutions are multiplied
together. If $P_{n}(t)$ and $P_{m}(t)$ are two polynomials as in the
statement of Theorem \ref{convergence}, then%
\begin{equation*}
tP_{n}(t)P_{m}(t)=\sum_{r=1}^{n+m-1}\left( \sum_{\substack{ 1\leq l_{1}\leq
n,1\leq l_{2}\leq m:  \\ l_{1}+l_{2}=r}}C_{n,l_{1}}C_{m,l_{2}}\right) t^{r-1}
\end{equation*}%
is a polynomial of order $n+m-1$. In order to write the sum between
parenthesis as $\left( \mathbf{C}_{n}\ast \mathbf{C}_{m}\right) _{r}$ we
extend the coefficients $C_{n,l}$, $1\leq l\leq n$, as an infinite sequence $%
\mathbf{C}_{n}=\left( C_{n,l}\right) _{l\geq 1}$ with $C_{n,l}=0$ for $l>n$.
To make our expressions shorter, we introduce a double (weighted)
convolution product of two sequences $\left( C_{n,l}\right) $ and $\left(
D_{n,l}\right) $ as a sequence $\left( \left( \mathbf{C}\circledast \mathbf{D%
}\right) _{n,l}\right) $ defined by%
\begin{equation}
\left( \mathbf{C}\circledast \mathbf{D}\right) _{n,r}=\sum_{\substack{ %
n_{1},n_{2}\geq 1:  \\ n_{1}+n_{2}=n}}n_{1}^{2}n_{2}^{2}\sum_{\substack{ %
1\leq l_{1}\leq n,1\leq l_{2}\leq m:  \\ l_{1}+l_{2}=r}}%
C_{n_{1},l_{1}}D_{n_{2},l_{2}}  \label{CD}
\end{equation}

Now, by induction, we construct an equation for the majorant polynomials $%
P_{n}(t)$. Suppose we have already had $k-1$ polynomials $P_{1}(t),\ldots
P_{k-1}(t)$, with $P_{j}$ of order $j-1$, whose coefficients are positive.
Then, writing%
\begin{eqnarray*}
V_{n} &=&nU_{n} \\
W_{n} &=&n^{2}U_{n} \\
U_{n}(t) &=&e^{-\varepsilon (1+a)nt}P_{n}(t)
\end{eqnarray*}%
and using $k\leq k^{2}$ and $\left\vert -A_{n,m}\right\vert \leq C\left(
t\eta \right) ^{n+m-1}$ on the r.h.s of (\ref{fut}), we have%
\begin{eqnarray}
\left\vert -A_{n,m}\right\vert (s)\left( \underset{n}{\underbrace{\mathbf{V}%
\ast \cdots \ast \mathbf{V}}}\ast \underset{m}{\underbrace{\mathbf{W}\ast
\cdots \ast \mathbf{W}}}\right) _{k}(s) &\leq &C\eta ^{n+m}e^{-\varepsilon
(1+a)ks}s^{n+m-1}  \notag \\
&&\times \sum_{\substack{ k_{1},\ldots ,k_{n+m}\geq 1  \\ k_{1}+\cdots
+k_{n+m}=k}}k_{1}^{2}P_{k_{1}}\cdots k_{n+m}^{2}P_{k_{n+m}}  \label{A}
\end{eqnarray}%
where 
\begin{equation}
s^{n+m-1}P_{k_{1}}\cdots P_{k_{n+m}}=\sum_{r=1}^{k}\left( \mathbf{C}%
_{k_{1}}\ast \cdots \ast \mathbf{C}_{k_{n+m}}\right) _{r}s^{r-1}  \label{PP}
\end{equation}%
is a polynomial of order $k-1$ whose coefficients are positive. The
integration needed to be performed to each term of (\ref{ukt}) is of the form%
\begin{eqnarray}
e^{-\varepsilon (k^{2}+ak)t}\int_{0}^{t}s^{r-1}e^{\varepsilon (k^{2}-k)s}ds
&=&e^{-\varepsilon (k^{2}+ak)t}\frac{1}{\varepsilon ^{r}(k^{2}-k)^{r}}%
\int_{0}^{\varepsilon (k^{2}-k)t}s^{r-1}e^{s}ds  \notag \\
&\leq &\frac{1}{\varepsilon k(k-1)}e^{-\varepsilon (1+a)kt}t^{r-1}~,
\label{integral}
\end{eqnarray}%
by (\ref{parts}).

The result of summing (\ref{A}) over $2\leq n,m\leq k$, together with (\ref%
{PP}) and (\ref{integral}), is bounded by a polynomial of order $k-1$,
denoted by $P_{k}(t)$, whose coefficients are positive. Defining $%
U_{k}(t)=e^{-\varepsilon (1+a)kt}P_{k}(t)$ together with (\ref{CD}), the
coefficients $C_{k,l}$ of $P_{k}$ satisfy the recursion relation%
\begin{equation}
C_{k,l}=\frac{C}{\varepsilon k(k-1)}\sum_{2\leq n,m\leq k}\eta ^{n+m}\left( 
\underset{n+m}{\underbrace{\mathbf{C}\circledast \cdots \circledast \mathbf{C%
}}}\right) _{k,l}~  \label{CCC}
\end{equation}

For $k=1$, we have%
\begin{equation*}
u_{1}(t)=e^{-\varepsilon (1+a)t}=U_{1}(t)
\end{equation*}%
with $P_{1}\equiv 1$. Suppose that 
\begin{equation}
\left\vert u_{n}(t)\right\vert \leq U_{n}(t)  \label{uU}
\end{equation}%
holds for $1\leq n\leq k-1$ and $t\in \mathbb{R}_{+}$ and let $\tilde{f}_{k}$
be defined by (\ref{fut}) with $-A_{n,m}$ replaced by $\left\vert
A_{n,m}\right\vert $. Then, 
\begin{eqnarray*}
\left\vert u_{k}(t)\right\vert &\leq &e^{-\varepsilon
(k^{2}+ak)t}\int_{0}^{t}e^{\varepsilon (k^{2}+ak)s}\tilde{f}_{k}\left(
U_{1},\ldots ,U_{k-1};s\right) ds \\
&\leq &e^{-\varepsilon (k^{2}+ak)t}\frac{C}{\varepsilon k(k-1)}\sum 
_{\substack{ 2\leq n,m\leq k;  \\ 1\leq l\leq k}}\eta ^{n+m}\left( \underset{%
n+m}{\underbrace{\mathbf{C}\circledast \cdots \circledast \mathbf{C}}}%
\right) _{k,l}t^{l-1}=U_{k}(t)
\end{eqnarray*}%
which establishes the second statement.

To conclude the proof of Theorem \ref{convergence}, we note that estimate (%
\ref{Cnk}) can be written as%
\begin{equation}
C_{n,l}\leq \delta \frac{1}{n^{2}}c_{n}\tilde{c}_{l}\equiv \tilde{C}_{n,l}
\label{Ccc}
\end{equation}%
where $c_{n}=AB^{n}/n^{2}$ and $\tilde{c}_{l}=A/l^{2}$ are sequences
satisfying the hypothesis of Lemma \ref{conv}. Clearly (\ref{Ccc}) holds for 
$n=l=1$%
\begin{equation*}
C_{1,1}=1\leq \delta c_{1}\tilde{c}_{1}=\delta A^{2}B
\end{equation*}%
provided $B=\left( \delta A^{2}\right) ^{-1}$. Assuming that (\ref{Ccc})
holds for $1\leq n\leq k-1$ and $0\leq l\leq n$ with $k\geq 2$, then
plugging (\ref{Ccc}) into (\ref{CCC}) together with Lemma \ref{conv}, yields%
\begin{equation}
C_{k,l}\leq \frac{C}{\varepsilon k(k-1)}\frac{\delta ^{2}\eta ^{2}}{\left(
1-\delta \eta \right) ^{2}}c_{k}\tilde{c}_{l}\leq \delta \frac{1}{k^{2}}c_{k}%
\tilde{c}_{l}  \label{CC}
\end{equation}%
provided $\delta $ is chosen to be the smaller solution, $\delta =1/\eta
+C/\varepsilon -\sqrt{2/\eta +C^{2}/\varepsilon ^{2}}$, of%
\begin{equation*}
\frac{2\delta \eta ^{2}C}{\varepsilon \left( 1-\delta \eta \right) ^{2}}=1~.
\end{equation*}%
The power series solution converges if 
\begin{equation*}
\lim_{n\rightarrow \infty }\frac{\tilde{P}_{n+1}(t)}{\tilde{P}_{n}(t)}%
e^{-\varepsilon (1+a)t}\left\vert z\right\vert <1
\end{equation*}%
where $\tilde{P}_{n}$ is the polynomial whose coefficients $\tilde{C}_{n,l}$
are given by r.h.s. of (\ref{Cnk}), concluding the proof.

\hfill $\Box $

\section{Polynomials with Positive Coefficients \label{PPI}}

\setcounter{equation}{0} \setcounter{theorem}{0}

We devote this appendix to the proof of Lemma \ref{Teta} and Proposition \ref%
{contra}.

Lemma \ref{Teta} is consequence of the following

\begin{proposition}
\label{Integral} For any $k\geq 2$, let $\mathcal{Q}_{k}(t)$ be given by (%
\ref{Qk}) 
\begin{eqnarray}
\mathcal{Q}_{k}(t) &=&\left( Q_{k}\circ \eta \right) (t),\qquad \eta
(t)=e^{-2\varepsilon t}  \notag \\
Q_{k}(\eta ) &=&1-(1-\eta )P_{k}(\eta )\   \label{Pk}
\end{eqnarray}%
is such that

\begin{enumerate}
\item[i.] $P_{k}(\eta )=1+\displaystyle\sum_{n=1}^{D_{k}}p_{k,n}\eta ^{n}$,
is a polynomial in $\eta $ of order $D_{k}=(k-2)(k+1)/2$ whose the
coefficients $p_{k,n}$ are positive and satisfy%
\begin{equation*}
p_{k,n-1}-p_{k,n}>0\ ,\qquad 1\leq n\leq D_{k}
\end{equation*}

\item[ii.] $Q_{k}(\eta )=\displaystyle\sum_{n=1}^{D_{k}+1}q_{k,n}\eta ^{n}$,
is a polynomial in $\eta $ of order $D_{k}+1$ with $%
q_{k,n}=p_{k,n-1}-p_{k,n} $ if $1\leq n\leq D_{k}$ and $%
q_{k,D_{k}+1}=p_{k,D_{k}}$ positive, by \textbf{i.}, satisfying (with $%
p_{k,D_{k}+1}=0$)%
\begin{equation*}
q_{k,n}-q_{k,n+1}=p_{k,n-1}+p_{k,n+1}-2p_{k,n}>0\ ,\qquad 1\leq n\leq D_{k}
\end{equation*}

\item[iii.] It follows by \textbf{i.} and \textbf{ii.} that $Q_{k}(\eta )$
is a convex function such that $Q_{k}(0)=0$, $Q_{k}(1)=1$ and 
\begin{equation*}
\lim_{k\rightarrow \infty }Q_{k}(\eta )=0\quad \text{if}\quad \ 0\leq \eta
<1~
\end{equation*}

\item[iv.] 
\begin{equation}
T_{k}(\eta ):=Q_{k}(\eta )+(k-1)(1-\eta )Q_{k+1}(\eta )\geq \eta
\label{Deltaeta}
\end{equation}%
holds for every $k\geq 2$ and$\ 0\leq \eta \leq 1$ (equality only for $\eta
=0$ and $1$).
\end{enumerate}
\end{proposition}

\noindent \textit{Proof. }It follows by (\ref{Qk}), (\ref{Pk}) and $\eta
=1-2\varepsilon \lambda $ that%
\begin{equation*}
P_{k}(\eta )=\frac{k(k+1)}{2}\frac{\eta ^{k(k+1)/2}}{(1-\eta )^{k+1}}%
\int_{\eta }^{1}\xi ^{-k(k+1)/2}(1-\xi )^{k}\frac{d\xi }{\xi }~.
\end{equation*}%
Note that $N_{k}=k(k+1)/2$ is an integer for any $k\in \mathbb{N}$ and, as
shown in the calculation performed below, $P_{k}$ is a polynomial whose
degree $D_{k}=k(k-1)/2-1=(k-2)(k+1)/2$ is non--negative for all $k\geq 2$.
Observe that 
\begin{equation}
N_{k}=D_{k}+k+1~.  \label{ND}
\end{equation}

We change variable $y=(1-\xi )/\xi $ so 
\begin{equation*}
P_{k}(\eta )=\frac{k(k+1)}{2}\frac{\eta ^{\frac{k(k+1)}{2}}}{(1-\eta )^{k+1}}%
\int_{0}^{(1-\eta )/\eta }y^{k}(1+y)^{k(k-1)/2-1}dy~,
\end{equation*}%
by the binomial theorem and explicit integration can be written as 
\begin{equation*}
P_{k}(\eta )=\sum_{l=0}^{D_{k}}\frac{N_{k}}{k+l+1}\dbinom{D_{k}}{l}\eta
^{D_{k}-l}\left( 1-\eta \right) ^{l}~.
\end{equation*}%
Applying the binomial theorem once again, we have 
\begin{eqnarray*}
P_{k}(\eta ) &=&\sum_{l=0}^{D_{k}}\sum_{m=0}^{l}\left( -1\right) ^{l-m}%
\dbinom{l}{m}\frac{N_{k}}{k+l+1}\dbinom{D_{k}}{l}\eta ^{D_{k}-m} \\
&=&\sum_{m=0}^{D_{k}}\dbinom{D_{k}}{m}\left( \sum_{l=0}^{D_{k}-m}(-1)^{l}%
\dbinom{D_{k}-m}{l}\frac{N_{k}}{k+l+m+1}\right) \eta ^{D_{k}-m}~
\end{eqnarray*}%
which, by inserting the identity%
\begin{equation*}
\frac{1}{k+l+m+1}=\int_{0}^{\infty }e^{-(k+l+m+1)s}ds
\end{equation*}%
together with the binomial theorem and (\ref{ND}), yields%
\begin{eqnarray}
P_{k}(\eta ) &=&\sum_{m=0}^{D_{k}}\dbinom{D_{k}}{m}\left(
N_{k}\int_{0}^{\infty }e^{-(k+m+1)s}\left( 1-e^{-s}\right)
^{D_{k}-m}ds\right) \eta ^{D_{k}-m}  \notag \\
&=&\sum_{n=0}^{D_{k}}\dbinom{D_{k}}{n}\left( N_{k}\int_{0}^{\infty
}e^{-N_{k}s}\left( e^{s}-1\right) ^{n}ds\right) \eta ^{n}  \notag \\
&\equiv &1+\sum_{n=1}^{D_{k}}p_{k,n}\eta ^{n}  \label{pkm}
\end{eqnarray}%
establishing the first part of the statement \textbf{i.}: $p_{k,n}>0$.

Successive partial integrations on%
\begin{eqnarray}
I_{k,n} &=&N_{k}\int_{0}^{\infty }e^{-N_{k}s}\left( e^{s}-1\right) ^{n}ds 
\notag \\
&=&N_{k}\frac{n}{N_{k}}\int_{0}^{\infty }e^{-(N_{k}-1)s}\left(
e^{s}-1\right) ^{n-1}ds  \notag \\
&=&\qquad \vdots  \notag \\
&=&\frac{n}{N_{k}-1}\frac{n-1}{N_{k}-2}\cdots \frac{2}{N_{k}-n+1}\frac{1}{%
N_{k}-n}  \label{Ikm}
\end{eqnarray}%
yields%
\begin{equation}
p_{k,n}=\dbinom{D_{k}}{n}I_{k,n}=\frac{D_{k}}{N_{k}-1}\frac{D_{k}-1}{N_{k}-2}%
\cdots \frac{D_{k}-n+1}{N_{k}-n}  \label{pkn}
\end{equation}%
and thus (with $p_{k,0}=1$) 
\begin{eqnarray}
\left( -\nabla p_{k,\cdot }\right) _{n+1} &=&p_{k,n}-p_{k,n+1}  \notag \\
&=&p_{k,n}\left( 1-\frac{D_{k}-n}{N_{k}-n-1}\right) >0~  \label{nablap}
\end{eqnarray}%
for $n=0,\ldots ,D_{k}-1$, $k>2$, by (\ref{ND}). This concludes the proof of
item \textbf{i.}.

Now, since 
\begin{equation*}
Q_{k}(\eta )=1-(1-\eta )P_{k}(\eta
)=\sum_{n=1}^{D_{k}}(p_{k,n-1}-p_{k,n})\eta ^{n}+p_{k,D_{k}}\eta ^{D_{k}+1}
\end{equation*}
and $Q_{k}(\eta )=\displaystyle\sum_{n=1}^{D_{k}+1}q_{k,n}\eta ^{n}$,
equations (\ref{pkm}) and (\ref{nablap}) imply the first part of statement 
\textbf{ii.}. By (\ref{Ikm}), we have

\begin{eqnarray}
\left( \nabla ^{\ast }q_{k,\cdot }\right) _{n} &=&q_{k,n}-q_{k,n+1}  \notag
\\
&=&p_{k,n-1}+p_{k,n+1}-2p_{k,n}  \notag \\
&=&p_{k,n}\left( \frac{N_{k}-n}{D_{k}-n+1}+\frac{D_{k}-n}{N_{k}-n-1}-2\right)
\notag \\
&=&p_{k,n}\frac{k\left( k-1\right) }{\left( N_{k}-n-1\right) \left(
D_{k}-n+1\right) }>0  \label{nablaq}
\end{eqnarray}%
for $1\leq n\leq D_{k}$ and $k>2$ (for $k=2$, $D_{2}=0$ and $%
Q_{2}=p_{2,0}\eta =\eta $), establishing item \textbf{ii.}.

Clearly, $Q_{k}(0)=0$,%
\begin{equation*}
Q_{k}(1)=\sum_{n=1}^{D_{k}}(p_{k,n-1}-p_{k,n})+p_{k,D_{k}}=p_{k,0}=1~
\end{equation*}%
and, as $q_{k,n}>0$, $Q_{k}^{\prime \prime }(\eta )>0$ for $0<\eta <1$,
verifying the first statements of \textbf{iii.}.

Finally, (\ref{Deltaeta}) is consequence of the following

\begin{claim}
\label{T}For any $k\geq 2$, the numerical sequence $\left( t_{k,n}\right)
_{1\leq n\leq D_{k+1}+1}$ defined by%
\begin{equation*}
T_{k}(\eta )=Q_{k}(\eta )+(k-1)(1-\eta )Q_{k+1}(\eta )=t_{k,1}\eta
-\sum_{n=2}^{D_{k+1}+1}t_{k,n}\eta ^{n}
\end{equation*}%
satisfies 
\begin{eqnarray*}
t_{k,1} &>&1 \\
t_{k,n} &>&0\qquad \ \text{for }2\leq n\leq D_{k+1}+1\ .
\end{eqnarray*}
\end{claim}

Since $T_{k}(1)=Q_{k}(1)=1$, it follows by Claim \ref{T}, together with $%
\eta ^{n}\leq \eta $ for $0\leq \eta \leq 1$, that 
\begin{equation}
T_{k}(\eta )\geq T_{k}(1)\eta =\eta  \label{TT}
\end{equation}%
which is exactly the statement (\ref{Deltaeta}), concluding the proof of
Proposition \ref{Integral}.

\hfill $\Box $

\noindent \textit{Proof of Claim \ref{T}.} The coefficients of terms in

\begin{equation}
(1-\eta )Q_{k+1}(\eta )=q_{k+1,1}\eta -\sum_{n=1}^{D_{k+1}-1}\left( \nabla
^{\ast }q_{k+1,\cdot }\right) _{n}\eta ^{n+1}-p_{k+1,D_{k+1}}\eta
^{D_{k+1}+1}  \label{Qqq}
\end{equation}%
with degree larger than $2$ are all negative in view of (\ref{pkn}) and (\ref%
{nablaq}). Since $Q_{k}(\eta )$ has degree $D_{k}+1$, this implies $%
t_{k,n+1}>0$ for $D_{k}+1\leq n\leq D_{k+1}$ and we thus need to prove: 
\textbf{(a)}%
\begin{eqnarray}
t_{k,1} &=&q_{k,1}-(k-1)q_{k+1,1}  \notag \\
&=&1-p_{k,1}+(k-1)(1-p_{k+1,1})  \notag \\
&=&k-\left( \frac{D_{k}}{N_{k}-1}+(k-1)\frac{D_{k+1}}{N_{k+1}-1}\right) >1
\label{t1}
\end{eqnarray}%
and \textbf{(b)}%
\begin{equation}
t_{k,n+1}=-\left( p_{k,n}-p_{k,n+1}\right) +(k-1)\left(
p_{k+1,n-1}+p_{k+1,n+1}-2p_{k+1,n}\right) >0  \label{tn}
\end{equation}%
for $k\geq 3$ and $1\leq n\leq D_{k}$. Note that, since $Q_{2}(\eta )=\eta $
has degree $1$, $t_{2,n+1}>0$, $1\leq n\leq 2$ follows from (\ref{Qqq}).

By (\ref{ND}),%
\begin{equation*}
\frac{D_{k+1}}{N_{k+1}-1}-\frac{\ D_{k}}{N_{k}-1}=\left( 1-\frac{1}{k}%
\right) \left( 1-\frac{1}{k+3}\right) -\left( 1-\frac{1}{k-1}\right) \left(
1-\frac{1}{k+2}\right) >0
\end{equation*}%
which, together with (\ref{t1}), yields 
\begin{eqnarray*}
t_{k,1} &>&k\left( 1-\frac{D_{k+1}}{N_{k+1}-1}\right) \\
&=&k\left( 1-\left( 1-\frac{1}{k}\right) \left( 1-\frac{1}{k+3}\right)
\right) =1+\frac{k-1}{k+3}
\end{eqnarray*}%
proving \textbf{(a)}.

Each factor $\dfrac{D_{k}-m+1}{N_{k}-m}=\dfrac{k(k-1)-2m}{k(k+1)-2m}$
involved in $p_{k,n}$ (see (\ref{pkn})) is an increasing function of $k$ and
decreasing function of $m$. For this, let $f(x,y)=\left( x^{2}-x-2y\right)
/(x^{2}+x-2y)$ and note that $f_{x}=2(x^{2}+y)/(x^{2}+x-2y)^{2}>0$ and $%
f_{y}=-4/(x^{2}+x-2y)^{2}<0$. This implies, in particular, that $p_{k,n}$ is
monotone increasing function of $k$ and%
\begin{equation}
\frac{p_{k+1,n}}{p_{k,n}}>\frac{D_{k+1}}{N_{k+1}-1}\frac{N_{k}-1}{D_{k}}=%
\frac{\left( k^{2}+k-2\right) ^{2}}{k(k+1)\left( k^{2}+k-6\right) }>1~
\label{pp}
\end{equation}%
for $k\geq 3$. By (\ref{nablap}), (\ref{nablaq}) and (\ref{pp}), (\ref{tn})
can be estimated by%
\begin{equation*}
t_{k,n+1}\geq \frac{p_{k+1,n}}{(N_{k}-n-1)(N_{k+1}-n-1)(D_{k+1}-n+1)}\frac{%
h_{k,n}}{k^{2}+k-6}
\end{equation*}%
where%
\begin{eqnarray*}
h_{k,n} &=&-k(k^{2}+k-6)(N_{k+1}-n-1)(D_{k+1}-n+1)+(k-1)\left(
k^{2}+k-2\right) ^{2}(N_{k}-n-1) \\
&=&-\frac{1}{4}k\left( k^{2}+k-6\right) \left( k^{2}+3k-2n\right) \left(
k^{2}+k-2n\right) +\frac{1}{2}\left( k-1\right) ^{3}\left( k+2\right)
^{2}\left( k^{2}+k-2(n+1)\right)
\end{eqnarray*}%
\textbf{\ }for $k\geq 3$ and $1\leq n\leq D_{k}$, satisfies 
\begin{equation*}
h_{k,n}\geq \min \left( h_{k,1},h_{k,D_{k}}\right) \geq h_{3,1}=80
\end{equation*}%
proves \textbf{(b)} and concludes the proof of Claim \ref{T}.

\hfill $\Box $

\noindent \textit{Proof of Proposition \ref{contra}.} The limit of $c_{k}(t)$
as $t$ goes to $0$ exists by continuity, is finite by (\ref{elz}) and
satisfies equation (\ref{ck0}) whose solution is uniquely determined by the
recursive equation (\ref{ck0}): $c_{0}(0)=1$, $c_{1}(0)=c_{0}^{2}(0)=1$, $%
c_{2}(0)=3c_{0}(0)c_{1}(0)/2=3/2$, $c_{3}(0)=8/3$, $c_{4}(0)=125/24$ and so
on. The solution's explicit form (\ref{Wic}) is a consequence of the
following

\begin{lemma}
\label{BK}%
\begin{equation}
\frac{1}{2}\sum_{j=1}^{n-1}\dbinom{n}{j}j^{j-1}(n-j)^{n-j-1}=(n-1)n^{n-2}
\label{sum2}
\end{equation}
\end{lemma}

\noindent \textit{Proof.} See Lemma 4.2 of \cite{Brydges-Kennedy}.

\hfill $\Box $

Setting $k=n-1$ and $l=j-1$, equation (\ref{sum2}) can be written as%
\begin{equation*}
\frac{k+1}{2k}\sum_{l=0}^{k-1}\frac{(l+1)^{j}}{(l+1)!}\frac{(n-l-1)^{n-l}}{%
(n-l-1)!}=\frac{(k+1)^{k}}{(k+1)!}
\end{equation*}%
which, together with the definition of convolution product, proves that (\ref%
{Wic}) solves (\ref{ck0}).

Let us now suppose that $t_{0}$ is the largest \textquotedblleft
time\textquotedblright\ for which $c_{k}(t)<c_{k}(t^{\prime })$ is satisfied
for $0\leq t<t^{\prime }\leq t_{0}$ and $k\geq 2$. Then, for all $t\leq
t_{0} $,%
\begin{equation}
c_{k}(t)>\mathcal{R}_{k}(t)\frac{1}{k}\left( \mathbf{c}(t)\ast \mathbf{c}%
(t)\right) _{k-1}\ ,  \label{ckt}
\end{equation}%
where%
\begin{equation*}
\mathcal{R}_{k}(t)=\frac{k+1}{2}\frac{1}{\lambda ^{k}(t)}\int_{0}^{t}e^{-%
\varepsilon k(k+1)(t-s)}\lambda ^{k-1}(s)ds\equiv \left( R_{k}\circ \eta
\right) (t)
\end{equation*}%
and, with $N_{k}=k(k+1)/2$ and $D_{k}=k(k-1)/2$, 
\begin{equation}
R_{k}(\eta )=\sum_{n=0}^{D_{k}}\frac{N_{k}}{k+l}\dbinom{D_{k}}{l}\eta
^{D_{k}-l}\left( 1-\eta \right) ^{l}\equiv 1+\sum_{n=1}^{D_{k}}p_{k,n}\eta
^{n}  \label{kRk}
\end{equation}%
is a polynomial of degree $D_{k}$ in $\eta $, with $p_{k,n}>0$ given by (\ref%
{pkn}), satisfying%
\begin{equation}
1\leq R_{k}(\eta )\leq \frac{k+1}{2}~.  \label{Rk}
\end{equation}%
Equation (\ref{ckt}) together with (\ref{ckdot}) and $2\varepsilon \lambda
=1-\eta $, yields%
\begin{eqnarray}
\dot{c}_{k} &=&\frac{-k\left( k+1\right) }{2\lambda }\left( (1-\eta )c_{k}+%
\frac{2\eta }{1+k}c_{k}-\frac{1}{k}\left( \mathbf{c}\ast \mathbf{c}\right)
_{k-1}\right)  \notag \\
&<&\frac{-k\left( k+1\right) }{2\lambda }\frac{L_{k}(\eta )}{R_{k}(\eta )}%
c_{k}~  \label{cdotk}
\end{eqnarray}%
where 
\begin{equation}
L_{k}(\eta )=\sum_{n=1}^{D_{k}+1}l_{k,n}\eta ^{n}=\left( 1-\frac{k-1}{k+1}%
\eta \right) R_{k}(\eta )-1~,  \label{L}
\end{equation}%
is a polynomial in $\eta $ of degree $D_{k}+1$ whose coefficients (with $%
p_{k,D_{k}+1}=0$)%
\begin{equation}
l_{k,n}=p_{k,n}-\frac{k-1}{k+1}p_{k,n-1}~,  \label{l}
\end{equation}%
the first few of them are positive. By (\ref{pkn}), 
\begin{eqnarray*}
p_{k,n}-\frac{k-1}{k+1}p_{k,n-1} &=&\left( 1-\frac{k-1}{k+1}\frac{N_{k}-n}{%
D_{k}-n+1}\right) p_{k,n} \\
&=&2\left( k+1-2n\right) \frac{1}{k+1}\frac{p_{k,n}}{D_{k}-n+1}>0
\end{eqnarray*}%
iff $n<(k+1)/2$, which is not an empty set for $k>2$ as one can see for $k=3$
($L_{3}(\eta )$ has degree $D_{3}+1=4$): $l_{3,1}$ is strictly positive, $%
l_{3,2}=0$ and $l_{3,3},~l_{3,4}<0$.

By (\ref{L}), (\ref{Rk}) and since $R_{k}$ attains its minimum and maximum
values at $\eta =0$ and $\eta =1$, resp., we have 
\begin{equation}
L_{k}(0)=L_{k}(1)=0~.  \label{LL}
\end{equation}%
which, together with and $\eta ^{n^{\prime }}<\eta ^{n}$ if $n<n^{\prime }$,
imply that 
\begin{equation}
L_{k}(\eta )>0\ ,\qquad \text{for }0<\eta <1.  \label{Lk}
\end{equation}

\begin{figure}[tbp]
\centering\includegraphics[scale=0.55]{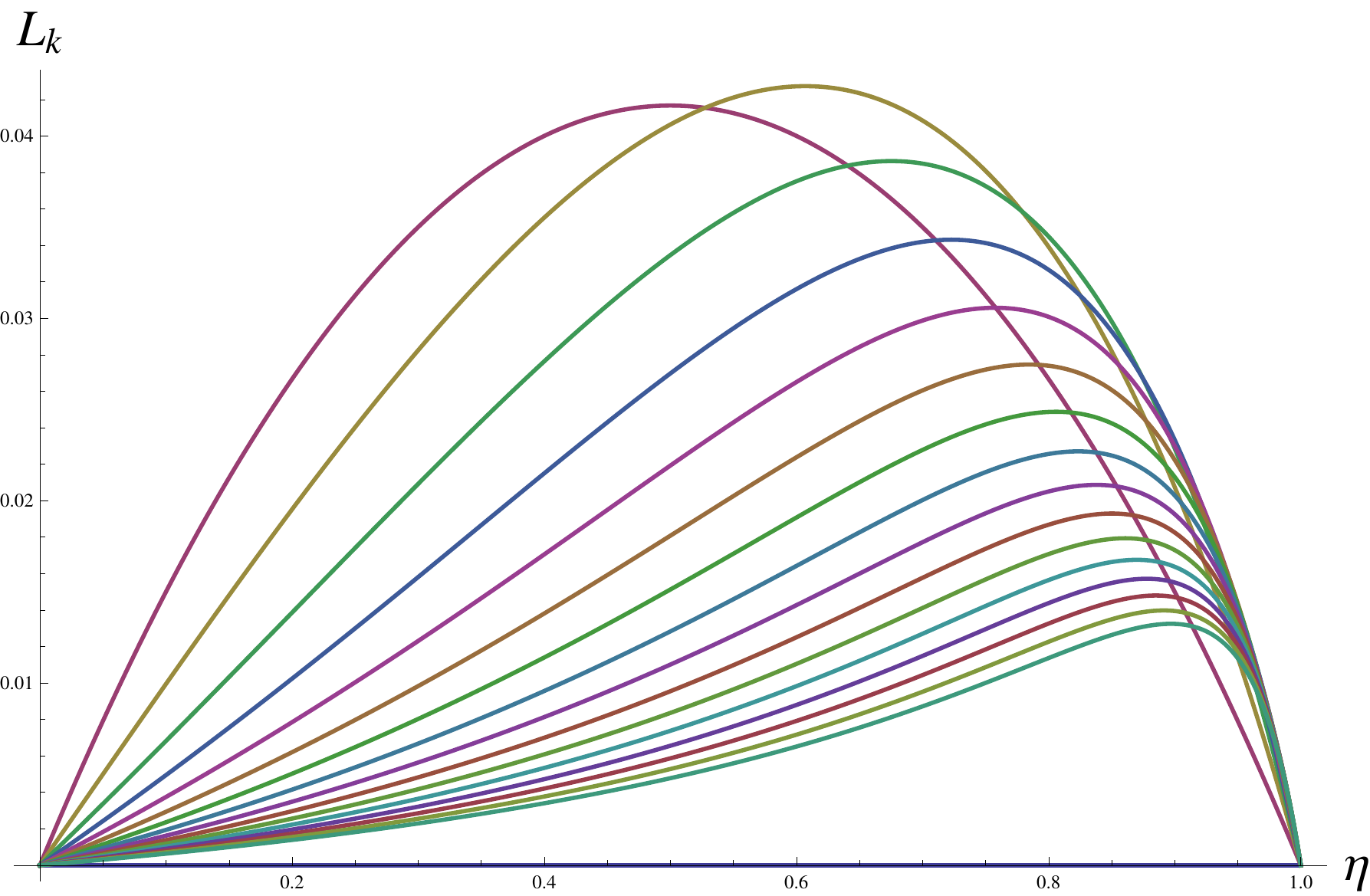}
\caption{$L_{k} $ as a function of $\protect\eta $ for several $k$'s.}
\label{figL}
\end{figure}

(\ref{cdotk}) and (\ref{Lk}) together imply that $c_{k}(t)$, for each $k\geq
2$, is monotone decreasing in $t$; (\ref{LL}) and (\ref{Lk}) are not
faithful for large $\eta $ and can be improved from equation (\ref{cdotk})
in the limit $t\rightarrow 0$. In Fig. \ref{figL} we plot $L_{k}(\eta )$ for
several values of $k$.

So far $c_{k}(t)<c_{k}(t^{\prime })$ has been established for $0\leq
t<t^{\prime }\leq t_{0}$ and $k\geq 2$ where $t_{0}$ is arbitrarily large.
As $t$ goes to $\infty $, $\eta \rightarrow 1$, $\lambda \rightarrow
(2\varepsilon )^{-1}$ and equation (\ref{ckdot}), reads%
\begin{equation*}
\dot{c}_{k}(\infty )=-\varepsilon \left( k+1\right) \left( kc_{k}(\infty
)-\left( \mathbf{c}(\infty )\ast \mathbf{c}(\infty )\right) _{k-1}\right) ~.
\end{equation*}%
The sequence $\mathbf{\dot{c}}(\infty )=\left( \dot{c}_{k}(\infty )\right)
_{k\geq 0}$ vanishes iff%
\begin{equation*}
c_{k}(\infty )=\frac{1}{k}\left( \mathbf{c}(\infty )\ast \mathbf{c}(\infty
)\right) _{k-1}=\frac{1}{k}\sum_{j=0}^{k-1}c_{j}(\infty )c_{n-1-j}(\infty
)~,\qquad k\geq 0
\end{equation*}%
and this holds only for $\mathbf{c}(\infty )\equiv 1$. Estimate (\ref{ckt})
is, however, sharp enough for $t$ close to $\infty $; Equations (\ref{cdotk}%
), (\ref{Lk}) imply that $\dot{c}_{k}(t)<0$ for all $t\geq 0$ and, as $%
c_{k}(t)$ is bounded from below, $\lim_{t\rightarrow \infty }c_{k}(t)=1$,
concluding the proof of Proposition \ref{contra}.

\hfill $\Box $

\end{document}